\documentclass[manuscript]{aastex}
\usepackage{emulateapj5}
\usepackage{lscape}

\slugcomment{}

\shorttitle{SQ dust emission and star formation}
\shortauthors{Natale et al.}

\begin{document}


\title{Dust emission and star formation in Stephan's Quintet}


\author{G. Natale\altaffilmark{1}, R. J. Tuffs\altaffilmark{1}, C. K. Xu\altaffilmark{2}, C. Popescu\altaffilmark{3}, J. Fischera\altaffilmark{4}, U. Lisenfeld\altaffilmark{5}, N. Lu\altaffilmark{2},P. Appleton\altaffilmark{6}, M. Dopita\altaffilmark{7}, P.-A. Duc\altaffilmark{8},
Y. Gao\altaffilmark{9}, W. Reach\altaffilmark{10}, J. Sulentic\altaffilmark{11}, M. Yun\altaffilmark{12} }






\altaffiltext{1}{Max Planck Institute f\"{u}r Kernphysik, Saupfercheckweg 1, D-69117 Heidelberg, Germany; giovanni.natale@mpi-hd.mpg.de, richard.tuffs@mpi-hd.mpg.de}
\altaffiltext{2}{Infrared Processing and Analysis Center, California Institute of Technology 100-22, Pasadena, CA 91125, USA}
\altaffiltext{3}{University of Central Lancashire, Preston, PR1 2HE, UK}
\altaffiltext{4}{Canadian Institute for Theoretical Astrophysics, University of Toronto, 60 Saint George Street, Toronto, ON, M5S 3H8, Canada.}
\altaffiltext{5}{Department de F\'{\i}sica Te\'{o}rica y del Cosmos, Universidad de Granada, Granada, Spain}
\altaffiltext{6}{NASA Herschel Science Center, IPAC, California Institute of Technology, Pasadena, CA 91125, USA}
\altaffiltext{7}{Research School of Astronomy \& Astrophysics, The Australian National University, Cotter Road,Weston Creek, ACT 2611, Australia}
\altaffiltext{8}{Laboratoire AIM, CEA/DSM-CNRS-Universit\'{e} Paris Diderot, Dapnia/Service d’Astrophysique, CEA-Saclay, 91191 Gif-sur-Yvette Cedex, France}
\altaffiltext{9}{Purple Mountain Observatory, Chinese Academy of Sciences, 2 West Beijing Road, Nanjing 210008, China}
\altaffiltext{10}{Spitzer Science Center, IPAC, California Institute of Technology, Pasadena, CA 91125, USA}
\altaffiltext{11}{Instituto de Astrof\'{\i}sica de Andaluc\'{\i}a, CSIC, Apdo. 3004, 18080, Granada, Spain}
\altaffiltext{12}{Department of Astronomy, University of Massachusetts, Amherst, MA 01003, USA}


\begin{abstract}
We analyse a comprehensive set of MIR/FIR observations of Stephan's Quintet (SQ), taken with the Spitzer Space Observatory. Our study reveals the presence of a luminous ($L_{IR}\approx4.6\times 10^{43}~{\rm erg/s}$) and extended component of infrared dust emission, not connected with the main bodies of the galaxies, but
roughly coincident with the X-ray halo of the group. We fitted the inferred dust emission spectral energy distribution of this extended source and the other main infrared emission components of SQ, including the intergalactic shock, to elucidate the mechanisms powering the dust and PAH emission, taking into account collisional heating by the plasma and heating through UV and optical photons. Combining the inferred direct and dust-processed UV emission to estimate the star formation rate (SFR) for each source we obtain a total SFR for SQ of $7.5~{\rm M_\odot/yr}$, similar to that expected for non-interacting galaxies with stellar mass comparable to the SQ galaxies. Although star formation in SQ is mainly occurring at, or external to the periphery of the galaxies, the relation of SFR per unit physical area to gas column density for the brightest sources is similar to that seen for star-formation regions in galactic disks. We also show that available sources of dust in the group halo can provide enough dust to produce up to $L_{IR}\approx10^{42}~{\rm erg/s}$ powered by collisional heating. Though a minority of the total infrared
emission (which we infer to trace distributed star-formation), this is several times higher than the X-ray luminosity of the halo, so could indicate an important cooling mechanism for the hot IGM and account for the overall correspondence between FIR and X-ray emission. We investigate two potential modes of star-formation in SQ consistent with the data, fuelled either by gas from a virialised hot intergalactic medium continuously accreting onto the group, whose cooling is enhanced by grains injected from an in-situ population of intermediate mass stars, or by interstellar gas stripped from the galaxies. The former mode offers a natural explanation for the observed baryon deficiency in the IGM of SQ as well as for the steep $L_{\rm X}$--$T_{\rm X}$ relation of groups such as SQ with lower velocity dispersions.
\end{abstract}


\section{Introduction} 
Physical processes occurring in the environments of groups of galaxies
play a fundamental role in determining the star formation history of the Universe. Galaxy groups are associated with intermediate mass dark matter haloes (DMH) which occupy a pivotal position in the formation of structures, acting
as a centre of aggregation of lower mass DMHs and their associated galaxies
while being the building blocks for the most massive clusters of galaxies that form at later epochs (\citealt{Springel05}). In the present universe about 50 percent of all stellar mass is contained within groups of total mass higher than $log_{10}M/M_\odot \gtrsim12.5$ (\citealt{Eke05}). The group environment affects the modality in which baryonic gas in the intergalactic medium (IGM) is being converted into stars. Whereas at early epochs gas fuelling of galaxies is thought to proceed via cold gas accretion in low mass DMHs,
this mechanism is predicted to be inhibited by the higher virial temperature of the IGM in high mass DMHs hosting galaxy groups (\citealt{Dekel06}). Furthermore, because of the high densities, galaxy-galaxy and galaxy-IGM interactions should become effective in removing interstellar gas from galaxies in groups, leading ultimately to a quenching of star formation in the already existing galaxies which have fallen into the groups. On the other hand, these same interactions lead to a chemical enrichment as the IGM becomes mixed with the stripped galaxian ISMs which, potentially, could enhance the cooling of the IGM and its ability to accrete onto existing galaxies and to form new star forming systems. The relative importance of all these processes in determining the star formation activity which is observed to be taking place in groups is an open question. 

The Stephan's Quintet compact group of galaxies (SQ) \footnote{Among the five galaxies that were observed for the first time by \'{E}douard Stephan in 1877, NGC 7320 was later found to be a foreground dwarf galaxy. A nearby sixth galaxy, NGC 7320c, shows radial velocities compatible with SQ. Therefore the group is still a quintet even if the original definition has changed} presents a natural laboratory with which all these phenomena affecting star formation in groups can be studied. 
As shown by \cite{T05} (T05) and \cite{Osul09} (OS9), this group presents a diffuse halo of X-ray emission extending in radius at least as far as $40~{\rm kpc}$ with the bulk of the gas radiating at temperatures $\approx 6\times10^6{\rm K}$. Assuming hydrostatic equilibrium, this indicates a dark matter halo mass of $\approx10^{12}{\rm M_\odot}$, intermediate between galaxies and clusters. The metal abundance of this hot gas is rather poorly constrained, consistent with a primordial and/or galaxian origin.

One galaxy, NGC 7318b, is apparently unbound, entering the group at a high relative velocity of $\approx 1000~{\rm km/s}$ and colliding with the group IGM, as evidenced by a $\sim 32~{\rm kpc}$ north-south long ridge prominent in radio continuum (\citealt{Xu03}), optical line emission (\citealt{Xu99}), X-ray (\citealt{T05}) and recently also powerful mid-infrared rotational hydrogen lines (\citealt{A06}; \citealt{Cluv10}). The galaxies of the group present extended tidal tails that have been used to constrain their recent interaction history (\citealt{Moles98}, \citealt{S01}). Potential consequences of these interactions is the presence of an AGN in the galaxy NGC 7319 (\citealt{Huch82}) and the presence of neutral and molecular gas located mainly outside the galaxies (\citealt{Wi02}, \citealt{Lis02}). UV observations with the GALEX satellite have provided a detailed picture of the unobscured component of recent star-formation in the group, including the tidal features (\citealt{Xu05}). 

In the context of understanding the physical processes controlling star formation in groups in general and in SQ in particular the FIR spectral region contains crucial information. 
Specifically, IR emission from dust is fundamental for an understanding of the amplitude and distribution of star formation since it traces the luminosity of young massive stars highly obscured by their parent molecular clouds at UV and optical wavelengths. The first infrared image of SQ, taken by the ISOCAM instrument on board the Infrared Space Observatory revealed the presence of
a starburst, SQ A, in the IGM (\citealt{Xu99}), possibly triggered by the ongoing collision between the intruder galaxy and the IGM (\citealt{Xu03}). The later GALEX imaging of SQ (\citealt{Xu05}) also detected SQ\,A in the UV. Overall, the distribution of UV emission in SQ measured by GALEX shows that most of the recent unobscured star formation has preferentially occured in the periphery regions or in the IGM. Further evidence for this is the discovery using HST of widespread young star clusters distributed over the tidal debris and surrounding area by \cite{gall06}. It is important to check whether this apparent shift from the
main disk of the galaxies (where most star formation occurs in isolated galaxies, see e.g. \citealt{Leon08})
is also shown in the obscured component of star formation, and
how the pattern of total star formation in the IGM is related to
the morphological distribution of gas in the different temperature 
ranges. 

A further motivation for studying SQ in the infrared are theoretical studies by \cite{Dwek90} and  \cite{Mon04} which have predicted that even small amounts of dust in the hot virialised IGM could provide an important cooling mechanism via inelastic gas-grain collisions, with the radiation appearing in the FIR. This has prompted searches for a FIR counterpart to the X-ray emitting intracluster medium in several rich clusters (\citealt{Stickel98}, \citealt{Stickel02}, \citealt{Bai07}, \citealt{Kita09}) which however have thus far yielded no unambiguous detection.
This may be attributed to the very low abundance of grains predicted on the basis of realistic estimates of sources and sinks of grains in the IGM (\citealt{Popescu00}), the expected similarity of the FIR colors of the collisionally heated dust emission component with photon-heated diffuse dust in foreground cirrus (\citealt{Popescu00}), and the problem of spatial confusion with star forming galaxies in the cluster (\citealt{Quill99}). SQ is a good object to search for this phenomena since its angular size is small enough for accurate photometric measurements of extended emission and, at the same time, it is possible to separate emission from discrete star-forming sources, such as the constituent galaxies and objects like SQ A. Finally, the detection of dust FIR emission, combined with gas mass measurements to obtain the dust to gas ratio, is a sensitive way of probing the metallicity, and hence the origin, of the various gas components seen in the group in situations where optical nebular line diagnostic are weak or absent. 

The most extensive previous study of the FIR emission from SQ was that by \cite{Xu03} in the wavelengths range $11$ to $100{\rm \mu m}$ using the Infrared Space Observatory. Apart from strong detections of the AGN galaxy NGC 7319 and the foreground galaxy NGC 7320, significant detections were obtained in the FIR of the intergalactic star formation region SQ A and the source SQ B located on the tidal feature associated with NGC 7319. Intriguingly the ISO images also hinted at the presence of FIR emission associated with the X-ray emitting shock ridge, which was argued to be evidence for collisionally heated dust embedded in the hot shocked gas. 

In this paper we present deep imaging of the FIR emission from SQ taken with the Spitzer Space Telescope. These data have superior angular resolution and sensitivity compared to the previous ISO study and moreover extend the wavelength coverage longwards to $160{\rm \mu m}$, thus providing much deeper investigation of any cold dust components, embracing the expected spectral peak of the emission from photon- and collisional-heated emissions. The data is used to characterize the pattern of star formation in SQ and to investigate possible infrared counterparts of the X-ray emitting structures. The Spitzer observations and data reduction are described in Sect. 2, where we also collate multiwavelength data needed for our analysis. In Sect. 3 we described the morphology of the dust emission of SQ in relation to the multiwavelength data set and also describe a novel fitting technique to the lower angular resolution FIR maps which we use to separate the emission from the main emitting structures. After extracting photometry from these structures across MIR/FIR range in Sect. 4, we fit the MIR/FIR SEDs with models in Sect. 5 to elucidate the mechanisms powering the observed dust emission of the sources in SQ. This information is then used to quantify star formation rates in Sect. 6 where we also quantify corresponding gas reservoir available to fuel the star formation. In Sect. 7 we quantitatively discuss the nature of star formation in and outside the constituent galaxies of SQ, together with the related question of the extent to which collisional heating plays a role in determing the thermodynamic properties of the IGM in SQ. 
In this paper, we assume a distance from the group equal to 94Mpc, corresponding to a systemic velocity of $v=6600~{\rm km/s}$ and assuming $H_o=70 {\rm km/s/Mpc}$. 

\section{Observations and Data Reduction}
In this work we have used Spitzer maps from two Guest Observer (GO) programs making use of continuum data at $24$, $70$ and $160{\rm \mu m}$ from the MIPS instrument \citep{Rieke04}. From the GO \#40142 (PI: Appleton, P.) we took the $24{\rm \mu m}$ map and from the GO \#3440 (PI: Xu, K.) the $70$ and $160 {\rm \mu m}$ maps. In addition, an image of SQ at $8{\rm \mu m}$, taken with the IRAC instrument \citep{Fazio04} and downloaded from the Spitzer Science Center (SSC) archive, was also used. In the following we provide details of the data reduction for the Spitzer FIR maps. The $24{\rm \mu m}$ map has been already presented by \cite{Cluv10} and we refer to that paper for technical details about the data preparation. 

\subsection{MIPS $70{\rm \mu m}$ and $160 {\rm \mu m}$ data reduction} 
The MIPS $70{\rm \mu m}$ and $160{\rm \mu m}$ data are from Spitzer pipeline version S11.0.2. In the pipeline default mosaic images, there are a few residual instrumental artifacts that are particularly noticeable at $70{\rm \mu m}$. Because of this, these pipeline images were not used in this work. In their place, new images were used in which the instrumental artifacts were reduced by performing additional data reduction steps on the ``basic calibrated data'' (BCD) frames. 

Specifically, for the $70{\rm \mu m}$ data, we observed an overall signal drift in time. To remove this, we masked out those pixels of each of the 468 non-stim
BCD frames that are within a radius of 125" of the SQ center (RA=339.0181d, DEC=33.969183d; J2000).  For each BCD frame, a median was calculated from all unmasked pixels.  These medians were plotted as a function of the BCD frame index (1 to 468). The resulting plot shows a clear discontinuity at frame index 313.  We fit the sections prior to and post this discontinuity separately with a cubic spline function of order 1 or 2. The two fitted curves were connected to form one curve covering all the BCD frames.  After being normalized by its mean, this curve was divided into the index-ordered BCD frames to remove the signal drift in time.  The next step was to create a sky flat image by median filtering only unmasked data for a given detector pixel.  The resulting sky flat image was normalized by its mean, and
subsequently divided into each unmasked BCD frames. Finally, we used Spitzer MOPEX tool to mosaic these improved BCD frames into our final image used in this paper.  With a much flatter sky background, our own mosaic image is significantly better than the pipeline counterpart. 

For the MIPS $160 {\rm \mu m}$ data, a similar procedure was used with all 522 non-stim BCD frames. In this case, the detector signal drifts in time differ significantly among individual readout modules. As a result, our signal drift removal was attempted on per readout module basis. 

Background subtraction has been performed by fitting a tilted plane to the maps, after masking a large area covering the main group and, in the case of the $160{\rm \mu m}$ map, the galaxy NGC 7320c (that lies outside the field of view shown in Fig. \ref{SQ_mwave}). The total area used to estimate the background is about $50\%$ of the entire maps at both wavelengths. These areas are well outside the group emission, allowing good background measurements. 

\subsection{Collected multiwavelength data}
In this paper we have also made use of a large set of multiwavelength data to derive physical quantities or for morphological comparison with the emission seen on the infrared maps. Specifically we utilized the SDSS maps of SQ, obtained from the SDSS Data Archive (data release 7), the GALEX FUV map, which have been presented in \cite{Xu05}, the XMM-NEWTON soft X-ray map from \cite{T05}, the VLA $21~cm$ line map from \cite{Wi02}, the $H\alpha$ maps from \cite{Xu99}, the IRAC $3.6\mu m$ map, obtained from the SSC archive, and the CO radio observations presented in \cite{Lis02}.  

\section{Morphology of dust emission in SQ}
\label{qual_descr}
In Fig. \ref{SQ_mwave} (lower panel) we show all the Spitzer broad bands (from IRAC and MIPS) where the signal is dominated by dust emission\footnote{Throughout this paper all the $8{\rm \mu m}$ images presented have had the stellar component of the emission subtracted using the relation $F_\nu(8{\rm \mu m,dust})= F_\nu(8{\rm \mu m})-0.232F_\nu(3.6{\rm \mu m})$ (\citealt{Hel04}). The $8{\rm \mu m}$ Spitzer band also contains emission from rotational hydrogen lines which however we estimate in Sect. 5 to be unimportant in relation to the PAH and dust continuum emission.}: $8{\rm \mu m}$, $24{\rm \mu m}$, $70{\rm \mu m}$ and $160{\rm \mu m}$. These can be compared with the upper panel showing the SDSS r-band map, the GALEX FUV map, the VLA radio 21 cm line map and the XMM NEWTON soft X-ray map. On each map, crosses identify the galaxy centres. 

The IR morphology of the galaxies seen in the IR maps is markedly different from the optical morphology. This is true not only for the early type galaxies but also for the late type galaxies which exhibit remarkably little infrared emission from the main bodies of the galaxies. The only exception is the emission from the foreground galaxy NGC 7320 which is quite symmetrical in all the Spitzer bands, having a filled disk of emission commonly seen in local universe field galaxies, compatible with the optical/UV appearance not showing any sign of interactions (compatible with its not being a member of SQ). 

The infrared emission from NGC 7319 is dominated by an unresolved nuclear source, presumably from the Seyfert 2 AGN, that is particularly prominent at $24$ and $70{\rm \mu m}$. At $8{\rm \mu m}$ one can also clearly see emission from the disk of the AGN host galaxy.  As can be seen from the MIR maps, dust emission from compact star formation regions are detected all over the group and especially on the elongated features of  the intruder galaxy NGC 7318b which are most prominently delineated in the UV. Here the similarity is strongest between the UV and the $8{\rm \mu m}$ band though the most prominent discrete sources are also clearly seen at $24$ and $70{\rm \mu m}$. The most prominent such MIR/FIR source is SQ A, the star formation region located to the north of NGC 7318b, already detected by ISO, which can also be seen at $160{\rm \mu m}$. From optical spectra (\citealt{Xu03}) and radio observations (\citealt{Lis02},\citealt{Wi02}) it is known that star formation in this region is associated with gas at radial velocities corresponding to both the intruder galaxy and the IGM of the group. Several further compact star formation regions are located on the southern arms of NGC 7318b. The brighter sources, HII SE and HII SW, are also detected on the $70{\rm \mu m}$ map but not clearly seen on the lower resolution $160{\rm \mu m}$ map. Two other bright MIR/FIR emitting regions detected on the Spitzer maps are SQ B, a star formation region located on the optical ``young '' tidal tail (see \citealt{S01}), and a source, HII N, located about $40''$  towards the north of the AGN galaxy. 

There is no clear morphological counterpart in the infrared to the shock region, defined here by the ridge of emission that can be seen on the X-ray map. Nevertheless the $70{\rm \mu m}$ map and more particularly the $160{\rm \mu m}$ map show enhanced emission towards the peak of the X-ray emission. The ratio of the MIR $8$ and $24{\rm \mu m}$ emission to the $160{\rm \mu m}$ emission at the same position appears rather low compared to the discrete sources associated with star formation such as SQ A. 
A previously undetected feature is an extended FIR emission component, spatially coincident with the main part of the group X-ray halo (as defined in \citealt{T05}). The appearance of the corresponding MIR emission on the higher resolution $8$ and $24{\rm \mu m}$ images suggests that the extended FIR emission may at least in part be clumpy rather than uniform, possibly indicating the presence of faint star formation regions far away from the centers of the galaxies. 

In order to quantify the morphology and brightness of the FIR emission seen towards the X-ray emitting halo and shock regions, it is necessary to subtract from the FIR maps the most prominent discrete sources associated with star formation regions and galaxies. To do this we devised a FIR source fitting technique which we describe in subsection \ref{map_fit_par}. This source fitting technique also serves to fix the photometry and the extent of discrete sources, information that  cannot be directly extracted from the maps due to the unknown level of mutual confusion.

\subsection{The FIR map fitting technique}
\label{map_fit_par}

The fitting technique models a preselected set of the most prominent discrete sources as a sum of elliptical gaussian convolved with the PSF. Seven parameters are calculated for each source: amplitude, the peak coordinates, the two gaussian widths, the axis rotation angle and the local background (included to avoid removal of any diffuse emission components).
The fit is performed simultaneously for the 10 brightest sources seen on the $70{\rm \mu m}$ maps, (see Fig. \ref{70mfit}): five compact sources (SQ A, HII SE, HII SW, SQ B, HII N), two sources to fit the emission from the AGN galaxy NGC 7319 (one for the central emission and one for a peripherical star formation region visible after the removal of the first component), two for the fit of the foreground galaxy NGC 7320 (one for the fit of the diffuse emission and one for a compact source) and one to fit the emission peaked in the middle of the shock region.
The fit to the $70{\rm \mu m}$ map is performed first, keeping all the fitting parameters as free variables. This is followed by a constrained fit to the $160{\rm \mu m}$ map in which the relative position of the sources are fixed to the values obtained at $70{\rm \mu m}$. In addition, a further constraint was that we kept the same shape and axis orientation for the five compact sources, as inferred by the $70{\rm \mu m}$ fit, allowing only a size change (to take into account the potentially more extended distribution of cold dust emission). This strategy was adopted because the higher resolution $70{\rm \mu m}$ map places the strongest constraints to the position and morphology of sources in the FIR. We did not use the highest resolution $8\mu m$ map for this purpose since, whereas the emission detected at $70$ and $160\mu m$ is produced by the same kind of solid dust grains, the emission at 8 $\mu m$ is dominated by line emission from PAH molecules.
Model images from the fitting technique are shown in Figs. \ref{70mfit} and \ref{160mfit}. The best fit parameters and the inferred total flux densities are shown in Table \ref{FIR_sources}. At both $70{\rm \mu m}$ and $160{\rm \mu m}$ the best fit model images are remarkably similar to the original maps. On the ``deconvolved'' maps in each figure, it is possible to see the contribution from each gaussian to the final map. Interestingly enough, the source at the position of the shock is much more predominant at $160{\rm \mu m}$ than at $70{\rm \mu m}$, confirming the original impression that the emission in the shock region is brighter at $160{\rm \mu m}$. It is also noteworthy that the position angle of the model source at $160{\rm \mu m}$ is aligned with the north-south orientation of the X-ray emitting ridge whereas at $70{\rm \mu m}$ no such alignment is apparent. At both $70$ and $160{\rm \mu m}$ the east-west width of the fitted elliptical gaussians at the shock position ($FWHM\approx 60''$) is larger than that of the X-ray shock ridge ($FWHM\approx 20''$). This indicates that the integrated emission is not necessarily entirely composed of emission from the shock ridge. 
A full description of the source fitting technique is given in Appendix \ref{ap1}.

\subsection{The FIR residual maps}
\label{fir_res_map_par}
To understand how the emission peaked in the shock region and the extended emission are distributed on the maps, we created FIR residual maps where the emission from all the sources fitted by the FIR map fitting technique, with the exception of the source associated with the shock, have been subtracted. The contours of these FIR residual maps are shown in Fig. \ref{resmaps} overlaid on HI, X-ray and FUV maps.
As one can see, the emission on the FIR residual maps is uncorrelated and perhaps even anticorrelated with the HI distribution but well correlated with the soft X-ray flux. The FIR emission, as already seen on the original maps, peaks in the middle of the shock region and its overall extent is similar to the X-ray halo emission (similar to what was seen in the low excitation pure rotational lines of $H_2$ by \citealt{Cluv10}). At first sight this finding supports the idea that collisional heating is producing the observed FIR emission. However a large part of the residual FIR emission covers areas emitting significant luminosity at UV wavelengths. The presence of these radiation sources complicates the interpretation of the dust emission in the shock region as well as for the extended emission (see Sect.\ref{discuss_sec}).   

\section{SQ Spitzer map photometry}
\label{phot_section}
Precise photometry of all these different emitting regions in SQ is required in order to elucidate the physical mechanisms that power dust emission and the related scientific implications. In the FIR the photometry is derived from the source fitting procedure (for the discrete sources) and from the residual maps (for the shock and the extended emission components). In the MIR the high resolution of the MIR $8$ and $24{\rm \mu m}$ maps allows a straightforward extraction of fluxes of the corresponding regions in SQ using aperture photometry. All results are summarized in Table \ref{phot_spitzer}. 

\subsection{Star formation regions and galaxies}
\label{sfr_phot}
The FIR map fitting technique, described previously, allows a precise measure of the flux coming from the compact star formation regions, the AGN galaxy NGC 7319 and the foreground galaxy NGC 7320, all well modelled by convolved elliptical gaussians (although some sources are not detected at $160{\rm \mu m}$). This technique allowed us not only to obtain the total source fluxes but also to derive accurately the source extent. In Fig. \ref{cont_deconv}, as an example, we show the contours of the ``deconvolved'' $70{\rm \mu m}$ and $160{\rm \mu m}$ emission at the position of SQ A overlaid on the $8{\rm \mu m}$ and $24{\rm \mu m}$ maps. The FIR emission has a MIR counterpart that peaks in the areas where FIR is higher. This is generally true for all the fitted FIR sources and it validates the use of  apertures for the photometry of the compact sources at MIR $8$ and $24{\rm \mu m}$ whose sizes are derived by the FIR fitting technique. Specifically, at $8$ and $24{\rm \mu m}$ we used elliptical apertures having the same axial ratio and orientation as the $70{\rm \mu m}$ best fit elliptical gaussian axis and semi-axis lengths equal to $2.17\sigma_{70{\rm \mu m}}$ (this area includes $90\%$ of an elliptical gaussian total flux). We took sizes based on the $70{\rm \mu m}$ map fit because all the compact sources are clearly seen on that map. These apertures are depicted in red in Figs. \ref{aperture8} and \ref{aperture24}. 
The local background for the aperture photometry has been estimated on regions located nearby the central source, where other peaks of emission are not clearly seen (see green circles on Figs. \ref{aperture8} and \ref{aperture24}). In this way we are confident that the background level has not been overestimated due to contamination by surrounding sources. We didn't apply aperture correction for the photometry at $8{\rm \mu m}$ because this is close to unity for the chosen apertures (typical aperture size about 10 pixels), as reported by the IRAC data Handbook. At $24{\rm \mu m}$ the chosen apertures typically delimit the first outer ring of the PSF. In this case we applied an aperture correction equal to 1.16 (MIPS data Handbook). For the MIR photometry of the galaxies NGC 7319 and NGC 7320 we used large apertures covering most of the emission from these objects and we didn't apply aperture corrections because of the large integration area. 

The uncertainties on the MIR photometry are given by the quadratic sum of the following contributions: 1) error on the aperture correction; 2) flux calibration uncertainty; 3) background fluctuations. Since we used elliptical apertures, instead of typical circular apertures, we rather conservatively assumed that the relative error on the aperture correction is 10\% (note that this error is applied only to aperture photometry at $24{\rm \mu m}$). The flux calibration relative uncertainty is equal to 4\% (IRAC and MIPS Handbook) while the error introduced by background fluctuations is derived from the variance of the background mean values in the several areas we selected nearby the sources. The measured fluxes and uncertainties for both MIR and FIR wavelengths are given in Table \ref{phot_spitzer}.

Since we are using different methods to derive the source fluxes on the MIR maps and the FIR maps, it is important to check that these two photometric techniques give consistent results. 
To verify this, we performed the source fitting technique on convolved MIR maps and derived the source fluxes exactly as we did for the FIR maps. Details of this test are shown in Appendix \ref{ap1}. In Fig. \ref{ap_vs_fit24} we plot the ratio of the $24{\rm \mu m}$ fluxes inferred by the fitting technique and by aperture photometry for each source. For all the sources, except for HII SW, the difference between the fluxes obtained using different methods is less than $20\%$. The discrepancy between the fluxes for HII SW arises because of the presence of a compact $24{\rm \mu m}$ source at the nearby nucleus of NGC 7318a\footnote{The presence of a compact X-ray source (see Fig. \ref{SQ_mwave}) and a compact radio source (see \citealt{Xu03}) suggest that this source is a low luminosity AGN.}. This source is included in footprint of the gaussian returned by the fitting procedure but is outside the integration area used for the aperture photometry. 

\subsection{Shock region}
\label{shock_phot_par}

Fig. \ref{EW_prof} show the east-west profiles, along a line passing through the center of the shock ridge, extracted from the FIR residual maps, the X-ray and the VLA 21 cm radio continuum maps convolved to the resolution of the $160 {\rm \mu m}$ map ($FWHM \approx 40 ''$). From these profiles, one can see that the FIR width is significantly larger than the shock ridge width as seen both in the X-ray and Radio. Using the current data it is impossible to say if this discrepancy is due to confusion with fainter unrelated infrared sources or to a systematic change in the width of the emitting region of the shock between the X-ray and the FIR. 
Nevertheless we estimated the flux coming from the shock ridge by fitting the FIR residual maps with a simple two component model: a PSF convolved uniform ridge, used to fit the dust emission in the shock region and whose size ($20''\times80''$) were derived from X-ray data, plus a uniform component. In this procedure the level of the fitted uniform component is influenced by the fact that the actual FIR source is more extended than the X-ray source. Taking into account the ambiguity in identifying all the flux from the more extended FIR emission with the shock, we have assigned extremely conservative flux uncertainties. Specifically, the upper and lower limits defined by the quoted error bars in Table \ref{phot_spitzer} are the fluxes contributed by the uniform emission component underlying the solid angle of the shock ridge (convolved with the PSF) in the cases the uniform emission component has a brightness equal to twice or zero times the value given by the fit. 

To measure the MIR $8$ and $24{\rm \mu m}$ emission from the shock region, we simply integrated the emission in the same rectangular area used before to define the shock ridge component in the fit of the FIR residual maps. The adopted apertures are shown in Figs. \ref{aperture8} and \ref{aperture24}. Before the integration we masked all the areas inside the apertures we used for the photometry of other sources, because the MIR emission in that areas is mainly connected with the corresponding sources that we fitted and subtracted from the FIR maps before the measure of the shock region flux. For the fraction of the masked regions that falls inside the shock region integration area we assumed that the surface brightness is equal to the average brightness on the other regions inside the aperture. The background level has been measured on areas around the rectangular aperture where no peaks of emission are clearly seen. The quoted error on the integrated fluxes is the sum of the contributions due to background fluctuation and flux calibration error. 

\subsection{Extended FIR emission}
\label{fir_res_map_par2}
We have measured the amount of extended flux on the FIR residual maps within a radius of $90''$ from the shock center, an area approximately equal to that covered by the X-ray HALO region as defined in \cite{T05}. We performed this flux measurement in the following way. First we constructed radial curves of growth of the integrated emission on the FIR residual maps starting from the shock center. These curves of growth are shown in Fig. \ref{radial_int_prof}. As one can see, the integrated emission continues to grow somewhat beyond the $90''$ radius. However we considered only the flux within this limit because it can be directly related to the X-ray Halo extent. We took the curve of growth values for the integrated fluxes at $90''$ and we subtracted the contribution from the shock region, estimated in Sect. \ref{shock_phot_par}. Similarly as before, the uncertainties on the fluxes are mainly due to the mutual contamination between shock ridge and extended emission. Therefore we assumed the same conservative errors that we assigned to the shock region FIR fluxes.
For the estimate of the extended MIR emission we have used an analogous method. We constructed radial curves of growth, starting from the shock region center, after having masked all the compact sources and galaxies whose photometry has been described in Sect. \ref{sfr_phot} (note that the shock region is not masked). For the calculation of the MIR curves of growth we took into account the missing areas, those that we masked, assuming that their brightness is equal to the average brightness inside the circular annuli passing through them. The derived curves are shown in Fig. \ref{radial_int_prof}. Exactly as for the FIR measurements, we took the value of the curve of growth at $90''$ and subtracted the emission from the shock region in order to obtain the integrated emission from the extended area corresponding to the X-ray halo. The flux uncertainties, in this case, are derived summing quadratically background fluctuation, calibration errors and the error on the shock ridge flux. 

\section{Modelling the Infrared SEDs}
\label{IF_SED_fit_par}

The spectral energy distribution (SED) of dust emission is determined by the heating mechanism, 
the intensity and color of the radiation fields (for the case of photon heating), the temperature and density of the hot plasma (for the case of collisional heating) and by the amount, size distribution and chemical composition of the emitting grains. Ideally one would solve for the distribution of photon sources, hot plasma and grains using a self consistent radiation transfer analysis to fit the entire X-ray/UV/Optical-MIR/FIR SED for each source, analogous to the treatment of photon heated dust in disk galaxies and starburst galaxies (see \cite{Popescu_malta} for a recent review). However, due to the extra dimension of collisional heating and the unusual geometry of the optical/UV and infrared emission from SQ, this approach would require a very individualized treatment which is well beyond the scope of this paper. We therefore adopt here a hybrid approach fitting just the MIR/FIR SED with superpositions of dust emission templates where each template is appropriate for specific dust emission regions: HII/photodissociation regions (PDR), diffuse photon-powered dust emission, AGN torus emission, collisionally heated dust embedded in hot X-ray plasma. Each of these templates is calculated self consistently in terms of physical input parameters, as described in detail in Appendix \ref{sed_temp_mod}.  In this approach the amplitude of the HII/PDR template, derived from the model of \cite{Dopita05} and \cite{Groves08} and shown in Fig. \ref{agn_templ} (left panel), quantifies the obscured component of on-going star formation, whereas the amplitude of the diffuse dust emission template (shown in Fig. \ref{diff_sed_int}) quantifies dust emission powered by longer range photons to which older stellar populations can also contribute. In the latter case a family of SEDs is calculated according to the strength and color of the radiation field which are both expected to vary according to radiation transfer effects and the relative contribution of young and old stellar populations. A family of SEDs is also calculated for the collisionally-heated dust emission template (Fig. \ref{coll_sed_k}) corresponding to a range of plasma parameters and to different grain size distributions thus accounting for the expected effect of the plasma on the grain size distribution. For both the photon heated and collisionally heated dust emission templates the stochastic emission from impulsively heated grains, which is important to determine the MIR emission, is calculated. In the case of the AGN template SED, we have used the existing self consistent model of \cite{fritz06} (right panel of Fig. \ref{agn_templ}). 

The fits were done by minimizing $\chi^2$ for each source, defined as: 
\begin{equation}
\chi^2=\sum\left(\frac{F_{\nu i}^{\rm model}-F^{\rm obs}_i}{\sigma_i}\right)^2 
\end{equation}
where $F^{\rm obs}_i$ and $\sigma_i$ respectively are the observed flux densities and their associated uncertainties , and $F_{\nu}^{\rm model}$ is the corresponding model prediction which is related to a given theoretical spectra $F_\nu$ through $F_\nu^{\rm model}=KF_\nu$. The color correction $K$ for each Spitzer band is calculated according to the formulas provided by the IRAC and MIPS data handbooks (see Appendix \ref{ap4}). It is particularly important for the $8{\rm \mu m}$ band where the emission spectra varies quickly along the bandwidth because of PAH line emission. The color correction can be up to $80\%$ at $8{\rm \mu m}$, while it is generally less than $10\%$ for the other bands.

\subsection{Star formation regions}
\label{star_form_sed_par}
In this section, we describe the SED fits for the star formation regions SQ A, HII SE, HII SW and SQ B. We fitted all the observed SEDs as the superposition of two components. The first component is the PDR/HII region dust emission template, that helps to fit the emission from warm dust located very close to young stars. The second component is the diffuse photon heated dust emission template that fits the emission from diffuse dust near star formation regions. This diffuse dust is heated by a combination of UV photons escaping from the PDR/HII region and any ambient large scale UV/optical radiation field pervading the region. We performed the fit varying four free parameters: the amplitude of the HII region/PDR SED template (determined by the parameter $F_{\rm 24}^{\rm HII}$, the fraction of $24{\rm \mu m}$ flux contributed by this component); the diffuse dust mass $M_{\rm dust}$; the strength and the color of the diffuse radiation field, determined by two parameters, $\chi_{\rm isrf}$ and $\chi_{\rm color}$ (see Appendix \ref{sed_temp_mod} for details).
As one can notice, there are not degrees of freedom in these fits. Therefore it is not possible to estimate the goodness of the fit (in the sense of the fidelity of the model) from a chi--square test. Nonetheless we estimated the error on the best fit dust masses and total dust luminosities from a multidimensional analysis of $\chi^2$ near to its minimum. The uncertainty on the dust mass is the minimum dust mass variation that gives $\Delta\chi^2=\chi^2-\chi^2_{\rm min}$ values always greater than one, independently from the values of all the other parameters. The total luminosity error bar is determined by the lowest and highest values of the total luminosity in the subspace of fitting parameters determined by $\Delta\chi^2 \leq1$. 

The SED fits are shown in Fig. \ref{SQA_SED} (note that for HII SE and SQ B we used the $160{\rm \mu m}$ fluxes within the $70{\rm \mu m}$ source size to limit contamination from neighbours). Table \ref{sedfitres} shows the best fit parameters, together with the estimate of the uncertainty on the dust mass, the total infrared dust luminosity and the contribution to this total infrared luminosity from optical heating of diffuse dust by the diffuse radiation field, UV heating of diffuse dust by the diffuse radiation field and localized UV heating of dust in PDR/HII regions. The predicted radiation fields needed to account for diffuse emission component are at least as strong as the local ISRF in the Milky Way with a strong variation in color. From the plots one can see that the $24{\rm \mu m}$ emission is dominated by HII region/PDR emission, as expected, while $8{\rm \mu m}$ and FIR points are generally dominated by diffuse emission, in accordance with fully self-consistent radiative transfer models of galaxies (\citealt{Popescu10}, submitted). This is consistent with a picture where MIR emission is produced by regions very close to young stars whilst FIR emission and PAHs emission come from dust illuminated by dilute radiation fields, as observed in galaxy disks (e.g. \citealt{bendo08}) 

\subsection{AGN galaxy NGC 7319}
\label{AGN_fit_par}
The observed SED of the Seyfert 2 galaxy NGC 7319 differs markedly from the SEDs we obtained for the other sources in SQ, for which the power emitted at $24{\rm \mu m}$ is typically a factor ten or more smaller than that observed at $70$ and $160{\rm \mu m}$. For NGC 7319 the amount of energy emitted at MIR wavelengths is comparable to the FIR luminosity. We first tried to fit the SED of this galaxy using the same HII region/PDR plus diffuse emission components as we have done previously for the star formation regions (we didn't include a component of synchrotron emission, originating from the AGN, because an extrapolation of its infrared luminosity from the radio measurements of \cite{Aoki99} gives values which are six orders of magnitudes lower than the observed infrared luminosities). The best fit is shown in the left panel of Fig. \ref{NGC7319_SED}. The fit tries to reproduce the high $24{\rm \mu m}$ flux with the HII region template but, doing so, completely overestimates the $70{\rm \mu m}$ flux. A second attempt has been performed substituting the HII region/PDR template with an AGN torus template (\citealt{fritz06}, see Appendix \ref{sed_temp_mod}). As one can see from the right panel of Fig. \ref{NGC7319_SED}, the observed data are reproduced much better in this case. Of course the limited number of data points does not allow any inference of the physical properties of the dusty torus or provide precise information on the radiation field. However the fit does provide an indication that the majority of the $24{\rm \mu m}$ emission in this galaxy is powered directly by the AGN power law radiation. The fitted parameters and the total dust luminosity derived from the AGN torus plus diffuse emission fit are shown in Table \ref{sedfitres}.  

\subsection{Shock region emission}
\label{shock_sed_fit_par}

The dust emission SED of the shock region contains MIR and FIR emission components which indicate an inhomogeneous structure of the emitting region. As one can see from the middle panels of Fig. \ref{resmaps}, the emission on the FIR residual maps shows a rough correlation with X-ray emission, especially near the center of the shock ridge, suggesting, as mentioned before in Sect. \ref{fir_res_map_par}, that collisional heating 
of dust embedded in the hot ($T\approx 3\times 10^6~{\rm K}$) shocked plasma powers at least some of the observed emission. However the presence of significant $8{\rm \mu m}$ emission rules out the possibility that the observed dust emission is purely collisionally heated because PAH molecules are not expected to survive in a medium shocked by a wave travelling at more than 125 km/s (\citealt{Mic10}). The PAH emission must therefore either be physically unrelated to the shock, coming from another region along the line of sight, or, if associated with the shock, belonging to colder gas phases embedded in the X-ray plasma. The latter scenario would be consistent with the detection of $H_2$ molecular line\footnote{Comparing the surface brightnesses in the $H_2~0-0 S(4)$ and $H_2~0-0 S(5)$ lines, derived from the fluxes given in Table 2 of \cite{Cluv10} (``Main Shock'' region), with the $8{\rm \mu m}$ surface brightness derived from the data given in Table \ref{phot_spitzer} of this paper, we estimate the contamination of the Spitzer $8{\rm \mu m}$ band by $H_2$ line emission to be $\approx10\%$.} (\citealt{A06}, \citealt{Cluv10}) and $H\alpha$ (\citealt{Xu99}) emission from the shock region. Peculiar ratios between the PAH line features have been detected towards the shock, which suggests an enhanced fraction of neutral and large PAHs compared to typical galaxy environments (\cite{guillard10}). 

The multiphase nature of the gas in the shock region implies that there are at least four potential sources of dust emission in this region: 1) diffuse dust collisionally heated in hot plasma; 2) dust in a colder medium and heated by a diffuse radiation field; 3) HII/PDR dust emission from optically thick clouds with embedded star formation regions; 4) cold dust emission from optically thick clouds without embedded star formation. 

In practice, it is not possible to distinguish model prediction for 2) and 4) over the wavelength range of the currently available data, since the only difference would correspond to a very cold emission component from the interior, self--shielded regions of the optically thick clouds, which
will only become apparent at longer submm wavelengths. Therefore, we only consider explicitly here components 1) - 3), ignoring component 4). Furthermore, the four available data points are insufficient to simultaneously fit these three potential sources of dust emission, especially the diffuse photon heated and collisionally heated components, which are both predicted to peak at FIR wavelengths. 
Therefore, we decided to perform the SED fit in two different ways corresponding to the two opposite cases, where collisional heating is either responsible for the entire FIR emission or is completely negligible. We followed this approach in order to understand which mechanism is predominant in powering the diffuse dust emission.  
First, we fitted the shock region SED as a superposition of two components: the HII region/PDR template, to fit MIR emission possibly associated with star formation regions and a collisionally heated dust SED template (see Appendix \ref{dust_coll_code}). This model is appropriate for the case where photon heated diffuse FIR emission is negligible compared to the collisionally heated dust emission. Given the plasma physical properties (fixed by the X-ray emission, as described in Appendix \ref{ap3}), the free parameters in this fit are: the amplitude of the HII region template and the collisionally heated dust mass. We performed these fits for the highest and the lowest densities admitted by the X-ray data, however, the final results are quite similar. In the upper panels of Fig. \ref{shock_sed} we show the best fit obtained for the highest considered particle density $n=0.016~{\rm cm^{-3}}$ and temperature $T=3\times10^6~{\rm K}$. As one can see, the collisionally heated component (dashed curve) is sufficient to reproduce the observed data except the $8{\rm \mu m}$ point which, as already remarked, is too high to be due to PDR/HII regions, requiring a diffuse photon-heated contribution.
The model curve shown (fitted in this case just to the three longer wavelength data points) was  calculated assuming a power law grain size distribution with exponent $k=2.5$, the expected value in the case of equilibrium between dust injection (with a standard $k=3.5$ interstellar distribution; \citealt{Mathis77}) and sputtering of grains in hot plasmas (\citealt{Dwek90}). 
The dust mass required to produce the observed FIR flux densities for the case where collisional heating, rather than diffuse photon heating, is considered is $2\times10^7~{\rm M_\odot}$. 

The second fit was performed using the HII region/PDR template and the photon-heated diffuse dust component. This combination of SED templates represents the case where the entire emission is powered by photons and collisional heating is negligible. Using this combination of SEDs, one can fit all the data points, including the $8{\rm \mu m}$ flux. The best fit curve is shown in the lower panels of Fig. \ref{shock_sed}. In this fit, the diffuse dust component dominates the emission in all the Spitzer bands, even at $24{\rm \mu m}$, where the contribution of PDR/HII regions is normally predominant. The fitted parameters are given in Table \ref{sedfitres}. We note that the diffuse radiation fields needed for the fit are colder than those needed to fit the star formation regions in Sect. \ref{star_form_sed_par}. 

In conclusion, the FIR part of the SED from the shock can be fitted by a cold continuum component either from collisional heating, which is cold due to the limited ambient density of plasma particles and the underabundance of small stochastically heated grains, or from photon heating, which is cold due to the low ambient density of photons coupled with a low UV to optical ratio.
The only real discriminant favouring photon heating would be the $8{\rm \mu m}$ emission which, as a PAH tracer, cannot be explained by collisional heating. However it is not entirely clear, on the basis of the current data, that the $8{\rm \mu m}$ emission really does all originate from the shock region. 

\subsection{Extended emission}
\label{diff_sed_fit_par}
In a completely analogous way as for the shock region emission, we tried to model the observed extended emission SED by including a collisionally heated dust emission component. This is motivated by the fact that the FIR residual maps show the presence of extended emission in the area of the brightest component of the X-ray halo emission (the so called ``HALO'' in \citealt{T05}). In the upper panels of Fig. \ref{diff_sed} we show the SED fit performed adding two components: HII regions plus collisionally heated dust emission. In this fit the adopted plasma physical parameters, derived in Appendix \ref{ap3}, are $n=0.001~{\rm cm^{-3}}$ and $T=6\times10^6~{\rm K}$ and we again assumed $k=2.5$. Using this combination of SEDs, one can roughly reproduce the SED longward of $24{\rm \mu m}$, but the collisionally heated component dominates only the $160{\rm \mu m}$ flux. The dust mass inferred from this fit (constrained, as in the case for the model fit to the shock SED incorporating collisional heating, only by the three longer wavelength points) is $1.0\times10^8~{\rm M_\odot}$.

In the lower panel of Fig. \ref{diff_sed}, we show the fit performed adding HII region and diffuse photon heated emission. As before for the shock region, using these two components one can fit the entire spectra including the $8{\rm \mu m}$ point. In this fit, HII region emission dominates at $24$ and $70{\rm \mu m}$ while the diffuse emission is responsible for the $8$ and $160{\rm \mu m}$ fluxes. The fitted parameters are shown in Table \ref{sedfitres}. As for the shock emission, we note that the extended emission component requires a very cold FIR component from the fit even though, observationally speaking, this is a somewhat less robust conclusion, since it really needs to be confirmed with longer wavelength photometry. 

\subsection{Foreground galaxy NGC 7320}
For the SED fit of the foreground galaxy NGC 7320, we used again the usual combination of HII region plus diffuse photon heated dust emission components. The best fit is shown in the left panel of Fig. \ref{hii_n__SED}. The relative contribution of HII regions to the $24{\rm \mu m}$ flux is reduced compared to the star formation regions, whose SED fits have been shown before. This might be physically understandable since, in this case, we are fitting the emission of a galaxy as a whole, including all the diffuse interstellar medium containing small dust particles whose stochastically heated emission can account for a major part of the total MIR emission (e.g. \citealt{Popescu00sed}). 

\subsection{Infrared source north of NGC 7319 (HII-N)}
We attempted to fit the observed emission for this source with a combination of HII region plus diffuse photon-heated dust emission. However, this source shows a very peculiar MIR to FIR ratio and, as a consequence, we have not managed to fit the entire spectra. As shown in the right panel of Fig. \ref{hii_n__SED}, the FIR points are highly underestimated by the best fit curve. If no systematic errors are present in our measurements, it might be that HII N is a distant background source and the intrinsic colors are heavily redshifted. This possibility is strongly supported by the presence of a red spiral galaxy at the position of HII-N, clearly visible on recently released high resolution HST color maps of SQ (Hubble Servicing Mission 4 Early Release Observations, observers: K.Noll et al., available at http://hubblesite.org/newscenter/archive/ releases/2009/25/image/x/). 

\section{Quantifying star formation rates and gas masses in SQ}

\subsection{Star formation rates}
\label{sfr_par}
The measurements of UV, recombination line and dust emission from sources in SQ can in principle be used to derive star formation rates (SFR), provided proper account is taken of the absorption of the UV/optical photons by dust and subsequent re-emission in the MIR/FIR spectral regimes. Several authors have provided empirically-based relations achieving this for spiral galaxies on scales of kpc. \cite{Calz07} presented an $H\alpha$ luminosity based star formation rate relation: $SFR({\rm M_\odot yr^{-1}})=5.3\times10^{-42}~[L(H_\alpha)_{\rm obs}+(0.031\pm0.006)L(24{\rm \mu m})]$. In this relation the $24{\rm \mu m}$ luminosity is used to estimate the $H\alpha$ luminosity obscured by dust. \cite{Bigiel08} combined GALEX FUV and Spitzer $24{\rm \mu m}$ fluxes to obtain SFR per unit area maps of a sample of spiral galaxies. Like \cite{Calz07}, they used a UV SFR calibration and they used the $24{\rm \mu m}$ flux to measure the obscured UV flux.

Simple application of the Calzetti relation to SQ is hampered by the lack of pure measurement of $H\alpha$ emission driven by star formation. As one can see from Figs. 4 and 5 from \cite{S01}, the shock region $H\alpha$ emission shows a diffuse component that reflects the north-south ridge seen in the soft X-ray regime. This diffuse $H\alpha$ emission cannot be considered for SFR measurements because, as demonstrated by \cite{Xu03} from spectral line analysis, the $H\alpha$ emission in the shock region is dominated by shock-excitation rather than star formation. In addition, it is not clear that the semi-empirical relations derived for spiral galaxies should apply to the sources in SQ. 

Therefore, we adopted a new approach, using only observational indicators of star formation activity available for all sources in SQ, and utilizing the results of the fits to the dust emission SEDs given in Sect. \ref{IF_SED_fit_par}. Our method to estimate SFRs is based on a UV - SFR calibration\footnote{UV emission traces star formation on time scales of $\sim100{\rm Myr}$ and, therefore, UV derived SFRs could not be appropriate for burst of star formation on smaller time scales.}. Specifically we adopted the calibration by \cite{Salim07}: 
\begin{equation}
 SFR_{\rm UV}=1.08\times 10^{-28} F_{UV}~{\rm M_\odot/yr}
\end{equation}
where $F_{\rm UV}$ is the FUV luminosity density in units of erg/s/Hz which would be observed in the GALEX FUV band in the absence of dust. $F_{\rm UV}$ can be written as: 
\begin{equation}
F_{\rm UV}=F_{\rm UV}^{\rm direct}+F_{\rm UV}^{\rm abs,local}+F_{\rm UV}^{\rm abs,diffuse}
\label{F_UV}
\end{equation}
where $F_{\rm UV}^{\rm direct}$ is the directly observed unabsorbed component of FUV luminosity density, $F_{\rm UV}^{\rm abs,local}$ is the FUV luminosity density locally absorbed by dust in star formation regions and $F_{\rm UV}^{\rm abs,diffuse}$ is the FUV luminosity density absorbed by dust in the diffuse medium surrounding the star formation regions. 

We measured $F_{\rm UV}^{\rm direct}$ for each source by performing aperture photometry on the GALEX FUV map in a completely analogous way to the photometry we performed on the Spitzer MIR maps, including the construction of the curve of growth after masking the galaxies NGC 7319 and NGC 7320 and the star formation regions (see Fig. \ref{radial_int_prof}). The final flux densities, shown in col. 4 of Table \ref{sf_regions_tab}, include the correction for Galactic foreground extinction ($E(B-V)=0.079$, \citealt{Schleg98}; $A(\rm FUV)=8.24*E(B-V)$, \citealt{Wyder07}). Values for the obscured emission components $F_{\rm UV}^{\rm abs,local}$ and $F_{\rm UV}^{\rm abs,diffuse}$ were extracted from the fits to the dust emission SEDs by noting that the total infrared luminosity emitted by dust and powered by UV photons can be written as: 
\begin{equation}
L_{\rm dust,UV}=L_{\rm UV}^{\rm abs,local}+L_{\rm UV}^{\rm abs,diffuse}
\end{equation}
where $L_{\rm UV}^{\rm abs,local}$ is the luminosity of dust emission in star formation regions, dominated by UV photon heating, and $L_{\rm UV}^{\rm abs,diffuse}$ is the part of the diffuse dust luminosity powered by UV photons, respectively tabulated in cols. 6 and 8 of Table \ref{sedfitres}. Since the intrinsic SEDs of the young stellar population are rather flat at UV wavelengths (\citealt{Kennic98}), it then follows that: 
\begin{equation}
F_{\rm UV}^{\rm abs,local}=\frac{L_{\rm UV}^{\rm abs,local}}{\Delta \nu(UV)} 
\end{equation}
and 
\begin{equation}
F_{\rm UV}^{\rm abs,diffuse}=\frac{L_{\rm UV}^{\rm abs,diffuse}}{\Delta \nu(UV)} 
\end{equation}
where $\Delta \nu \approx 1.8\times 10^{15}~{\rm Hz}$ is the UV frequency width. The total obscured UV luminosity density $F_{\rm UV}^{\rm abs}=F_{\rm UV}^{\rm abs,local}+F_{\rm UV}^{\rm abs,diffuse}$ and SFRs are shown in cols. 6 and 7 of Table \ref{sf_regions_tab}. 

As a check of the consistency between our method to derive star formation rates and the $H\alpha--24{\rm \mu m}$ SFR relation of \cite{Calz07}, we used the latter to derive SFR for the compact star formation regions in SQ which are the objects closest resembling galaxies. The $H\alpha$ fluxes have been measured on the interference filter maps published in \cite{Xu99} while the $24{\rm \mu m}$ fluxes are those derived by aperture photometry (Sect. \ref{sfr_phot}). The results are shown in cols. 2 and 3 of Table \ref{sf_regions_tab}. As one can see comparing cols. 3 and 7 of that table, the SFR inferred with our method are consistent with the results found using the Calzetti relation (except for HII SW but in that case the SED fitting was performed without varying $\chi_{\rm color}$ because of the non detection of the source at $160{\rm \mu m}$.)

\subsection{Gas Masses}
In the absence of a homogeneous set of observations of gas tracers for the gas phases of interest, namely HI, H2 and X-ray emitting plasma, we proceeded differently for the several sources in determining gas masses. For the star formation regions SQ A, HII SE, HII SW and SQ B we measured gas mass column densities from the HI map published by \cite{Wi02} and the CO maps of \cite{Lis02}. Specifically we measured the average atomic and molecular hydrogen column density in the observed areas close to the position of the starburst regions. The inferred gas mass column densities are shown in col. 9 and 10 of Table \ref{sf_regions_tab} (note that for some values only upper limits are available and no CO observation are available for HII SE). 

Gas in the shock region is mainly in the form of X-ray emitting plasma and molecular gas (\citealt{guillard09}). From the X-ray luminosities, measured by \cite{T05}, we determine the X-ray gas mass $M_X\approx10^9{\rm M_\odot}$ (see Appendix \ref{ap3} for details). Dividing the total gas mass by the physical area covered by the shock ($330 {\rm kpc^2}$), we obtained the hot gas mass surface density shown in col. 11 of Table \ref{sf_regions_tab}. To determine the molecular gas mass surface density, we measured the average of several observed positions within the shock area on the CO maps by \cite{Lis02} to obtain $\Sigma(H_2)=9~ {\rm M_\odot/pc^2}$. As one can see from Fig. 1 of that paper, most of the observations were performed in the upper part of the shock region. \cite{guillard10} and \cite{A06} found a cold and warm molecular gas surface mass densities in the central parts of the shock region respectively equal to $5~ {\rm M_\odot/pc^2}$ and $3.2~ {\rm M_\odot/pc^2}$. Interestingly enough, the sum of these two values is very close to the cold molecular gas surface density we measured in the upper parts of the shock region. 

For the extended emission, the measurement of the corresponding neutral and molecular gas masses cannot be realistically performed because the extended emission cover irregular parts of the large HI distribution in SQ and there are no CO observations covering the whole group with enough sensitivity to detect a plausible extended molecular hydrogen distribution. The measurement of the corresponding X-ray emitting gas mass is instead rather straightforward since it can be derived by the X-ray Halo luminosity, as for the shock region (see Appendix \ref{ap3}). Dividing the total X-ray gas mass by the projected X-ray halo emission area (a circle of radius $\approx 40~{\rm kpc}$), we determined the hot gas mass surface density shown in col. 11 of Table \ref{sf_regions_tab}. 

\section{Discussion}
\label{discuss_sec}

If one were to view SQ at a greater distance such that the group would appear as a point--like source to Spitzer (i.e. at a redshift of $\gtrsim 0.5 - 2 $, in the main star--forming epoch of the Universe, and also the epoch when galaxy
groups were first forming), one would not regard this as a particularly unusual infrared source. On the basis of the shape of the total dust emission SED, plotted as the black curve in Fig. \ref{global_sed}, the only noteworthy points would be the quite warm MIR colours and moderately high FIR luminosities which would most likely lead our hypothetical observer to conclude that this source had an AGN, possibly combined with a mild starburst. In this he would be at least in part correct, as illustrated in Fig. \ref{global_sed} by the curves representing the constituent emission components from the AGN galaxy (red), from star formation regions (blue) and the emission from X-ray emitting regions (green), where one can immediately see the high relative contributions of the AGN galaxy to the $8{\rm \mu m}$ and $24{\rm \mu m}$ emission, respectively $\sim50\%$ and $\sim70\%$. In the FIR, although the AGN is still the most luminous individual source, there is emission at a comparable level from the combination of the distributed star--formation regions in SQ and the infrared counterparts of the X-ray emitting shock and halo structures; the combined FIR/submm SED resembles that of star--forming galaxies, with an amplitude similar to that of the local starburst galaxy M82.

However our analysis of the spatially resolved structures has 
shown that the characteristics of infrared emission in SQ are the very opposite of a nuclear starburst, with the star--formation activity enhanced in regions far away from the main bodies of the galaxies. In the following Sect. \ref{star_form_disc} and \ref{x_ray_em_disc} we discuss to what extent the nature of this distributed star formation in SQ may differ from the star formation in the disks of individual galaxies, in terms of sources of gas and the star--formation efficiency for the group as a whole, and consider the related issue of the extent to which collisional heating of grains in the IGM of SQ may be cooling the IGM and thus contributing to the fuelling of the star--formation.

\subsection{Star formation in SQ}
\label{star_form_disc}

The results of the SED fitting to the constituent components of SQ
indicate that star formation activity and photons from old stars are the major agents powering the observed global dust emission from the group, supplemented by photons produced by the accretion flows in the AGN and possibly, in the case of the X-ray emitting regions, by collisional heating. In our quantitative
discussion of star formation activity in SQ we will adopt an initial working hypothesis that the collisional heating mechanism is minor compared with photon heating in the X-ray sources. We will scrutinized this hypothesis in detail in Sect. \ref{x_ray_em_disc}, where we discuss physical constraints of the fraction of the infrared emission that can be collisionally powered. 

Under this working hypothesis we can gather together the information from Table \ref{sf_regions_tab} to obtain a total global star formation rate of $7.5~{\rm M_\odot/yr}$ for SQ in its entirety.
This global star formation rate does not seem particularly discrepant
from that typically found for most of the galaxies of similar mass in the local Universe
that are clearly not so strongly interacting as SQ galaxies. Using the empirical relation 
between galaxy stellar mass and SFR for field galaxies 
shown in Fig. 17 of \cite{Brinch04} we can estimate the typical
SFR of galaxies having the same stellar mass of SQ galaxies, of order of
$\approx 10^{11}~{\rm M_\odot}$ (see table \ref{sdss_tab}). For this value of the stellar mass,
the mode of the distribution is at $SFR\approx 1~{\rm M_\odot/yr}$. Therefore, since we estimated the SFR in a field containing four galaxies, the expected SFR would be $\approx4{\rm M_\odot/yr}$, comparable to the measured value.
Thus it seems that the star formation efficiency of SQ in relation to field galaxies
is largely independent of whether the gas is inside or outside the main stellar disk. 
In fact, of the total global star formation rate of $7.5~{\rm M_\odot/yr}$ for SQ,
$2.2$ and $5.3~{\rm M_\odot/yr}$ can respectively be ascribed to the SED components for star formation regions and X-ray sources in Fig. \ref{global_sed}. This is a very remarkable result indicating, as it does, that the bulk of star formation activity in SQ is apparently associated with X-ray emitting structures,
occurring far away from the galaxy centers, either at the peripheries of the galaxies or in the intergalactic medium. Furthermore, the total extragalactic SFR is well in excess of the SFRs of the
previously studied individual examples of extragalactic compact star formation regions, SQ A and SQ B. 

To quantify the efficiency of star formation in the various components of SQ in relation
to that of the disks of individual galaxies, we plotted in Fig. \ref{plot_sfr_sgas} the SFR per unit area for dust emission sources in SQ as a function of the gas mass surface density. On the same diagram, as reference, we plotted the relations found by \cite{Kennic07} (K07) and \cite{Bigiel08} (B08) for star formation regions inside nearby spiral galaxies. This figure shows a wide
range of star formation efficiencies, which we discuss below for each of the structural
components, with special emphasis on the extended component of star formation which dominates
the global SF activity in SQ.

\subsubsection{Discrete Star Formation Regions}
Fig. \ref{plot_sfr_sgas} shows that SQ A and SQ B, have star formation rates very similar to those observed for galaxian regions with the same gas column density, despite their being
located well outside the galaxies of the group. This result is consistent with the conclusions reached by \cite{Braine01}, studying a sample of tidal dwarf galaxies, and by \cite{Boquien09}, which performed a multiwavelength analysis of star formation regions in collisional debris. The situation is however different for the other two bright star formation regions, HII SE and HII SW, which present much higher SFRs than those predicted by the plotted relations. This might in principle be due to a more efficient mode of star formation happening in those regions. It would be of interest to acquire more complete information on the gas content of these sources to further investigate this conjecture. An alternative explanation for the high SFR found in HII SE and HII SW is that there is an additional component of UV emission produced by the radiative cooling of shocked gas. These UV photons, unrelated to star formation phenomena, can in principle contaminate the UV flux measurement but also power part of the observed dust emission, thus leading to an overestimation of the SFR. Optical spectra of these regions show evidences of line shock excitation (P.-A. Duc and collaborators, private communication), therefore one cannot rule out this possibility. 

\subsubsection{Star Formation associated with the Shock}
The average SFR in the shock region ($3.1\times10^{-3}~{\rm M_\odot/yr/kpc^2}$) seems to be well in agreement with the empirical SFR--gas surface density relation in Fig. \ref{plot_sfr_sgas}.
Thus, at first sight, the shock does not seem to have had much effect on the
star formation activity in the stripped interstellar gas, neither triggering enhanced
star formation through shock--compression of dense clouds of gas, nor suppressing star formation
through heating and dispersion of the clouds. The observational situation is however complex in
that the star formation observed towards this region could be happening inside the extended features connected with the incoming galaxy NGC 7318b, seen through, but not obviously physically co-existent with
the shock region. This is supported by the local morphology of the UV and MIR emission
that seems to follow the optical shape of the intruder, instead of showing a linear north--south ridge resembling the shocked gas emission morphology. These two scenarios could in principle be combined since the ISM of the intruder galaxy has presumably been shocked as well and this could have triggered the observed star formation out of the gas associated with the intruder galaxy. In addition, it is conceivable that some fraction of the UV luminosity, albeit probably a minority, could be due to gas cooling rather than from massive stars.


It is also of interest to compare our constraints on the SFR in the shock region with the luminosity of the radio synchrotron emission from the shocked gas. Traditionally radio synchrotron measurements are compared to infrared emission measurements in the context of the radio--FIR correlation for individual galaxies, for which we use here the relation given by \cite{Pierini03}: 
\begin{equation}
log~L_{1.4~{\rm GHz}}=1.10\times log~L_{\rm FIR} -18.53 
\end{equation}
where $L_{1.4~{\rm GHz}}$ is the luminosity at $\nu=1.4~{\rm GHz}$ in units of ${\rm W~Hz^{-1}}$ and $L_{\rm FIR}$ is the total FIR luminosity in units of ${\rm W}$. 
We used the relation from \cite{Pierini03} since it was derived from data covering cold dust emission longwards of $>100{\rm \mu m}$, where, as in the case for the SQ shock region, most of the power is radiated.
Using the value we measured for the shock region FIR luminosity, $L_{\rm FIR}=1.5\times10^{36} {\rm W}$, we
obtain for the predicted radio luminosity $L_{1.4~{\rm GHz}}=1.8\times 10^{21}~{\rm W~Hz}$. This value is 20 times smaller than the radio luminosity derived from the radio measurement in Xu et al. 03: $L_{1.4~{\rm GHz}}=3.7\times 10^{22}~{\rm W~Hz}$. The high radio/infrared luminosity ratio (which would be higher still
with respect to the photon--powered component if collisional heating was important) points to
the existence of an additional source of relativistic particles in the SQ shock, 
accelerated at the shock itself (see e.g. \citealt{Bland87}), which dominates the
population of particles accelerated in sources more directly linked to star formation regions
such as supernova remnants. This confirms and strengthens the preliminary results of \cite{Xu03}, and, bearing in mind that the total radio emission from SQ is dominated by the emission from
the shock region, suggests that caution is needed in using radio synchrotron measurements to infer star formation rates in groups involving strong dynamical interactions of galaxies with the IGM.

\subsubsection{Star Formation associated with the Extended Emission}
Our measurements have shown that both the obscured and visible components of
the star formation in SQ are distributed in a widespread pattern, loosely coincident
both with the overall dimensions of the group as well as with the extended ``halo'' of X-ray
emission. This current morphology of SQ star formation can readily be accounted for in terms of the several galaxy -- galaxy or galaxy -- IGM interactions which have occurred in the group. In this standard scenario, the gas that is currently converted into stars is
interstellar in origin (as we will discuss later in this section). However, the data are also consistent with a rather different scenario, in which the reservoir
of gas out of which stars are being formed is the hot gas phase of the intragroup medium.

Specifically we consider a scenario in 
which the hot intergalactic medium is cooling and condensing into clouds which are
the sites of the star formation providing the UV powered component of the
extended infrared emission. If this is occurring in a steady state, in which the rate at which
the removal of hot gas by cooling is balanced by accretion of further primordial gas (that is, we assume that variation in the accretion rate have longer time scales than the cooling time scale), we can write 
\begin{equation}
SFR_{\rm hot}\le\frac{M_{\rm hot}}{\tau_{\rm hot}}
\label{sfr_hot_eq}
\end{equation} 
where $M_{\rm hot}$ and $\tau_{\rm hot}$ respectively denote the mass and cooling timescale
of the X-ray emitting medium, and the inequality denotes the fraction of the
cooling gas that ultimately condenses into stars. If the observed X-ray emission
is the main component of luminosity of the hot medium, and taking observational
values for the X-ray emitting plasma from Trinchieri et al. (2005)
of $M_{\rm hot}=1.0\times10^{10}{\rm M_\odot}$ and  
$\tau_{\rm hot}\approx0.3\times10^{10}{\rm yr}$ (see Sect. 7.2.2) we obtain  $SFR_{\rm hot}\le3{\rm M_\odot yr^{-1}}$,
comparable to the observed value of $5.3{\rm M_\odot yr^{-1}}$ for star formation seen towards
the X-ray--emitting halo. In reality, however, one would expect
a large fraction of the cold gas resulting from the cooling
of the gas to be recycled into the hot medium due to the feedback of mechanical energy 
from the newly--formed stars, in which case $SFR_{\rm hot}$ would be substantially 
less the upper limit from Eqn.\ref{sfr_hot_eq}. Thus, if X-ray emission driven by gas--gas collisions 
is the dominant cooling path for the X-ray halo,
one would conclude that the bulk of the star formation would not be
directly fuelled out of the IGM. On the other hand, as shown in Sect.\,7.2.2, our Spitzer FIR
data, coupled with astrophysical constraints on the injection rate of grains into
the hot medium do not at present rule out the possibility that the dominant cooling mechanism
of the halo is FIR emission driven by gas--grain collisions. At present, therefore, we
cannot rule out star formation out of a primordial IGM purely on considerations of gas fuelling.

A more powerful constraint 
would be to consider the efficiency at which such a mode
would have to operate at. Unfortunately, the total mass of molecular cold gas, $M_{\rm cold}$
on a global scale in the halo is unknown, preventing a direct empirical measurement of
star formation efficiency of the halo on the KS diagram. One can however estimate
$M_{\rm cold}$ under our simple steady state scenario by writing

\begin{equation}
\frac{M_{\rm hot}}{\tau_{\rm hot}}\approx\frac{M_{\rm cold}}{\tau_{\rm cold}}
\end{equation} 

where $\tau_{\rm cold}$ is the typical time scale spent by the gas in the cold molecular phase
before being converted into stars, which is the timescale
for the collapse of a molecular clouds to form stars, multiplied by the mean number of
times the cold gas is recycled into the hot medium through mechanical feedback before
condensing into a star. Even allowing for several cooling cycles of the hot gas before
condensing into stars, the very short collapse timescale of a few million years for molecular
clouds - some three orders of magnitude shorter than $\tau_{\rm hot}$ -
makes it likely that $\tau_{\rm hot}>>\tau_{\rm cold}$, which in turn implies $M_{\rm hot}>>M_{\rm cold}$.
If this is the case, the cold gas amount, related to the extended emission, that we should add to $\Sigma_{\rm gas}$ would be negligible, and the position of the point for the SQ halo on the KS diagram
would be well to the left of the relation for star--forming galaxies, as shown in 
Fig. \ref{plot_sfr_sgas}. This would require a more efficient star--formation process in
the IGM than is typically observed to occur in the disks of star--forming field galaxies, 
which would seem to be counter--intuitive to the naive expectation that denser, colder 
reservoirs of gas closer to the minimum of the gravitational potential wells
associated with individual galaxies should form stars more easily. On the other hand 
a definition of efficiency in terms of gas surface density over scales much larger than the
star--formation regions may have limited predictive power in this context. Furthermore,
any star formation in a cooling hot IGM would not be expected to occurring in 
rotationally supported systems similar to the galaxies used to define the KS relation,
so may have a different relation to the local gas density.

The second, more conventional scenario to explain the widespread star formation in the
extended halo component of SQ, is to invoke interstellar gas as the source of cold 
gas fuelling most of the star formation. In this scenario tidal interactions between
galaxies or hydrodynamical interactions of the ISM with the IGM
remove cold galaxy material that, for SQ, produces a 
similar SFR as would have been the case for a similar amount of gas inside field galaxies.
For groups in general the statistical
relation of neutral gas observed within and outside the member galaxies to
the X-ray emission characteristics of the IGM indeed indicates that at least some
of the cold gas in the IGM medium originates in the galaxies, and that at least some
of the X-ray emitting gas also has an interstellar origin (eg. \citealt{Verder01}; \citealt{Rasmuss08}). 
More direct evidence of the removal of interstellar gas and dust
through galaxy interactions in high density environments is provided
by observations of interacting galaxies in clusters, for example
through Herschel imaging of extraplanar dust emission associated with
stripped atomic and molecular gas around the Virgo cluster galaxy NGC4438
(\citealt{Cortese10}).

In SQ, there is direct morphological evidence for bursting sources associated with
tidally removed interstellar gas in the form of SQ A and SQ B, and a hint that other 
less prominent components of the FIR emission in the IGM may also be associated with tidal
features, such as the enhancement of $H_\alpha$ and MIR/FIR emission seen towards the bridge
linking the intruder galaxy with the AGN host galaxy. In this picture the extended emission is a conglomeration of discrete sources similar in nature to SQ A and SQ B but much more numerous and of
lower power. The fact that SQ A and SQ B are likely to have a fundamentally interstellar origin despite their present location outside the galaxies is further supported by the rather high value of the dust to gas ratio we have found for these sources, $Z_d=0.002$, which is a signature of a high gas metallicity. The one puzzling aspect of the extension of this scenario to
the global star formation in the IGM of SQ is the lack of prominent
star formation associated with the bulk of the HI in the IGM located to the South and East of
the group. However, as shown by Fig. 5 of \cite{Wi02}, the gas column density in these clouds (generally less than $\approx3\times10^{20}~{\rm atoms~cm^{-3}}$) is rather low compared to the values found for the brightest star formation sites in SQ. Therefore it is plausible that star formation is inhibited in these clouds because of the low gas density. 

Finally, we remark that if stripped neutral and molecular interstellar gas
was the only source of cold gas powering star formation, one would expect a
rapid quenching of star formation activity in SQ in the future, whereas if the
star formation were fuelled by a cooling primordial IGM, the star formation activity
would simply follow the accretion rate onto the group of the primordial IGM. Thus,
measurement of the UV and MIR/FIR luminosity functions of well defined statistical
samples of galaxy groups, now being defined through
deep optical spectroscopical surveys such as the Galaxy And Mass Assembly survey (GAMA; 
\citealt{Driver09}), may offer a way of quantifying the relative importance of these
modes of star formation on the dynamical halo mass of groups in the local Universe.

\subsection{The nature of the X-ray correlated FIR emission in Stephan's Quintet}
\label{x_ray_em_disc}

Previous searches for infrared emission counterparts of the hot X-ray
emitting components of the intergalactic
medium of nearby objects have almost exclusively
targeted the intracluster medium (ICM) of rich
galaxy clusters, either using a stacking analysis for X-ray or optical
selected clusters from the IRAS all sky survey (\citealt{Giard08}; \citealt{Roncar10}) or in detailed imaging observations of the Coma Cluster
and other clusters with ISO or Spitzer 
(\citealt{Stickel98}; \citealt{Quill99}; \citealt{Stickel02};
\citealt{Bai07}; \citealt{Kita09}). All these studies have
either yielded upper limits or marginal apparent detections of the ICM
at far--IR brightness levels far fainter than those we have measured
towards the X-ray emitting IGM of SQ with Spitzer. Furthermore,
even the apparent detections of the ICM were susceptible to confusion
with foreground cirrus 
or the background galaxy population due to the accidental
similarity (\citealt{Popescu00}) in far--IR colour of the
collisionally-- and photon--heated emissions,
so realistically must be treated as upper limits
to any collisionally heated emission from the ICM (\citealt{Bai07}).
Although intracluster dust has been unambiguously detected at optical
wavelengths, through statistical studies
of the reddening of background sources through large numbers of clusters
(\citealt{Chel07}) this has been at a low abundance of
about $10^{-4}$ in the dust--to--gas mass ratio, consistent
both with the non--detections of the ICM in infrared emission and with
specific predictions by \cite{Popescu00} of the dust content of 
typical ICMs. 

In the present work we have been
able to sidestep the confusion problems afflicting the
previous attempts to detect infrared emission from the ICM in clusters
only by virtue of the
much higher far--IR brightness levels measured towards the 
X-ray emitting structures in SQ. These high brightness levels, together with
the correspondence of the infrared emission morphologies with those
seen in the UV, and the evidence from our fits to the infrared
emission SEDs described
in  Sect. \ref{shock_sed_fit_par} however suggest that
a major part of the dust emission is photon heated. This, in turn,
raises an apparent paradox that the closest spatial correlation 
between dust emission and gas column density is not, as might be
expected for photon heating, with the cool HI and molecular gas component,
but rather with the hot X-ray emitting component. In our discussion on
star formation in SQ we identified a possible way out of this
paradox by postulating that the extragalactic star formation in SQ
is fuelled by gas from a hot IGM, whose cooling is enhanced by 
grains injected into the hot IGM. Here, we use the relative levels of
detected emission at infrared, X-ray and UV/optical wavelengths to
constrain the possible sources of grains and the resulting collisionally
driven cooling of the hot plasmas,considering separately the two 
main X-ray emitting structures corresponding to the shock and the halo
regions.

\subsubsection{The shock region}
The main constraint on the collisional heating of dust in the X-ray emitting plasma downstream of the
shock is given by the high 
dust--to--gas mass ratio that would be required
to reproduce the FIR measurements in the case that grain heating by a diffuse
radiation field can be neglected. For
pure collisionally heated emission from grains in a homogeneous medium, 
a dust mass $M_{\rm d}=2\times 10^{7}~{\rm M_\odot}$ (see Sect. \ref{shock_sed_fit_par}) would be
required, which, taken together with the measured X-ray emitting gas mass, $M_{\rm X}\approx10^9~{\rm M_\odot}$ (see Appendix \ref{ap3}), would imply a dust--to--gas ratio for the hot plasma of $Z_d=0.02$. This value is high in comparison with
expectations based on a balance between injection of grains and their
removal through sputtering.
Assuming that the shock is propagating into a stripped interstellar medium or through the ISM of the intruder galaxy, characterized by a dust--to--gas ratio comparable to the Milky way value, we can derive the amount of dust mass per unit time $\dot{M_{\rm D}}$ that is injected downstream of the shock: 
\begin{equation}
\dot{M_{\rm D}}=V_{\rm sh}A\rho_{\rm up}Z_{\rm d,up}
\end{equation}
where $V_{\rm sh}=600~{\rm km/s}$ is the shock velocity (\citealt{guillard09}), $A=330~{\rm kpc^2}$ is the projected area of the shock front (corresponding to $20\times80 {\rm arcs^2}$ on the sky), $\rho_{\rm up}\approx 1.1n_{\rm H}m_{\rm H}= 4.5\times10^{-27}~{\rm g/cm^3}$ is the upstream gas density ($n_{\rm H}=0.01/4.~{\rm cm^{-3}}$) and $Z_{\rm d,up}=0.01$ is the assumed dust to gas ratio for the gas upstream of the shock front. The resulting dust injection rate is $\dot{M_{\rm D}}=0.14~{\rm M_\odot/yr}$. To estimate grain destruction 
through sputtering in collisions with heavy particles we use the
formula of \cite{Draine79} for the dust sputtering time scale:
\begin{equation}
\tau_{\rm sp}=2\times10^{6}\frac{a}{n_{\rm H}}~{\rm yr}
\label{sput_time}
\end{equation}
where $a$ is the dust size in ${\rm \mu m}$ and $n_{\rm H}$ is the proton number density in ${\rm cm^{-3}}$. Assuming $n_{\rm H}\approx0.01~{\rm cm^{-3}}$ and $a=0.1~{\rm \mu m}$, we get $\tau_{\rm sp}=2\times10^7~{\rm yr}$. Explicit calculation of the cooling of the shock gas due to gas--gas and gas--grain collisions, taking into account the evolution of the grain size distribution (under the assumption that there are no further sources of grains), show that the gas cooling time scale is $t_{\rm cool}\approx7\times10^7~{\rm yr}$, more than three times larger than $\tau_{\rm sp}$ (Natale G.,2010, PhD Thesis). Under these circumstances a conservative estimate for the dust to gas ratio downstream of the shock can be derived by dividing the total mass of dust predicted to be at any time downstream of the shock to the measured hot gas mass, that is:  
\begin{equation}
Z_{\rm d,down}=\frac{\dot{M_{\rm D}}\tau_{\rm sp}}{M_{\rm X}}
\end{equation}
Applying this formula, the expected value for the X-ray plasma dust to gas ratio is $Z_{\rm d,down}\approx3\times10^{-3}$, a factor $\approx 7$ smaller than the value inferred from the data, for the case where the entire FIR dust emission is powered by collisional heating. This
is only a crude estimate for the dust abundance, which might take higher values if the 
line of sight depth of the shock was assumed to be larger than the projected
width seen on the sky, but which on the other hand is an overestimate compared to
predictions taking into account the true evolution of grain size downstream of the shock.
In any case, it is evident that the measured dust to gas ratio is much larger than the predicted $Z_{\rm d,down}$, and a further source of dust grains would be required if collisional heating
were to be a significant driver of the FIR emission. 

The most obvious such source would be the
reservoir of grains in the cold component of the gas in the shock region, which could
be potentially released into the hot gas by ablation if the clouds were in relative
motion to the hot gas. As already mentioned the presence of cold gas is implied by the detection of $8{\rm \mu m}$ emission and has been directly demonstrated by the detection of molecular hydrogen in the MIR rotational lines (\citealt{Cluv10}). \cite{guillard09} modelled the $H_2$ emitting clouds as cooling pre-existing clouds in the upstream medium, a scenario which would predict the clouds to be in relative motion to the hot gas downstream of the shock. Another possibility which
might potentially account for a FIR counterpart to the shock is to invoke heating of
grains upstream of the shock by the UV radiation from the cooling clouds downstream of the
shock. A quantitative treatment of all these effects would however require a self consistent model
for the passage of the shock through a two phase medium which tracked the exchange of photons, gas
and dust between the phases, which is beyond the scope of this paper.

It's appropriate to emphasize here that even in the case that the collisional heating is relatively unimportant in bolometric terms, the effect on the gas cooling time scale is not negligible. Detailed calculations for a steady state 1D homogeneous model, considering dust sputtering and dust cooling, show that the cooling time scale is shorter by a factor $\approx2-3$ (Natale G.,2010,PhD thesis) for shock velocities and gas densities similar to those characteristic of the shock in SQ. From this perspective it would be important to have more direct empirical constraints on the gas cooling mechanism using the improved sensitivity, angular resolution and wavelength coverage of the Herschel Space Observatory. Using the simple approach described above, we estimate the surface brightness of the radiation emitted by collisionally heated dust to be equal to $1.4~{\rm MJy/sr}$ at $150{\rm \mu m}$, the wavelength where the flux density is predicted to peak according to the collisional heated dust model we used for the shock region SED fit (see Fig. \ref{shock_sed}). 

\subsubsection{The extended emission}
As for the shock, the detection of substantial $8{\rm \mu m}$ emission towards
the extended X-ray emitting halo indicates the presence of a component of
dust in cold gas
phases, shielded from collisions with electrons and ions in the hot
plasma, which can only be heated by photons. Rather interestingly, the curves of growth we constructed after removing or masking the galaxies and the compact star formation regions from the infrared and FUV maps (Fig. \ref{radial_int_prof}) show a very similar profile at all wavelengths. This is a strong indication that FUV sources are associated with the FIR extended emission and, therefore, star formation is the main mechanism powering the extended emission.

The total luminosity of the extended dust emission is $4\times10^{43} ~{\rm erg/s}$, a factor $\approx130$ higher than the X-ray Halo luminosity (Trinchieri et al.2005, see also Appendix \ref{ap3}). This large difference between the dust emission and X-ray luminosities implies that collisional heating of dust grains
could still be important for the cooling of the hot halo gas even if it only powers a small fraction of the
dust luminosity. We define the fraction $F_{\rm coll}$
of dust emission that is collisionally powered, by the relation:
\begin{equation}
L_{\rm IR}^{\rm coll}=F_{\rm coll}L_{\rm IR}
\label{l_ir_coll_equ}
\end{equation}
where $L_{\rm IR}^{\rm coll}$ is the dust luminosity that is powered by collisional heating.\\
A first physical constraint on $F_{\rm coll}$ that we can apply
is the requirement that the gas cooling time scale $\tau_{\rm cool}$ is not larger than the dynamical hot gas time scale, given by the sound crossing time $\tau_{\rm cross}$. Including dust cooling, $\tau_{\rm cool}$ is equal to: 
\begin{equation}
 \tau_{\rm cool}=\frac{3}{2}\frac{kT}{\mu m_{\rm H}}\left( \frac{M_{\rm X}}{L_{\rm X}+F_{\rm coll}L_{\rm IR}}\right)
\end{equation}
where $T=6\times10^6~{\rm K}$ is the gas temperature, $M_{\rm X}=10^{10}~{\rm M_\odot}$ is the X-ray emitting gas mass, and $L_{\rm X}$ and $L_{\rm IR}$ are the X-ray and infrared luminosities. Requiring that $\tau_{\rm cool} > \tau_{\rm cross}=2R_{\rm halo}/c_s$, where $R=35~{\rm kpc}$ is the halo radius and $c_s\approx300~{\rm km/s}$ is the sound speed, one obtains $F_{\rm coll}<0.05$.\\
Another constraint can be derived from an estimation of the dust injection rate into the hot gas. For equilibrium between dust injection and sputtering, one would have: 
\begin{equation}
\dot{M_{\rm d}}=F_{\rm coll}\frac{M_{\rm d}}{\tau_{\rm sp}}=2.4F_{\rm coll}~{\rm \frac{M_\odot}{yr}} 
\label{dust_inj_equ}
\end{equation}
where $\dot{M_{\rm d}}$ is the dust injection rate, $\tau_{\rm sp}= 2\times10^8~yr$ is the sputtering time scale (calculated using formula \ref{sput_time}
with $n_H=0.001 {\rm cm^{-3}}$ and $a=0.1{\rm \mu m}$), and $M_{\rm d}=4.7\times10^8~{\rm M_\odot}$ is the inferred dust mass producing the extended emission. 

The first dust injection sources we consider are the halo stars. As described in Appendix \ref{ap5}, we have derived an upper limit to the dust injection rate from halo stars based on the observed R-band halo surface brightness (\citealt{Moles98}) and the theoretical predictions for stardust injection by Zhukovska et al.08: $\dot{M_{\rm d}}=0.075 {\rm M_\odot/yr}$. Substituting this value in Eqn. \ref{dust_inj_equ}, we get $F_{\rm coll}<0.03$. A similar procedure could be followed considering the dust injection from type II supernovae given the inferred SFR associated with the extended emission. Empirical studies of Galactic supernova remnants have shown the yield of condensate per supernova event to be of the order of $0.01$ to $0.1~{\rm M_\odot}$ (\citealt{Green04}, \citealt{Fischera02}). Relating the frequency of core collapse supernova events to the SFR in the extended component of SQ, we estimated $\dot{M_{\rm d}}=7\times10^{-4}~{\rm M_\odot/yr}$ assuming $0.01~{\rm M_\odot}$ of dust produced in each supernova. Therefore supernovae dust injection cannot be considered important in replenishing dust in the hot medium. Finally we considered the mechanism proposed by \cite{Popescu00} where dust in IGM external to a cluster can be introduced into the hot intracluster medium in the supersonic accretion flow of barions onto the cluster. In the case of SQ the baryonic accretion rate can be estimated to be of the order of the current IGM seen in X-ray divided by a Hubble time which is $\approx 2~{\rm M_\odot/yr}$. Even if we consider that there is maybe a cooler component of the accreting IGM not visible in the X-ray (\citealt{Mulch00}), it does not seem plausible to achieve dust accretion rate for the particular case of SQ much greater than $0.01~{\rm M_\odot/yr}$ for reasonable values of the dust to gas ratio in the accreting material. 

Given these estimates, we can conclude that if there is a contribution of collisional heated dust to the total observed dust emission, this should only be of the order of a few percent and maintained by direct injection from the in situ stellar population. This however is sufficient to maintain a collisionally powered FIR luminosity of up to $\approx 10^{42}~{\rm erg/s}$, which is several times the X-ray luminosity. 

Such a level of FIR cooling could help to provide a natural
explanation for the rather low ratio of $\sim\,0.02$ for the ratio of the mass
of baryons in the hot X-ray emitting halo of $\sim\,M_{\rm X}=2\times10^{10}~{\rm M_\odot}$
to the dynamical mass of $\sim\,10^{12}~{\rm M_\odot}$\footnote{This value is derived from the X-ray temperature that is consistent with the velocity dispersion of the group, $\sigma=340{\rm km/s}$, as calculated by \cite{Osul09}. However including the more recent recession velocity measurement for NGC 7320c, quoted in Sect.4.1 of \cite{S01}, the group velocity dispersion would be $\sigma=80{\rm km/s}$ indicating a DMH mass of order of $10^{11}{\rm M_\odot}$. Since the number of group galaxies is very low for a good estimate of the velocity dispersion, we consider the X-ray temperature based estimate for the DMH mass more reliable in this case.}. 
For the concordance CDM cosmology, this ratio should be $\sim\,0.09$ in the absence of significant
dissipative effects (see e.g. Fig. 9 \citealt{Frenk99}). Recalling that the total baryonic mass currently
locked in stars in SQ of $\sim\,10^{11}~{\rm M_\odot}$ is comparable to the 
expected total mass of baryons, it is apparent that
this scenario is also quantitatively
consistent with the efficient condensation of the IGM accreting onto
the group into stars, as postulated in Sect.$\,$7.1.
Some empirical indication of efficient star formation in groups has 
been found by \cite{Tran09}, who showed that the group
environment is  more conducive to star formation
than either the cluster or field environment. 
In general, enhanced cooling due to a FIR collisionally powered luminosity 
of a few times $L_{\rm X}$ would, since it promotes the condensation of hot X-ray emitting gas into stars, also provide an alternative explanation for the
steeper $L_{\rm X}\,-\,T_{\rm X}$ relation observed for groups with lower virial velocities, an effect that is otherwise attributed to thermal or kinetic
feedback from AGNs (e.g. \citealt{Caval08}). 

In summary, the detection and recognition of the putative collisionally
heated component of FIR emission from SQ will be a crucial
measurement with wide ranging implications for the nature of star--formation
in the group, the division of Baryons between hot gas, cold gas and stars,
and the thermodynamic properties of the IGM in SQ.
Such a detection will however be very challenging,
requiring a combination of excellent surface brightness sensitivity and angular resolution at submm wavelengths, as illustrated by the following estimates:
assuming the upper limit $F_{\rm coll}=0.03$, we predict the SED of the collisionally powered
component of the extended emission to peak at a level of $0.15~{\rm MJy/sr}$ at $190{\rm \mu m}$.

\subsection{The AGN galaxy NGC 7319 dust emission}
\label{agn_gal_em_disc}
\begin{landscape}
\begin{figure}
\hspace{-2.cm}
\includegraphics[scale=1.4,angle=0]{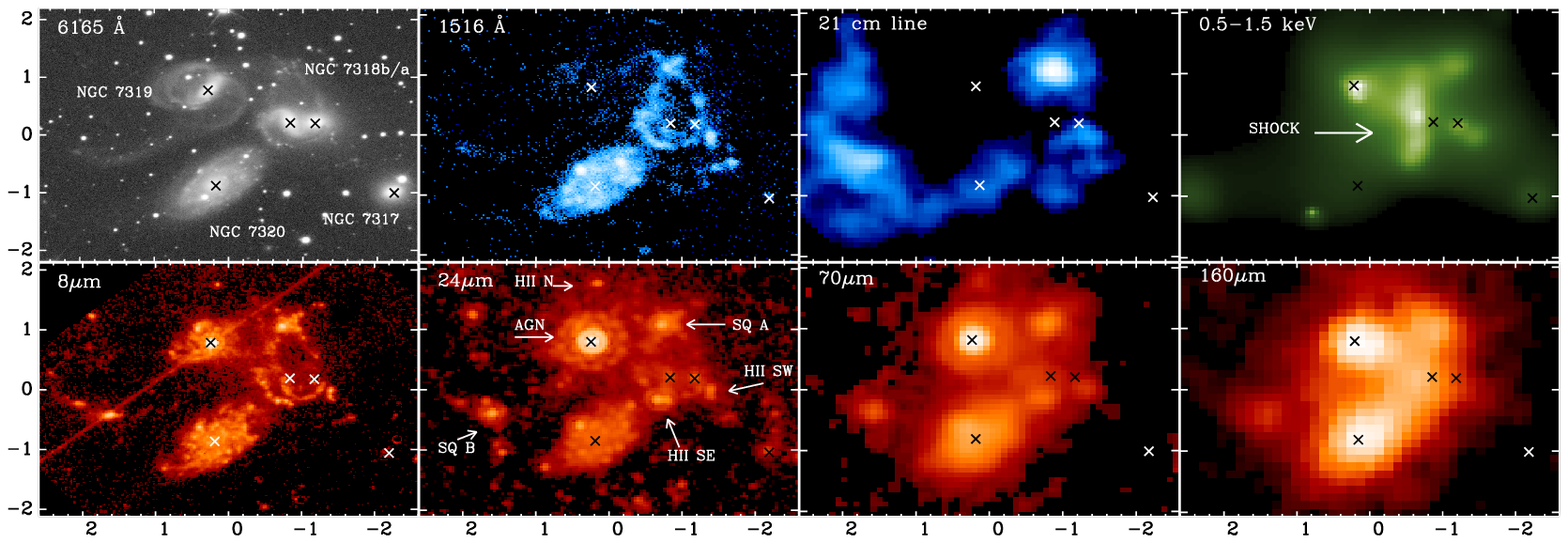}
\caption{Stephan's Quintet Multiwavelength Data. Upper row from left to right: SDDS r-band, GALEX FUV, VLA radio , XMM NEWTON soft X-ray. Lower row: SPITZER IRAC $8{\rm \mu m}$ and MIPS $24,70,160{\rm \mu m}$. Crosses on the maps identify the galaxy centers. The position (0,0) coincides with $RA=22^{\rm h}36'02.4''$,$Dec=+33^\circ57'46.0''$. The units on the axis are arcmin. At the distance of SQ ($94~{\rm Mpc}$) every arcmin corresponds to $24{\rm kpc}$. Note that NGC 7320 is a foreground galaxy at the distance of $10~{\rm Mpc}$ and its HI distribution, not shown in this figure, can be found in \cite{Wi02}.}
\label{SQ_mwave}
\end{figure}
\end{landscape}

The Seyfert2 galaxy NGC 7319 is the most powerful source of IR emission in SQ and is the only galaxy where we found enhanced infrared emission in the central regions. The total IR luminosity is $9.6\times10^{43}~{\rm erg/s}$. Of this, we deduced from the SED fitting using a combination of the AGN torus template and the diffuse dust emission component (see Sect. \ref{AGN_fit_par}) that $4.1\times10^{43}~{\rm erg/s}$ are emitted by dust in the torus and $5.5\times10^{43}~{\rm erg/s}$ by diffuse dust. As one can see from the SED best fit, shown in Fig. \ref{NGC7319_SED}, the $24{\rm \mu m}$ MIR emission is dominated by the torus emission which also contributes $\approx 25\%$ of the $8{\rm \mu m}$ emission. This is consistent with the observed morphology at $8$ and $24{\rm \mu m}$ emission, which is a combination of a central source plus a disk component in the case of the $8{\rm \mu m}$ map and a dominant central source in the case of the $24{\rm \mu m}$ map (see Fig. \ref{SQ_mwave}). According to the SED fit, the FIR $70$ and $160{\rm \mu m}$ emission is due to diffuse dust emission. This is also consistent with the brightness distributions on the Spitzer FIR maps which show the emission to be partially resolved, with extents inferred from the FIR map fitting technique of $5$ and $12~{\rm kpc}$ at $70$ and $160{\rm \mu m}$ respectively. This extended FIR emission is centred at the position of the AGN, potentially indicating the presence of star formation regions distributed on kpc scales in the surrounding areas. However, as one can see from Fig. 1 of \cite{Gao00}, the CO emission is predominantly from the northern regions of this galaxy. Therefore the locations of the bulk of cold gas, where the star formation sites should be, and the cold dust emission are not coincident suggesting that star formation may not be the main phenomena powering the FIR dust emission. 

Overall, the diffuse dust emission in the inner disk, together with the absence of HI (\citealt{Wi02}) and the displacement of the molecular hydrogen emission suggest that galaxy--galaxy interactions have removed gas from the central regions on kpc scales, as has plainly also happened in the other late--type SQ galaxies. In the case of NGC 7319, the interactions have apparently also induced gas flows that have triggered AGN activity. This raises the question whether the AGN itself might be the source of photons powering the surrounding diffuse dust emission in this galaxy. According to the standard AGN unification theory, the AGN torus of a Seyfert 2 galaxy is viewed edge on, which, in the case of NGC 7319, would imply that the plane of the torus is almost perpendicular to the galaxy disk, since NGC 7319 is seen basically face on (see SDSS g-band map in Fig. \ref{SQ_mwave}). In this configuration, it would indeed be plausible that accretion powered radiation, escaping into the polar direction from the AGN torus structure could propagate into the galaxy disk and contribute to the diffuse radiation field in the central parts of the galaxy. An especially efficient coupling between the dust and radiation would be expected if the polar axis of the AGN torus were to be parallel to the diffuse dusty disk of the host galaxy. 

One indirect indication that the radiation from an AGN with this orientation may be responsible for at least part of the diffuse FIR emission in NGC 7319 comes from the rather low observed UV/FIR luminosity ratio, which is qualitatively consistent with the expected strong attenuation of the direct UV light from an AGN to which the line--of--sight passes through the torus. From the measured FUV luminosity, $L_{\rm FUV}=4\times10^{42}~{\rm erg/s}$, and from the $60$ and $100{\rm \mu m}$ luminosities extrapolated from the SED best fit, $L_{60{\rm \mu m}}= 2\times10^{43}~{\rm erg/s}$ and $L_{100{\rm \mu m}}=3.7\times10^{43}~{\rm erg/s}$, we have found $L_{\rm FUV}/L_{60{\rm \mu m}}=0.2$ and $L_{\rm FUV}/L_{100{\rm \mu m}}=0.1$. These values are several factors lower than the average values found by \cite{Popescu02} for galaxies of similar morphological type ($L_{\rm UV}/L_{60{\rm \mu m}}=1$ and $L_{\rm UV}/L_{100{\rm \mu m}}=0.6$) and by \cite{Xu06} for galaxies of similar $L_{\rm FUV}+L_{60{\rm \mu m}}$ total luminosity ($L_{\rm UV}/L_{60{\rm \mu m}}=0.5$). 

A scenario in which the diffuse dust emission in NGC 7319 is powered by the AGN also seems to be broadly consistent with simple constraints on the UV luminosity of the AGN derived from the observed hard X-ray luminosity. Specifically, for $L_{2-10~{\rm keV}}=2.8\times10^{42}~{\rm erg/s}$ (\citealt{T05}) and adopting the bolometric corrections for AGN galaxies with similar $L_{2-10~{\rm keV}}$ by \cite{Vasu10}, $\kappa_{2-10~{\rm keV}}\approx 10-30$, one predicts a total UV luminosity from the AGN of $L_{\rm AGN}=\kappa_{2-10~{\rm keV}}L_{2-10~{\rm keV}}= 2.8-8.4\times10^{43}~{\rm erg/s}$. The upper end of this range is comparable to the total MIR/FIR luminosity of the AGN and AGN host galaxy. If the AGN were indeed powering the diffuse dust emission in the inner disk one would expect that future FIR observations of angular resolution sufficient to image the diffuse dust emission could be used to determine some basic geometrical properties of the AGN. Specifically, one would expect a cone structure for the component of diffuse dust heated by the AGN, with a strong radial colour gradient, and with a symmetry axis aligned with the polar axis of the torus.

Finally, we draw attention to a further FIR--emitting feature potentially connected to the AGN, whose origin has still to be clarified. This is the bridge which apparently connects the AGN to the center of the shock region, appearing in the X-ray (\citealt{T05}), $H\alpha$ (\citealt{Xu99}) and warm $H_2$ line emission (\citealt{Cluv10}). The presence of a cold dust emission from this feature is hinted by a slight asymmetry of the FIR emission on the $160{\rm \mu m}$ residual map (see Fig. \ref{resmaps}). High resolution FIR observations would also be useful to elucidate the correspondence between dust emission and the gas distribution in this bridge feature. 

\section{Summary}
\begin{figure}
 \includegraphics[scale=0.8,angle=0]{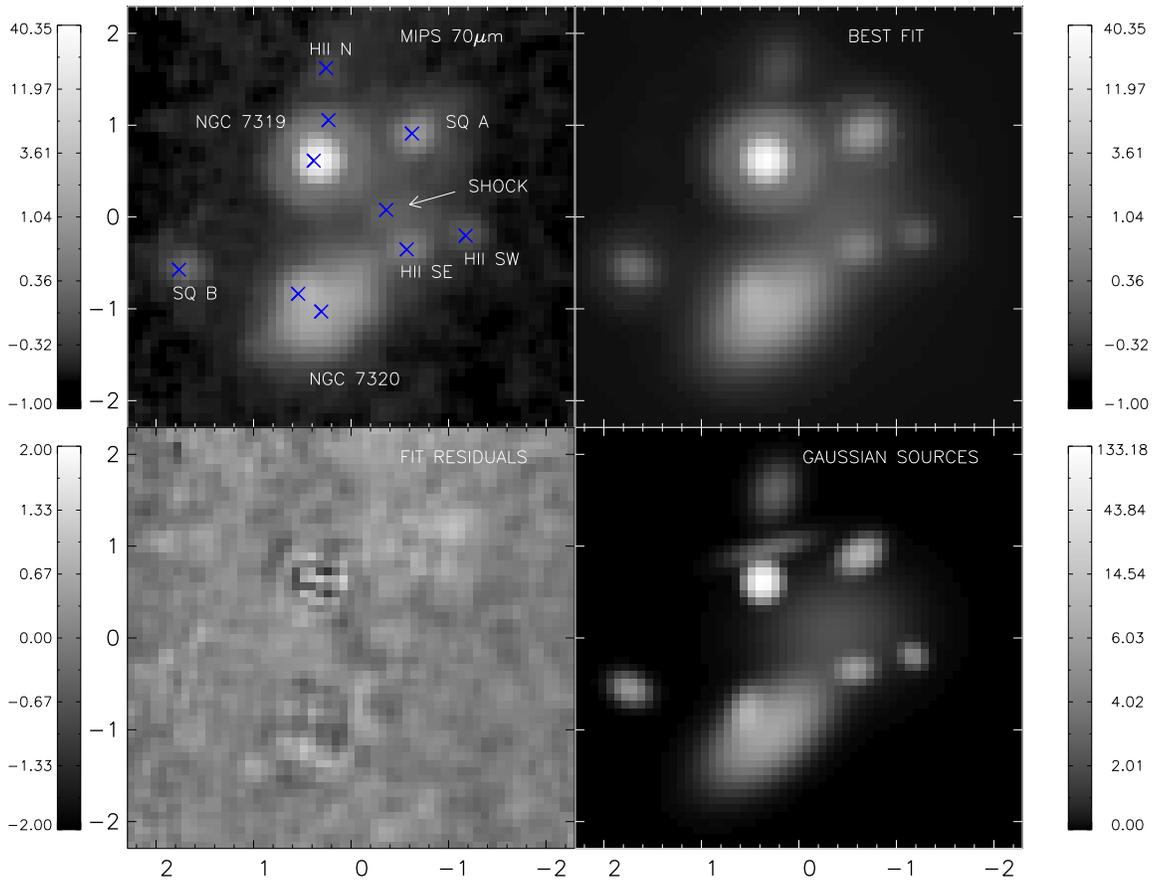}
\caption{Fit results for the $70{\rm \mu m}$ map. Top--left: original $70{\rm \mu m}$ map; Top--right: best fit map; Bottom--left: fit residuals; Bottom--right: ``deconvolved'' map. The crosses on the original map identify the centers of the fitted sources. Units of the values aside the color bars are MJy/sr. Note: The fit residual map, shown on the bottom--left panel, has all the fitted sources subtracted. It differs from the $70{\rm \mu m}$ ``FIR residual map'', whose contours are shown in the upper panels of Fig. \ref{resmaps}, where the SHOCK source has not been subtracted.}
\label{70mfit}
\end{figure}

\begin{figure}
 \includegraphics[scale=0.8,angle=0]{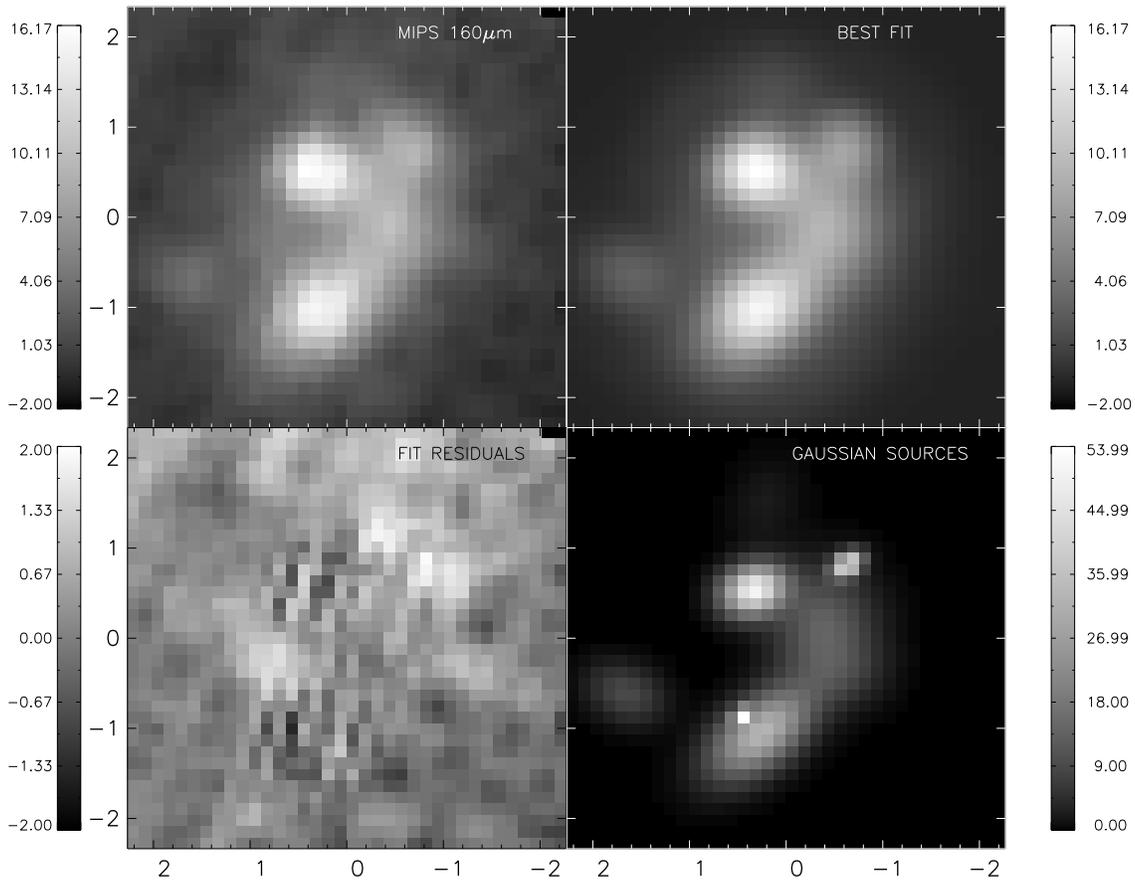}
\caption{Fit results for the $160{\rm \mu m}$ map. Top--left: original $160{\rm \mu m}$ map; Top--right: best fit map; Bottom--left: fit residuals; Bottom--right: ``deconvolved'' map. Units of the values aside the color bars are MJy/sr. Note: The fit residual map, shown on the bottom--left panel, has all the fitted sources subtracted. It differs from the $160{\rm \mu m}$ ``FIR residual map'', whose contours are shown in the lower panels of Fig. \ref{resmaps}, where the SHOCK source has not been subtracted.}
\label{160mfit}
\end{figure}

In this work we have performed an extensive analysis of the infrared emission from SQ, based on the Spitzer MIR and FIR maps and including multiwavelength data from optical/UV, Radio and X-ray. The main aim has been understanding the nature of the FIR emission seen on the Spitzer $70$ and $160{\rm \mu m}$ maps and its implications for star formation and hot gas cooling in this group of galaxies. The technical part of our analysis can be summarized in four points:

1) We have identified the FIR sources seen on the Spitzer maps and performed their photometry on both the MIR and FIR maps. For the FIR photometry we devised a new source fitting technique that we used to model the source emission on the maps (see Figs. \ref{70mfit} and \ref{160mfit}). We measured MIR fluxes using aperture photometry (see Figs. \ref{aperture8} and \ref{aperture24}). Photometry results are shown in Table \ref{phot_spitzer}. 

\begin{landscape}
\begin{figure}
 \includegraphics[scale=0.9,angle=0]{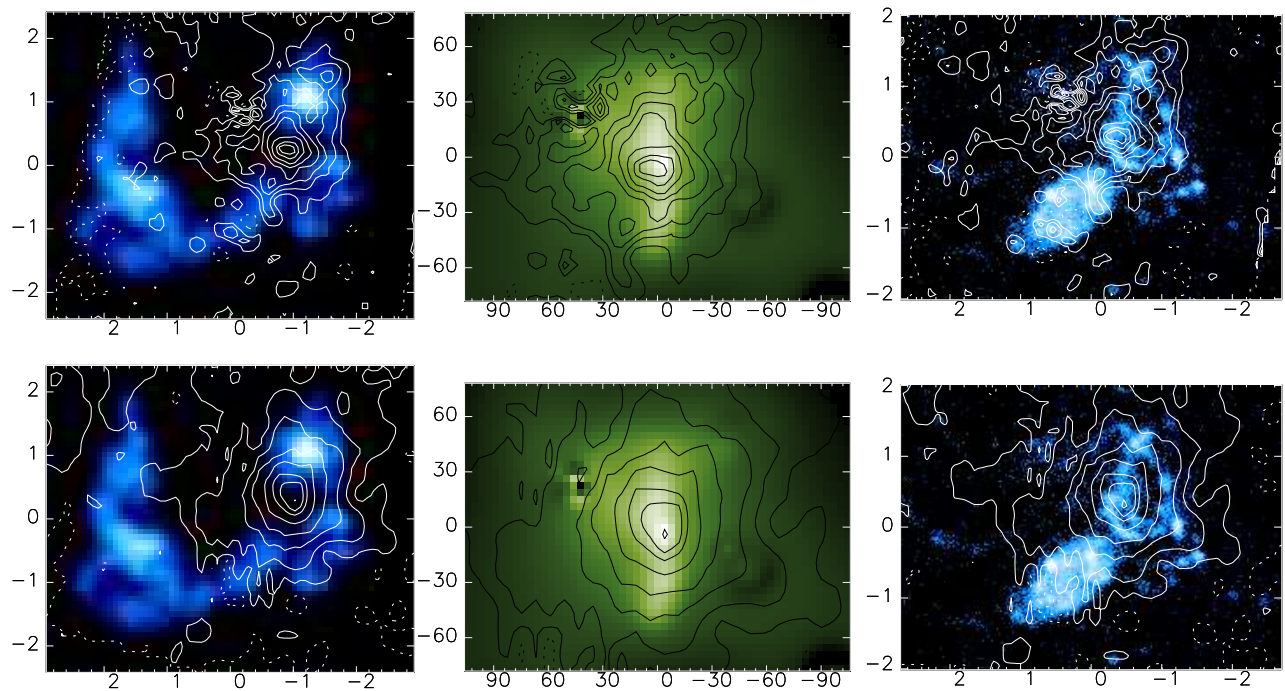}
\caption{\small FIR residual maps (see Sect. \ref{fir_res_map_par}) countours overlaid on HI (left panels), X-ray (middle panels) and FUV maps (right panels). The countours in the upper row are from the $70{\rm \mu m}$ residual map while those in the lower row are from the $160{\rm \mu m}$ residual map. The $70{\rm \mu m}$ residual map contour levels are $-0.3, 0.3, 0.6, 0.9, 1.2, 1.5, 1.8, 2.1, 2.25~{\rm MJy/sr}$ and those for the $160{\rm \mu m}$ residual map are $-0.4, 0.4, 1.2, 2.4, 3.6, 4.8, 6.0, 6.6, 7.2~{\rm MJy/sr}$ (the dashed lines are the negative contours, corresponding to $-1\sigma$ level). The field of view is different in the three maps. The coordinates corresponding to (0,0) are for the HI map  $RA=22^{\rm h}36'04.1''$ and $Dec=+33^\circ57'47.2''$, for the X-ray map $RA=22^h35'59.7''$ and $Dec=+33^\circ58'05.9''$, for the FUV map $RA=22^{\rm h}36'01.6''$ and $Dec=+33^\circ57'42.1''$. The axis units are arcminutes for the HI and FUV maps and arcseconds for the X-ray map. Note that point sources have been removed from the X-ray map (see Fig. \ref{SQ_mwave}).}
\label{resmaps}
\end{figure}
\end{landscape}

\begin{figure}
\includegraphics[scale=0.8,angle=0]{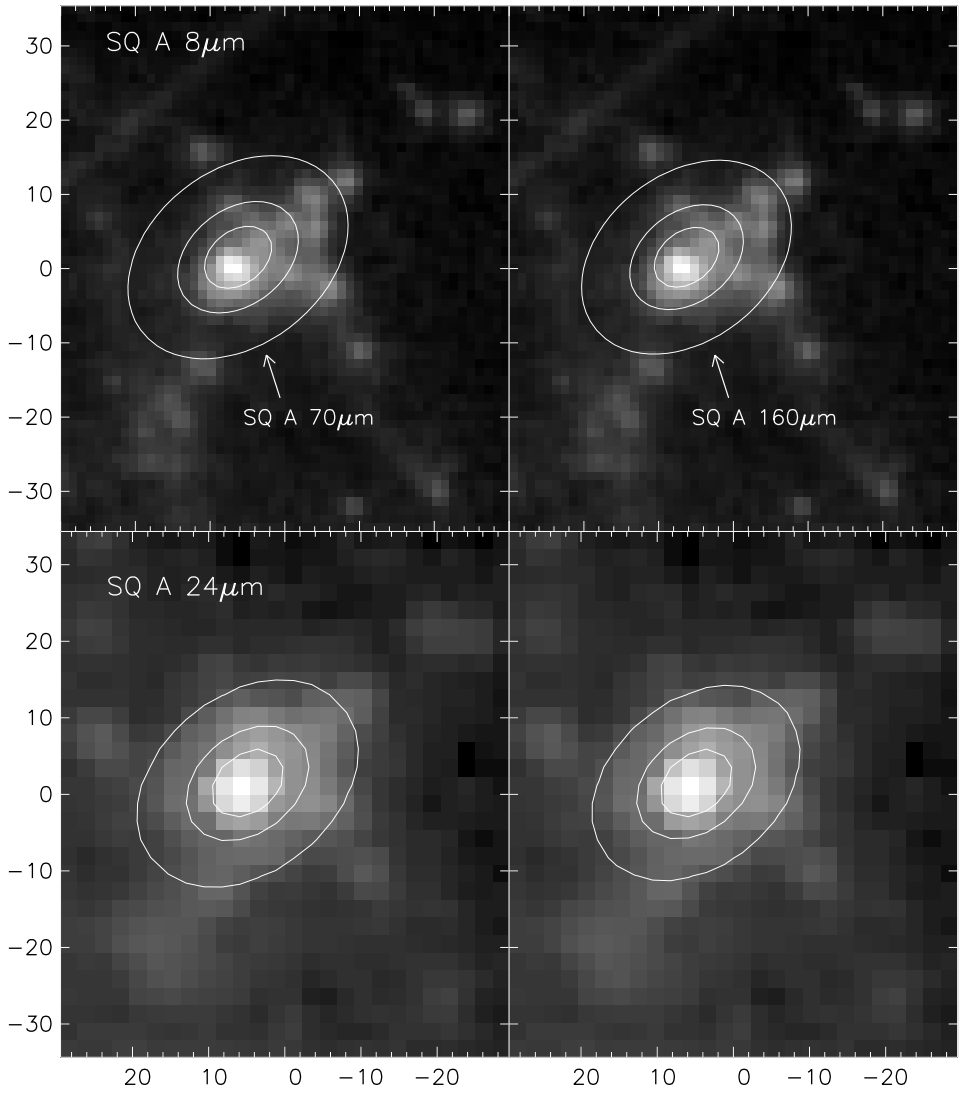}
\caption{``Deconvolved'' $70{\rm \mu m}$ and $160{\rm \mu m}$ SQ A emission contours overlaid on $8{\rm \mu m}$ (upper row) and $24{\rm \mu m}$ (lower row) maps. Units on the axis are arcseconds.}
\label{cont_deconv}
\end{figure}

\begin{figure}
\includegraphics[scale=0.4,angle=0]{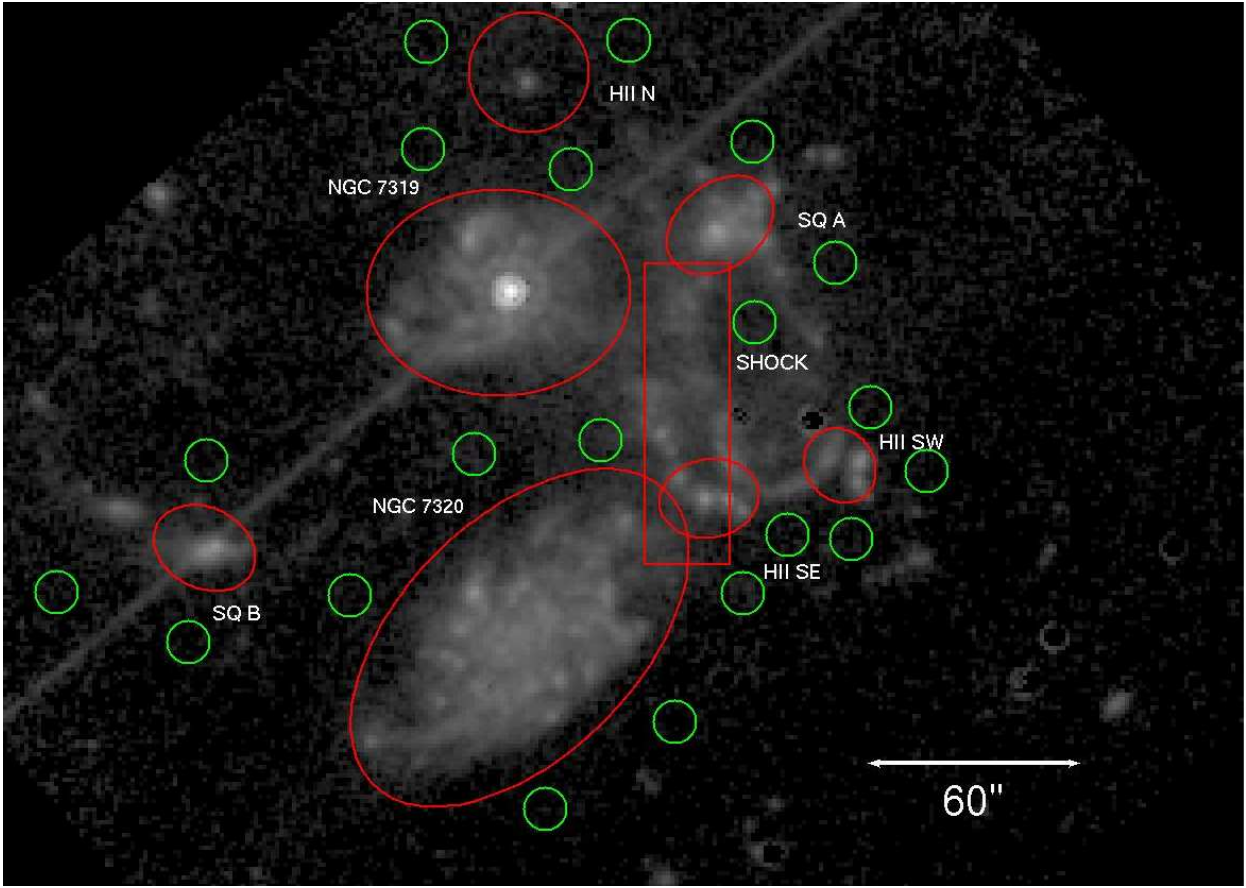}
\caption{Apertures for the photometry of each source at $8{\rm \mu m}$. Integration areas are delimited in red while background areas in green. The straight line passing through the AGN galaxy and SQB is a map artifact.}
\label{aperture8}
\end{figure}

\begin{figure}
\includegraphics[scale=0.4,angle=0]{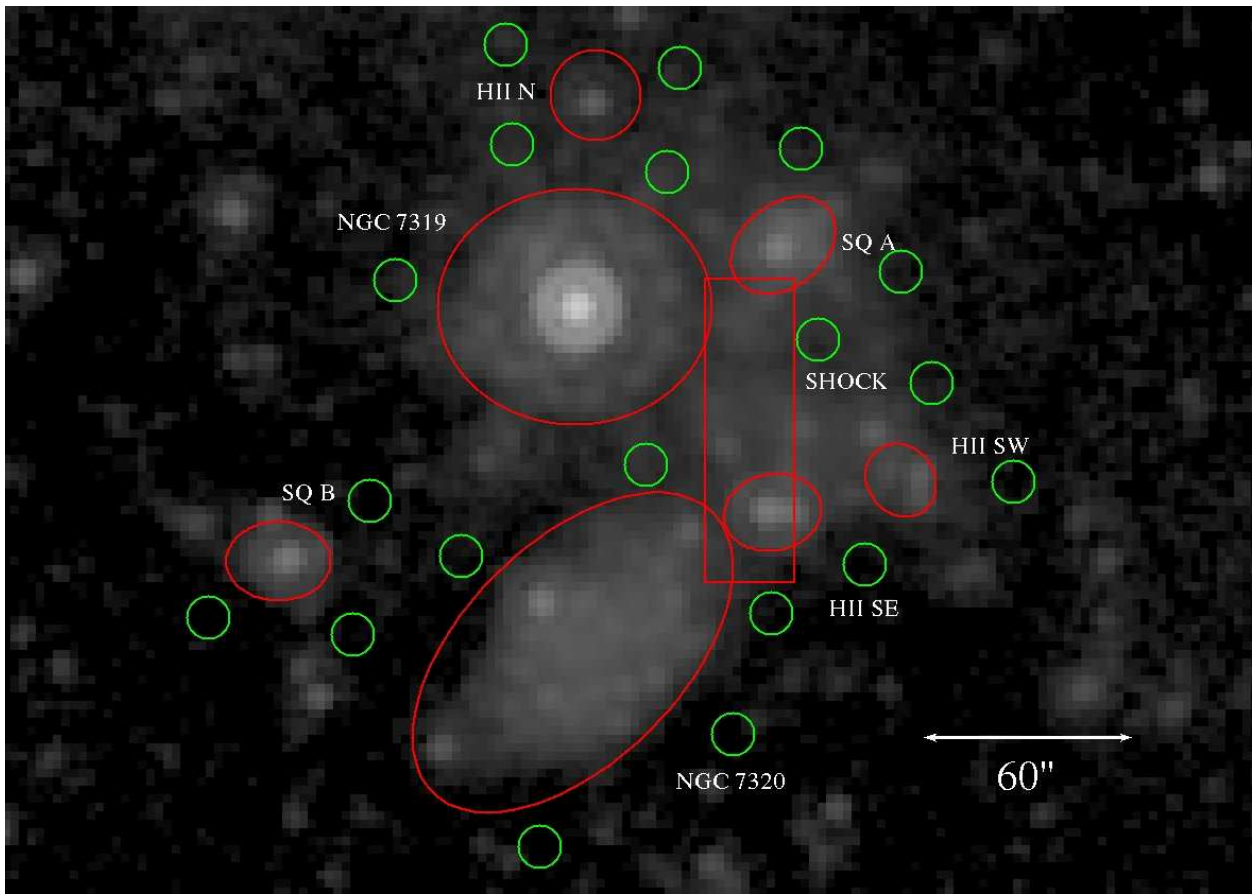}
\caption{Apertures for the photometry of each source at $24{\rm \mu m}$. Integration areas are delimited in red while background areas in green.}
\label{aperture24}
\end{figure}

2) We have compared the residual FIR emission (defined as the FIR emission after the subtraction of galaxies and discrete star formation regions) with HI distribution, X-ray emission and UV emission (see Fig. \ref{resmaps}). This comparison shows that the residual FIR emission, dominated by the emission from the shock region and a previously undetected extended component, is correlated with X-ray emission and anticorrelated with the HI distribution. At the same time, the area covered by the FIR residual emission is populated by an extended distribution of UV sources associated with tidal features and intergalactic star formation regions. 

3) We have modelled the inferred infrared source SEDs using combinations of dust emission templates. These templates, described in Appendix \ref{sed_temp_mod}, are used to reproduce emission from PDR/HII regions, diffuse photon--powered dust emission, collisionally powered dust emission and infrared emission from AGN torus. Among the immediate outcomes of the fitting procedure there are the dust mass required to produce the observed emission, the total dust infrared luminosity and the amount of UV powered dust emission (see Table \ref{sedfitres}).

4) Combining the amount of absorbed UV luminosity, obtained from the infrared SED fitting,  with the observed UV luminosity, measured directly on the GALEX FUV maps, we determined the star formation rate for each dust emitting source. We also measured gas masses/column densities from the observed radio HI and CO lines and soft X-ray fluxes and, combining these measurements with the dust masses inferred from the infrared SED fits, we estimated the dust to gas ratios for several sources. Results are summarized in Table \ref{sf_regions_tab}.
\\

From this set of measurements and morphological comparisons, together with simple predictions for the gas cooling time scale and the dust injection rate in hot plasma, we derived the following conclusions:

1) The total star formation rate, within a circular area of radius $90''$ containing the four group galaxies in close proximity, is $7.5{\rm M_{\odot}/yr}$. The star formation regions in the group are localized at the edges of the galaxies, on tidal features and in the intergalactic medium; no evidences for star formation have been found towards galaxy centers. 

2) The star formation efficiency for the star formation regions SQ-A and SQ-B, localized on tidal features, is close to that observed for star formation regions within nearby spiral galaxy disks (see Fig. \ref{plot_sfr_sgas}). Star formation in HII-SE and HII-SW seems to be much more efficient but the lack of complete gas measurements and the possible contribution to UV radiation from radiative shocks does not allow a definite conclusion for these sources.

3) The star formation efficiency in the shock region seems to be consistent with the star formation observed in nearby spiral galaxies. However, because of the elongated tidal feature of the intruder galaxy NGC 7318b seen through the shock region, it is not clear if the observed star formation in that area is associated with the shocked gas or not. The observed radio--IR luminosity ratio is 20 times higher than the value predicted by the radio--infrared relation derived by \cite{Pierini03}, given the observed amount of infrared luminosity. This reinforces the result obtained by \cite{Xu03}, suggesting that the acceleration of relativistic particles in the shock region is not associated with star formation.

4) $70\%$ of the total star formation in SQ is associated with the extended FIR emission. The star formation efficiency for this component is rather higher than that observed in spiral galaxies. This could in principle be due to a new mode of star formation operating in the group and involving cooling of the X-ray halo gas, possibly enhanced by dust--plasma particle collisions. However the lack of cold gas measurement for this distributed star formation component makes this hyphothesis rather uncertain. In addition, the evidence of tidal interactions and star formation in tidally displaced cold gas, like SQ A or SQ B, suggest that the distributed star formation observed in the group is in part occuring out of gas of galaxian origin.

5) For both the shock region and the extended emission we have measured substantial $8{\rm \mu m}$ flux, mainly due to PAH molecule line emission and tracer of cold gas phases. This, together with simple estimates of the dust injection rate across the shock front and in the halo of the group, demonstrates that the mechanism powering most of the dust emission seen on the FIR residual maps is not dust collisional heating. However, given the high FIR--X-ray luminosity ratio, the current data do not exclude that the majority of the hot gas cooling can still be due to dust--plasma particle collisions rather than X-ray emission. For dust in the group halo we estimated that dust collisional heating can power up to $10^{42}{\rm erg/s}$ of the dust luminosity. 

6) We have found that the morphology of the FIR dust emission in the Seyfert 2 galaxy NGC 7319 does not correspond to the cold gas distribution. Therefore, the FIR emission in this galaxy does not seem to be related to star formation. Given the AGN UV luminosity, estimated from the observed hard X-ray flux, and from considerations on the particular torus--galaxy disk geometry, we deduced that accretion powered radiation from the AGN, propagating through the disk of the galaxy, can power most of the observed dust infrared emission. 

\acknowledgments
We thank Pierre Guillard for fruitful discussions on the shock region of SQ and Francois Boulanger for discussions and comments on the manuscript. G.N. acknowledges support from the International Max--Planck Research School (IMPRS) Heidelberg. This work is based on observations made with the Spitzer Space Telescope, which is operated by the Jet Propulsion Laboratory, California Institute of Technology under a contract with NASA.

\clearpage

\clearpage

\begin{deluxetable}{lcccccc}
\tablecolumns{7} 
\tablecaption{Results of the FIR map fitting}
\tablewidth{0pt}
\tablehead{\colhead{Source}&\colhead{RA}&\colhead{Dec}&\colhead{$\Delta_{70{\rm \mu m}}$}&\colhead{$\Delta_{160{\rm \mu m}}$}&\colhead{$F_{70{\rm \mu m}}$}&\colhead{$F_{160{\rm \mu m}}$}\\
&\colhead{({h}\phn{m}\phn{s})} &\colhead{(\phn{\arcdeg}~\phn{\arcmin}~\phn{\arcsec})} & \colhead{(arcs)} & \colhead{(arcs)} & \colhead{(mJy)} & \colhead{(mJy)}}
\startdata
SQ A& 22 35 58.81& 33 58 51.1&     16&      15&      $134\pm 15$&      $265\pm 86$ \\ 
HII SE & 22 35 59.08& 33 57 34.0&      14&      40&      $61\pm 9$&      $186\pm 140$ \\ 
HII SW  & 22 35 56.10& 33 57 43.1&      11&     ND &      $31\pm5$&      ND  \\
SQ B& 22 36 10.54& 33 57 20.5&      14&      36&      $63\pm 8$&      $302\pm 34$ \\
N7319& 22 36 3.76& 33 58 33.0&      12&      30&      $616\pm 62$ &      $1192\pm 130$\\
N7319 (HII) & 22 36 3.01& 33 59 0.0&      24&      ND &      $50\pm8$&      ND\\
N7320& 22 36 3.38& 33 56 52.7&      44&      48&      $716\pm 73$ &      $1899\pm 196$ \\ 
N7320 (HII)& 22 36 4.54& 33 57 4.5&      16&      ND &      $111\pm 13$&      ND \\ 
HII N & 22 36 3.13& 33 59 34.6&      23&      42&      $52\pm 6$&      $98\pm 41$\\
SHOCK\tablenotemark{a} & 22 36 0.10& 33 58 0.3 &      61&      58&      $258\pm 30$&      $1218\pm 230$\\
\enddata
\tablecomments{Col. 1: FIR source; cols. 2-3: Source center RA and Dec coordinates; cols. 4-5: Averaged FWHMs of the sources at $70{\rm \mu m}$ and $160{\rm \mu m}$; cols. 6-7:Total flux densities at $70{\rm \mu m}$ and $160{\rm \mu m}$. Some of the sources are not detected at $160{\rm \mu m}$ (ND = ``non detected'')}
\label{FIR_sources}
\tablenotetext{a}{The SHOCK source on this table refers only to the gaussian used to fit the emission centered in the middle of the shock region. Therefore the quoted fluxes do not represent the actual fluxes coming from that area. The emission from the shock region is given in the Table 2.}
\end{deluxetable}

\clearpage
\clearpage

\begin{deluxetable}{lcccc}
\tablecolumns{5} 
\tablewidth{0pt} 
\tablecaption{Inferred infrared fluxes for the discrete sources detected on the Spitzer $70{\rm \mu m}$ map}
\tablehead{\colhead{Source}&\colhead{$F_{8{\rm \mu m}}$}&\colhead{$F_{24{\rm \mu m}}$}&\colhead{$F_{70{\rm \mu m}}$}&\colhead{$F_{160{\rm \mu m}}$}\\
& \colhead{(mJy)} & \colhead{(mJy)} & \colhead{(mJy)} & \colhead{(mJy)}} 
\startdata
SQ A&  $7.9\pm 0.5$ & $11\pm2$ &  $134\pm 15$&      $265\pm 86$    \\
HII SE & $3.4\pm 0.4$& $7.5 \pm 1$& $61\pm9$ & $186 (51)\pm142(40) $\\
HII SW & $2.5\pm0.1$ & $2.4\pm0.35$ & $31\pm5$ & ND \\
SQ B& $3.1\pm0.15$& $5.6\pm0.8$& $63\pm8$& $302(120)\pm 34(13)$\\
NGC 7319& $68 \pm3 $& $185\pm8$& $666\pm 67$& $1192\pm 130$\\
NGC 7320& $45\pm 3$& $38\pm 2$& $827\pm 85 $& $1899\pm 196$ \\
HII N & $0.69\pm0.07$& $1.1\pm0.2$& $52\pm6$& $98(78)\pm41(33)$\\
Shock region & $8.1\pm0.8$& $6\pm1.5$& $80\pm30$& $506\pm200$ \\
Extended emission & $29\pm2$& $40.6\pm5$ & $233\pm30$& $805\pm200$ \\
\enddata
\tablecomments{The $8{\rm \mu m}$ and $24{\rm \mu m}$ fluxes are obtained from aperture photometry while the $70{\rm \mu m}$ and $160{\rm \mu m}$ fluxes for the galaxies and compact star formation regions are the results of the FIR map fitting technique. The FIR fluxes from the shock region are obtained by the fitting of the FIR residual maps. The MIR and FIR values for the extended emission are derived from the curves of growth of the MIR and FIR maps (see Sect.\ref{phot_section}). The values in brackets are the $160{\rm \mu m}$ fluxes within the size of the $70{\rm \mu m}$ emission.}
\label{phot_spitzer}
\end{deluxetable}

\clearpage

\begin{figure}
\includegraphics[scale=0.5,angle=0]{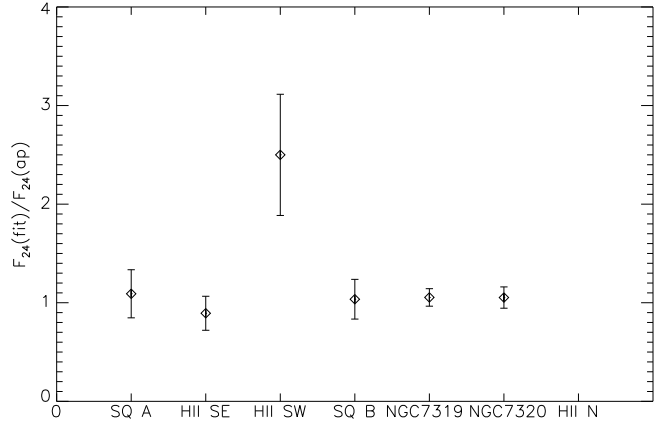}
\caption{Ratio between fluxes inferred from the fitting technique and from aperture photometry for the sources on the $24{\rm \mu m}$ map.}
\label{ap_vs_fit24}
\end{figure}

\begin{figure}
\includegraphics[scale=0.6,angle=0]{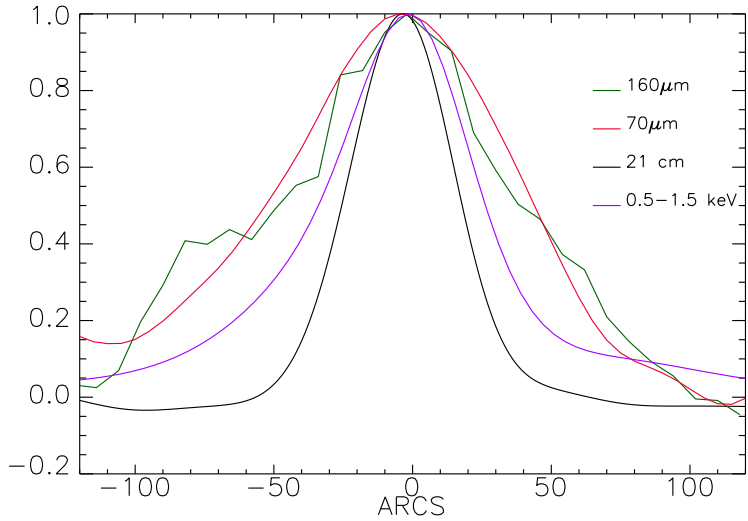}
\caption{East--West profile of the shock region at several wavelengths. The profiles for the $70{\rm \mu m}$, X-ray and Radio maps are derived after convolution to the resolution of the $160{\rm \mu m}$ map ($FWHM= 40''$). The profiles are derived from an horizontal strip passing through the center of the shock region ($RA=22^{\rm h}35'59.7$,$Dec=+33^\circ58'14.2'' $) and of vertical width equal to $10''$.}
\label{EW_prof}
\end{figure}

\begin{figure}
\includegraphics[scale=0.9,angle=0]{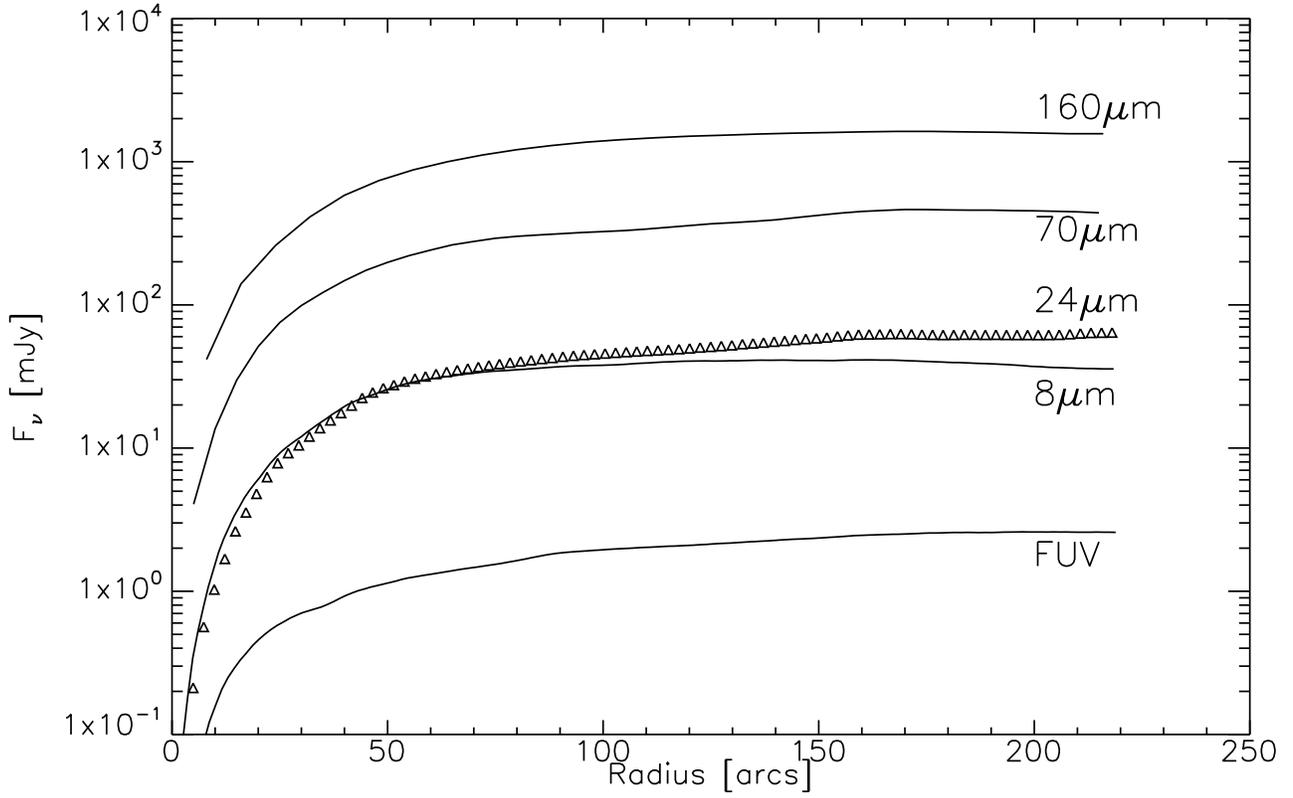}
\caption{Radial curves of growth of the FUV, MIR and FIR emission after masking or subtraction of star formation regions and the two galaxies NGC 7320 and NGC 7319 (see Sect.\ref{fir_res_map_par2}). The zero on the X-axis corresponds to the center of the shock region: $RA=22^{\rm h}35'59.7$,$Dec=+33^\circ58'14.2'' $}
\label{radial_int_prof}
\end{figure}

\begin{figure}
\includegraphics[scale=0.8,angle=0]{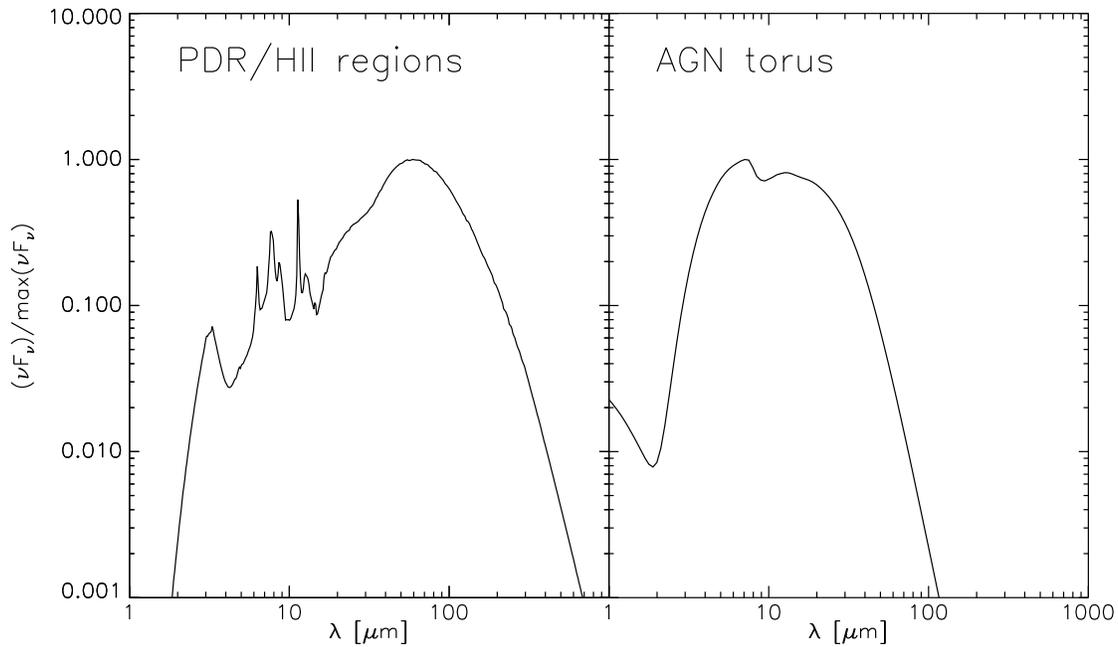}
\caption{PDR/HII region dust emission template (left panel) and AGN torus dust emission template (right panel)}
\label{agn_templ}
\end{figure}

\begin{figure}
\includegraphics[scale=0.8,angle=0]{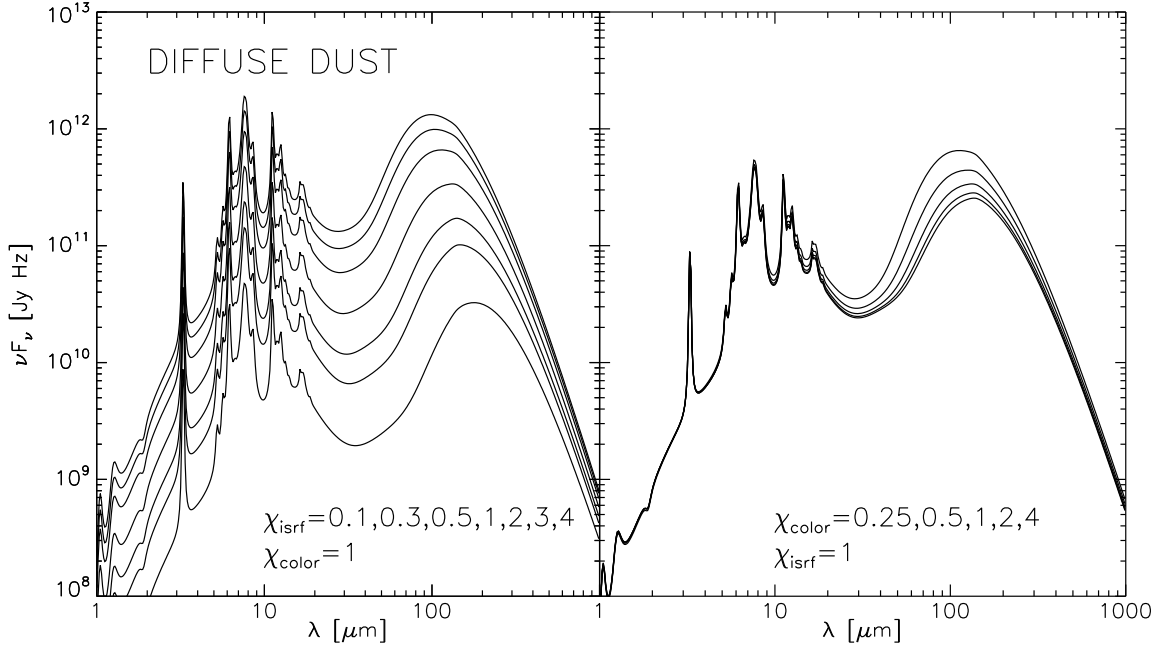}
\caption{Diffuse photon heated dust SEDs for different choices of $\chi_{\rm isrf}$ and $\chi_{\rm color}$. In the left panel the SEDs are calculated assuming $\chi_{\rm color}=1$ and several values of $\chi_{\rm isrf}$ (the higher curves correspond to higher values of $\chi_{\rm isrf}$). In the right panel we assumed $\chi_{\rm isrf}=1$ and different values of $\chi_{\rm color}$ (similarly, the higher curves correspond to higher values of $\chi_{\rm color}$). The SEDs are calculated assuming a distance equal to $94~{\rm Mpc}$ (SQ distance) and a dust mass $M_d=10^7~{\rm M_\odot}$.}
\label{diff_sed_int}
\end{figure}

\begin{figure}
\includegraphics[scale=0.5,angle=0]{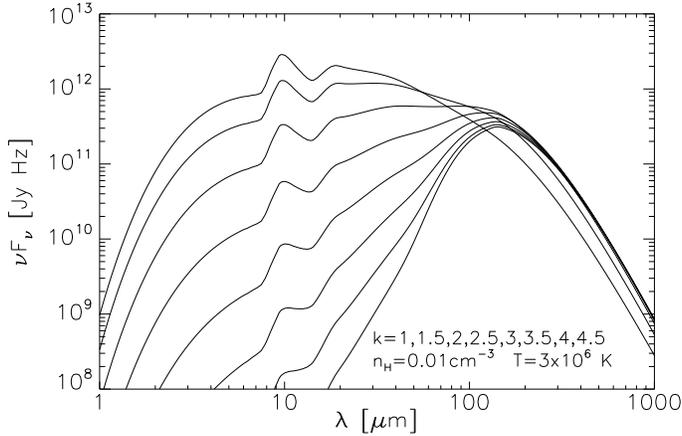}
\caption{Theoretical collisionally heated dust SEDs for several values of the size distribution exponent $k$. From the lowest to the highest curve the $k$ values are respectively $1, 1.5, 2, 2.5, 3, 3.5, 4, 4.5$. The SEDs are calculated assuming for the plasma density and temperature respectively $n_H=0.01~{\rm cm^{-3}}$ and $T=3\times10^6~{\rm K}$, a dust mass $M_{\rm d}=10^7~{\rm M_\odot}$ and a distance equal to $94~{\rm Mpc}$.  }
\label{coll_sed_k}
\end{figure}

\begin{figure}
 \includegraphics[scale=0.9,angle=0]{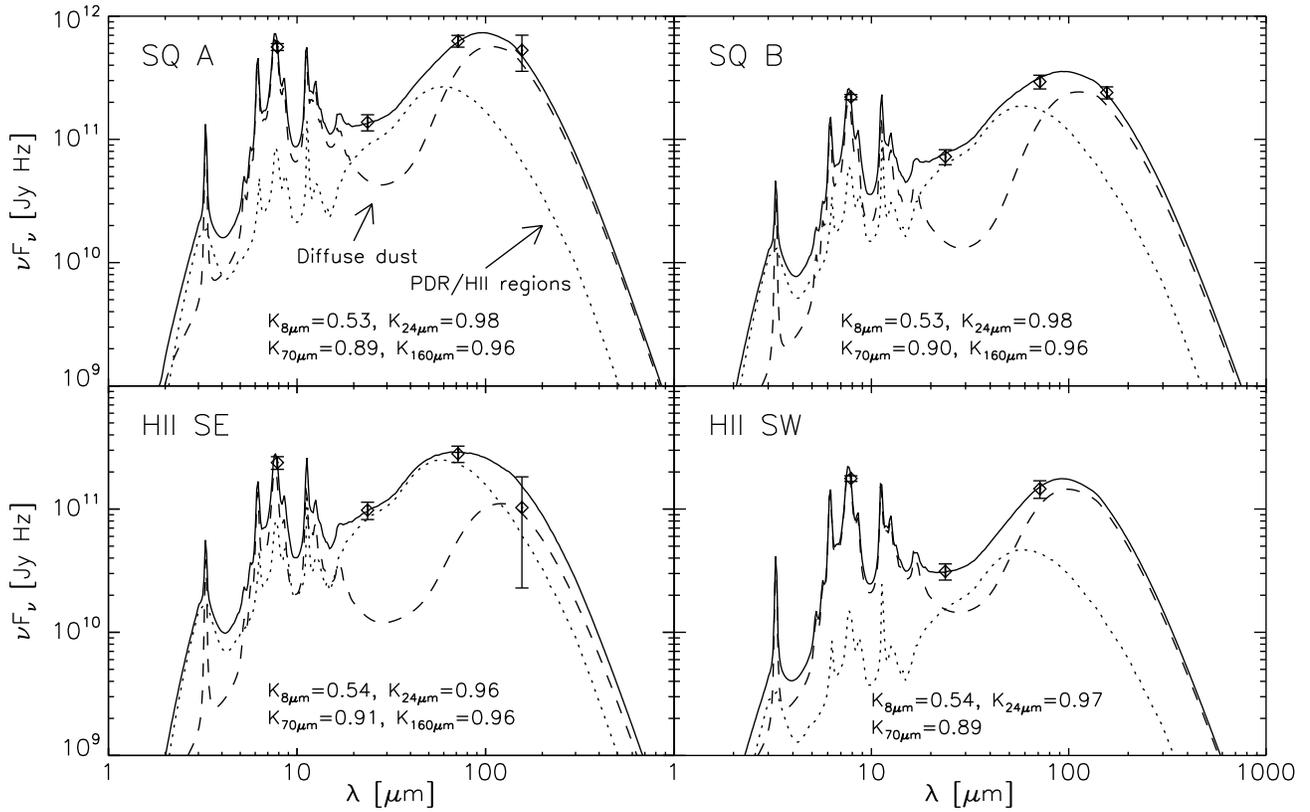}
\caption{SEDs of discrete SQ star formation regions: the plotted curves are the best fit SED (solid line), the contribution from PDR/HII regions (dotted curve) and diffuse dust (dashed line). The values $K_{8{\rm \mu m}}, K_{24{\rm \mu m}}, K_{70{\rm \mu m}}, K_{160{\rm \mu m}}$, shown in each plot of this and the following figures, are the color corrections applied to each fitted point.}
\label{SQA_SED}
\end{figure}

\clearpage

\begin{deluxetable}{lcccccccc}
\tablecolumns{9} 
\tablewidth{0pt} 
\tabletypesize{\scriptsize}
\tablecaption{Results from the infrared SED fitting}
\tablehead{\colhead{Source}& \colhead{$M_{\rm dust}$}& \colhead{$\chi_{\rm isrf}$}& \colhead{$\chi_{\rm color}$}& \colhead{$F_{24}$}&\colhead{$L_{\rm UV}^{\rm abs,local}$}  &\colhead{ $L_{\rm opt}^{\rm abs,diffuse}$}&\colhead{$L_{\rm UV}^{\rm abs,diffuse} $}  &\colhead{ $L_{\rm tot}$}\\
&\colhead{($10^7~{\rm M_\odot}$)} & & & & \colhead{($10^{43}~{\rm \frac{erg}{s}}$)} &\colhead{($10^{43}~{\rm \frac{erg}{s}}$)} &\colhead{($10^{43}~{\rm \frac{erg}{s}}$)} &\colhead{($10^{43}~{\rm \frac{erg}{s}}$)}} 
\startdata
 SQ A&  $0.6^{+0.1}_{-0.2}$ & $2$ & $2$& $0.65$ (HII)         & $ 0.45$               & $0.61$              & $0.36$                 & $1.4^{+0.2}_{-0.1}$ \\
 HII SE 7318b&  $0.22^{+0.04}_{-0.05} $ & $2$ & $0.25$& $0.86$(HII)     &$0.42$        & $0.10$             &$0.12$               & $0.64^{+0.06}_{-0.06} $\\
 HII SW 7318b& $0.11^{+0.01}_{-0.01}$ & $4$ & $1$ & $0.50$ (HII)     &$0.08$          & $0.14$             & $0.12$                 &$0.34^{+0.03}_{-0.01}$\\
 SQ B&  $0.37^{+0.09}_{-0.04}$ & $1$& $4$ & $0.80$ (HII)           &$0.31$             &$0.28$             &$0.10$                & $0.70^{+0.02}_{-0.03}$ \\
 NGC 7319& $2.7^{+0.4}_{-0.4} $ & $2$& $4$& $0.90$ (AGN)              & $4.13$          &$4.01$             & $1.47 $             & $9.6^{+0.3}_{-0.3} $ \\  
 NGC 7320& $0.08^{+0.02}_{-0.01}$ & $1$ & $4$ & $0.45$ (HII)         &$0.014$           &$0.059$           &$0.021$              & $0.094^{+0.005}_{-0.002}$ \\
 Shock region & $4.4^{+0.5}_{-1}$ & $0.3$ & $4$ & $0.35$ (HII)        & $0.14$        &$ 1.00$          & $0.37$                 & $1.5^{+0.1}_{-0.3}$ \\
 Extended emission& $47^{+3}_{-4}$ & $0.1 $ &$1$ & $0.71$(HII)        &$1.71$        &$1.56$           &$1.30$              &$4.6^{+0.2}_{-0.5}$ \\
\enddata
 \tablecomments{Col. 1: Fitted source; col. 2: Cold component dust mass; cols. 3-4: Radiation field $\chi_{\rm isrf}$ and $\chi_{\rm color}$ parameters (see Sect. 5); col. 5: Fraction of $24{\rm \mu m}$ flux contributed by HII regions or AGN torus templates; col. 6: Luminosity of dust in PDR/HII regions or in AGN torus in the case of NGC 7319; col. 7: Luminosity of diffuse dust powered by the absorption of optical photons of the diffuse radiation field; col. 8: Luminosity of diffuse dust powered by the absorption of UV photons of the diffuse radiation field; col. 9: Total dust luminosity.}
 \label{sedfitres}
 \end{deluxetable}

\begin{figure}
 \includegraphics[scale=0.8,angle=0]{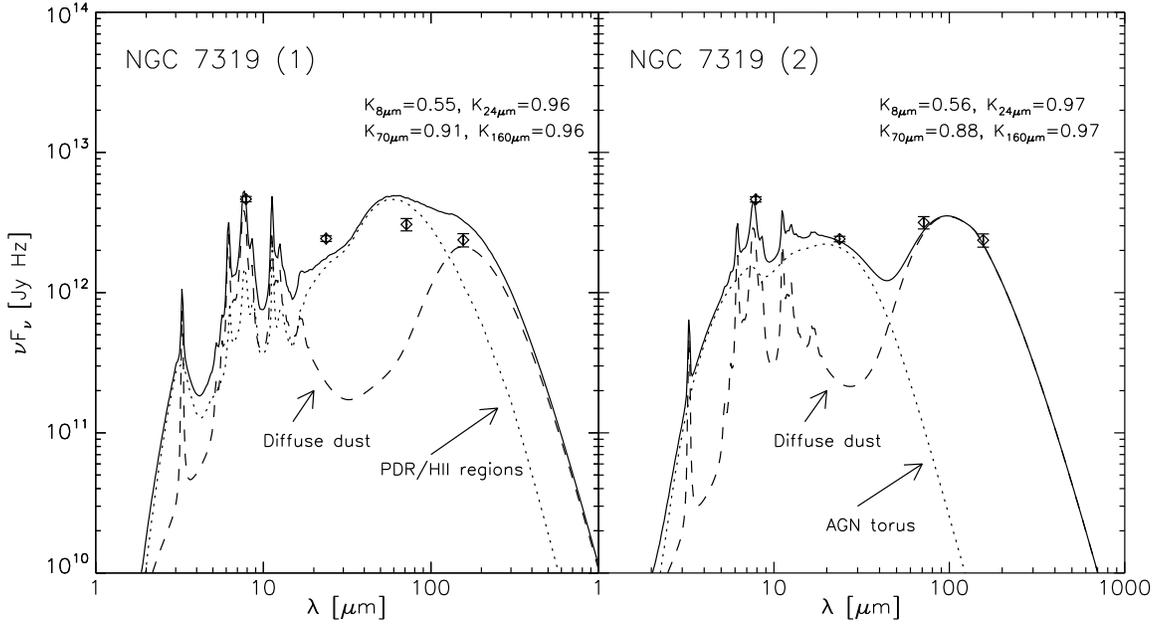}
\caption{SED of NGC 7319 performed using two different combinations of templates. In the left panel the best fit SED (solid line) is calculated using a PDR/HII regions template (dotted curve) and a diffuse dust component (dashed line). In the right panel the best fit is calculated using an AGN torus dust emission template (dotted) and a diffuse dust component (dashed).}
\label{NGC7319_SED}
\end{figure}

\begin{figure}
 \includegraphics[scale=0.8,angle=0]{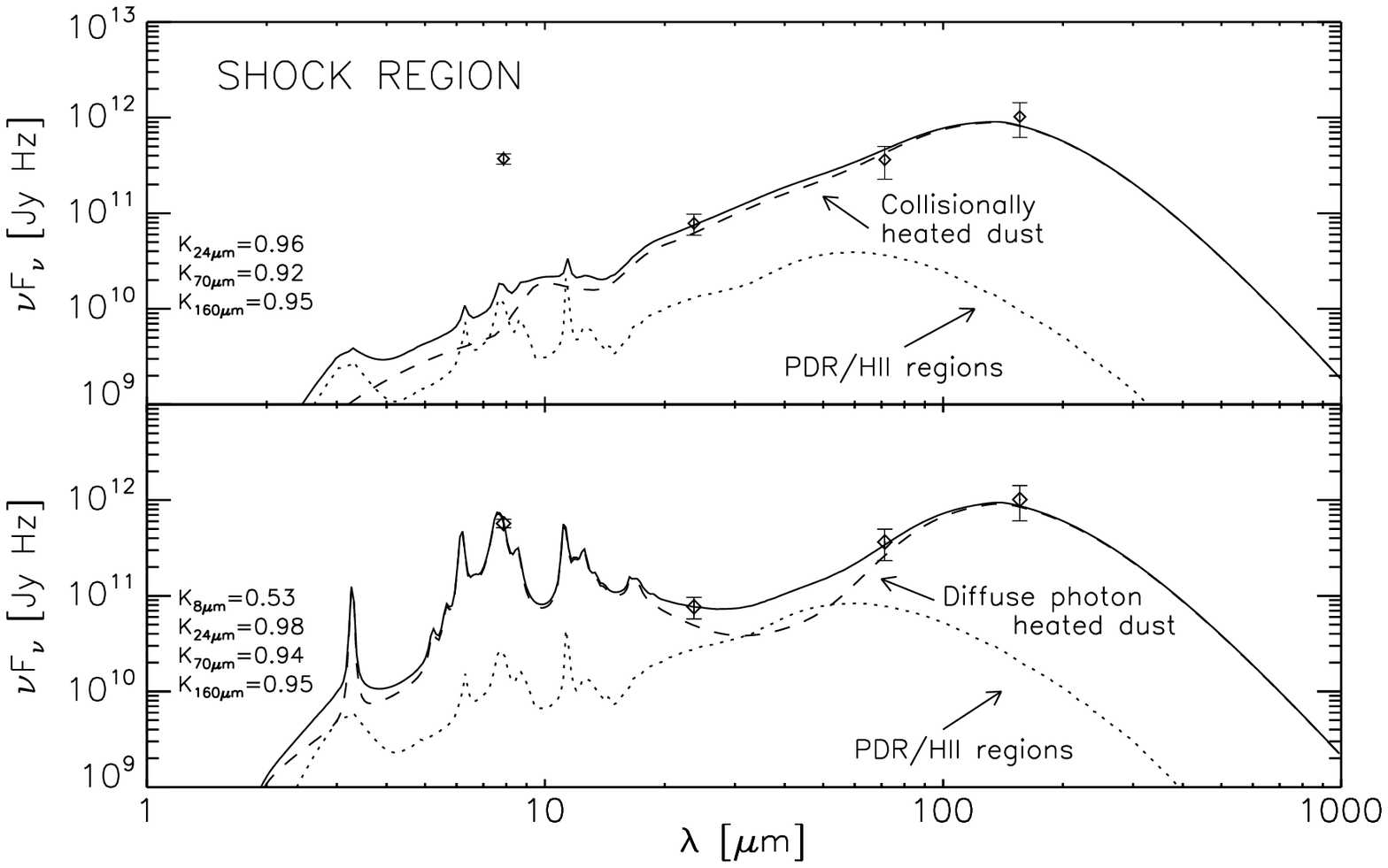}
\caption{SED of the shock region. The upper panel shows the fit performed using an PDR/HII region template plus a collisional heating component. The hot gas parameters adopted to calculate the collisionally heated dust SED are: $n=0.016~{\rm cm^{-3}}$ and $T=3\times10^6~{\rm K}$. The lower panel shows the fit performed using the PDR/HII region template and a uniformly photon--heated dust model.}
\label{shock_sed}
\end{figure}

\begin{figure}
 \includegraphics[scale=0.8,angle=0]{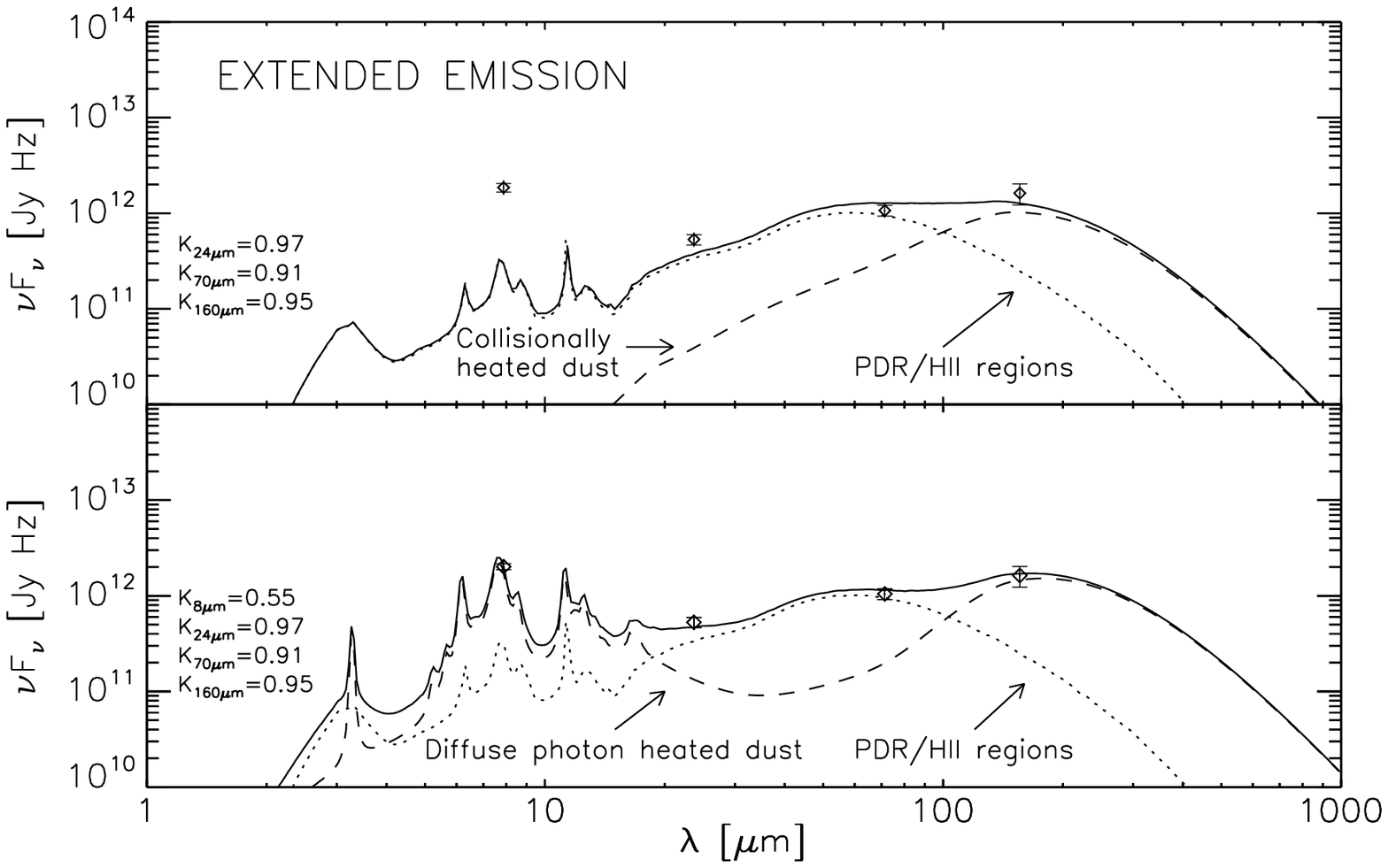}
\caption{SED of the extended emission. The upper panel shows the fit performed using a PDR/HII region template plus a collisional heating component. The hot gas parameters adopted to calculate the collisionally heated dust SED are: $n=0.001~{\rm cm^{-3}}$ and $T=6\times10^6~{\rm K}$.  The lower panel shows the fit performed using the HII region template and a uniformly photon--heated dust model.} 
\label{diff_sed}
\end{figure}

\begin{figure}
 \includegraphics[scale=0.8,angle=0]{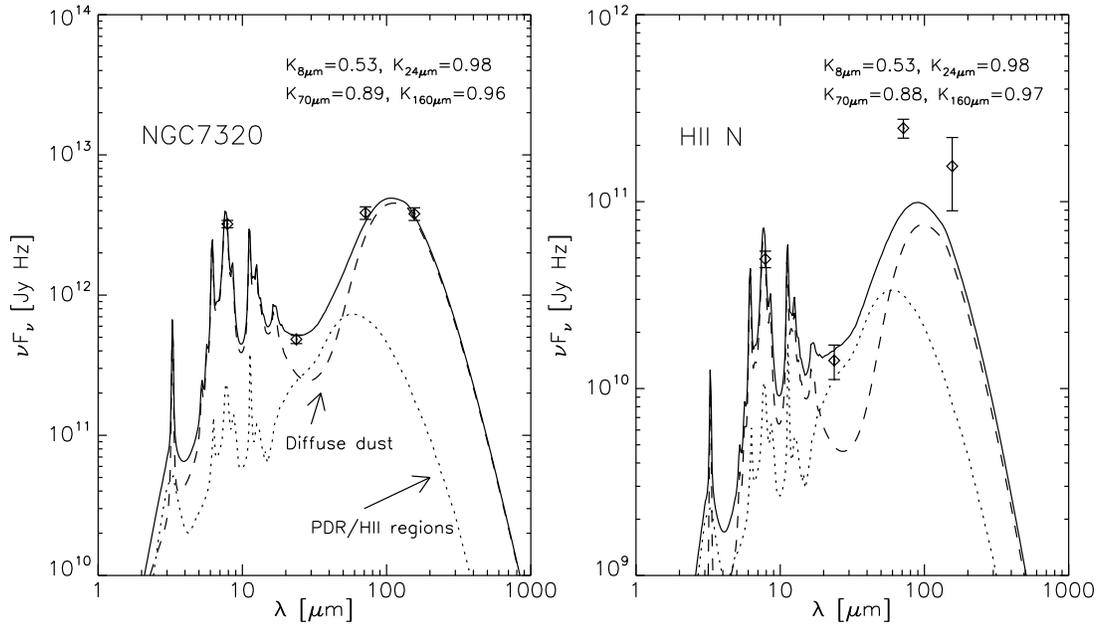}
\caption{SED of NGC 7320 and HII N: the plotted curves are the best fit SED (solid line), the contribution from PDR/HII regions (dotted curve) and diffuse dust (dashed line).}
\label{hii_n__SED}
\end{figure}

\begin{landscape}
\begin{deluxetable}{lccccccccccc}
\tablecolumns{12} 
\tablewidth{0pt} 
\tabletypesize{\scriptsize}
\tablecaption{Measured optical--UV luminosities, star formation rates and gas amount for the detected SQ dust emitting sources}
\tablehead{\colhead{Source}& \colhead{$L(H\alpha)$\tablenotemark{a}}& \colhead{$SFR_{H\alpha}$} & \colhead{$F_{\rm UV}^{\rm direct}$ }& \colhead{$L_{\rm dust,UV}$} & \colhead{$F^{\rm abs}_{\rm UV}$} & \colhead{$SFR_{\rm UV}$} & \colhead{$SFR_{\rm UV}/A$} & \colhead{$\Sigma(HI)$}& \colhead{$\Sigma(H_2)$}& \colhead{$\Sigma(H^+)$} & \colhead{$Z_d$\tablenotemark{b}} \\ 
& \colhead{($10^{40} {\rm \frac{erg}{s}}$)} & \colhead{(${\rm \frac{M_\odot}{yr}}$)} & \colhead{($10^{27}~{\rm \frac{erg}{s~Hz}}$)} &\colhead{($10^{42}~{\rm \frac{erg}{s}}$)} & \colhead{($10^{27}~{\rm \frac{erg}{s~Hz}}$)}&\colhead{(${\rm\frac{M_\odot}{yr}}$)} & \colhead{($10^{-3}{\rm \frac{M_\odot}{yr~kpc^{2}}}$)} & \colhead{(${\rm \frac{M_\odot}{pc^{2}}}$)} & \colhead{(${\rm \frac{M_\odot}{pc^{2}}}$)} & \colhead{(${\rm \frac{M_\odot}{pc^{2}}}$)}} 
\startdata
SQ A& $9.\pm2$ & $0.7\pm0.2$ & $2.75$ & $8.09 $ &  $4.49$ &  $0.78$  & $6.4$ &  $7.9 $ & $6.0$  & - &  $0.002$  \\
HII SE  & $6.6\pm1.5 $ & $0.5\pm0.1$ & $2.32$ & $5.44 $ & $3.02$ &$0.58 $ & $ 5.8 $ &  $<0.40$\tablenotemark{c}   & -  &  - & -  \\
HII SW  & $3.9\pm0.5$ & $0.2\pm0.04$ & $2.11$ & $1.99 $& $1.10$ & $0.35$ & $5.5 $  &  $3.2$   & $<1.7$\tablenotemark{c}  &  - & - \\
SQ B&  $1.3\pm0.3$ & $0.2\pm0.05$ &  $0.31$ & $4.18 $ & $2.32$ & $0.28 $ & $ 2.7$ & $5.2$& $4.0$  & - &   $0.002$ \\
NGC 7319 & - & - & $2.01$   & -   &  - & $\gtrsim0.22$ &  - & -   &- &- &-   \\ 
Shock  & - & - &  $6.98$   & $5.09 $ &  $2.83 $ & $1.05 $ & $ 3.1 $ & - & $9.1$ & $3.0$ & $0.007$  \\ 
Extended & - & - & $22.7$  & $ 30.21$& $16.78$  & $4.26$ & $ 1.0 $  &  - &  -   & $2.4$ & $0.01$  \\
\enddata
\tablecomments{Col. 1: source name; col. 2: $H\alpha$ luminosity; col. 3: star formation rate derived from the $H\alpha$ and $24{\rm \mu m}$ luminosity using the relation given by Calzetti et al.07; col. 4: observed GALEX FUV luminosity density corrected for Galactic extinction; col. 5: UV powered infrared dust luminosity; col. 6: absorbed UV luminosity density derived by dividing the UV powered dust luminosity $L_{\rm dust,UV}$ by $\Delta \nu (\rm UV)=1.8\times10^{15}~{\rm Hz}$; col. 7: SFR based on the inferred total (obscured and unobscured) UV luminosity density, using the relation by Salim et al.07; col. 8: star formation rates in col. 7 divided by the projected emission area; col. 9: average neutral hydrogen mass column density; col. 10: average molecular hydrogen mass column density; col. 11: average X-ray emitting plasma mass column density; col. 12: dust to gas ratio $Z_{\rm d}$ (note that for each source only the corresponding available gas masses are considered for the estimate of $Z_{\rm d}$)}
\label{sf_regions_tab}
\tablenotetext{a}{We measured the $H_\alpha$ fluxes from the interference filter maps published in Xu et al.1999. The quoted flux uncertainties are the quadratic sum of background fluctuations and flux calibration uncertainty ($\approx10\%$).}
\tablenotetext{b}{The dust to gas ratio is equal to $Z_{\rm d}=M_{\rm d}/(1.4\times M_{\rm gas})$ where $M_{\rm d}$ is the dust mass inferred by the SED fit, $M_{\rm gas}$ is equal to $N(H)\times \pi R^2$ with $R=FWHM(rad)\times distance$ the source size as derived from the $70{\rm \mu m}$ source fit (the factor $1.4$ is applied to take into account the contribution from noble gasses).} 
\tablenotetext{c}{The value of $\Sigma(HI)$ for HII SE is an upper limit because HI is not detected at that position on the maps by Williams et al.02. No value is given for $\Sigma(H2)$ because the area wasn't observed by Lisenfeld et al.2002. The value of $\Sigma(H2)$ for HII SW is an upper limit because CO has not been detected at that position.}
\end{deluxetable}
\end{landscape}
\clearpage

\begin{figure}
 \includegraphics[scale=0.8,angle=0]{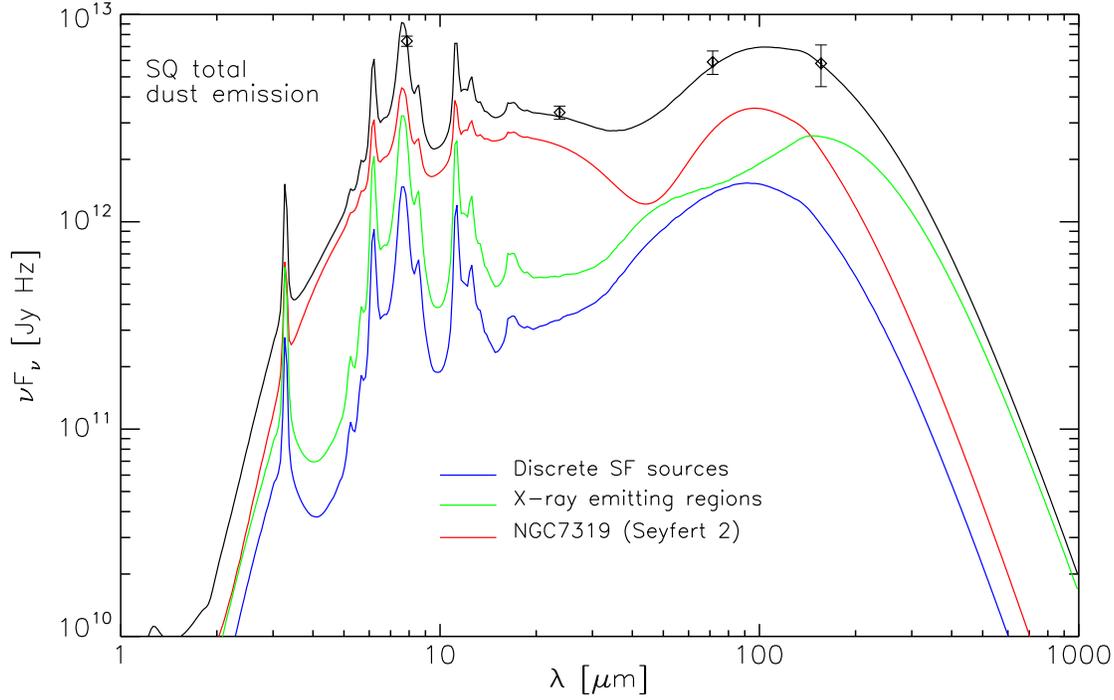}
\caption{SED of the total SQ dust emission. The plotted points are the sum of the fluxes of all the dust emitting sources in SQ: the discrete sources SQ-A, HII-SE, HII-SW, SQ-B, the AGN galaxy NGC 7319, the shock region and the extended emission (the applied color corrections are: $0.62$ ($8{\rm \mu m}$), $0.97$ ($24{\rm \mu m}$), $0.90$ ($70{\rm \mu m}$) and $0.97$ ($160{\rm \mu m}$)). The plotted curves are the total emission SED (black), the discrete star formation region total SED (blue), the AGN galaxy NGC 7319 SED (red) and the X-ray emitting region total SED (green). All the SEDs representing the total emission from more sources are derived summing the fitted SEDs as described in Sect.5. In the case of the X-ray emitting region SED, we combined the SED fits for the shock region and the extended emission performed using purely photon--heated dust emission components (see Sect. 5.3 and 5.4). } 
\label{global_sed}
\end{figure}

\begin{deluxetable}{llll}
\tablecolumns{4} 
\tablewidth{0pt} 
\tabletypesize{\scriptsize}
\tablecaption{Galaxy optical photometry and stellar masses}
\tablehead{\colhead{Source}& \colhead{$g$} & \colhead{$r$} & \colhead{$M_*$\tablenotemark{a}}} 
\startdata
NGC 7319 & 13.58 & 12.83 & 1.6 \\
NGC 7318a & 15.00 & 14.06 & 0.8 \\
NGC 7318b & 13.31 & 12.55 & 2.1 \\ 
NGC 7317 & 14.58 & 13.78 & 0.8 \\ 
\enddata
\tablecomments{Col. 1: source name; cols. 2-3: SDSS $g$ and $r$ apparent magnitudes; col. 4: galaxy stellar mass ($10^{11}{\rm M_\odot}$)}
\label{sdss_tab}
\tablenotetext{a}{The stellar mass has been estimated using one of the relations given by Bell et al. 2003 to calculate the stellar mass to luminosity ratio from SDSS magnitudes: $log_{10}(M/L_g)=-0.499+1.519\times (g-r)$ (note that $M/L_g$ is in solar units).}
\end{deluxetable}

\begin{figure}
 \includegraphics[scale=0.8,angle=0]{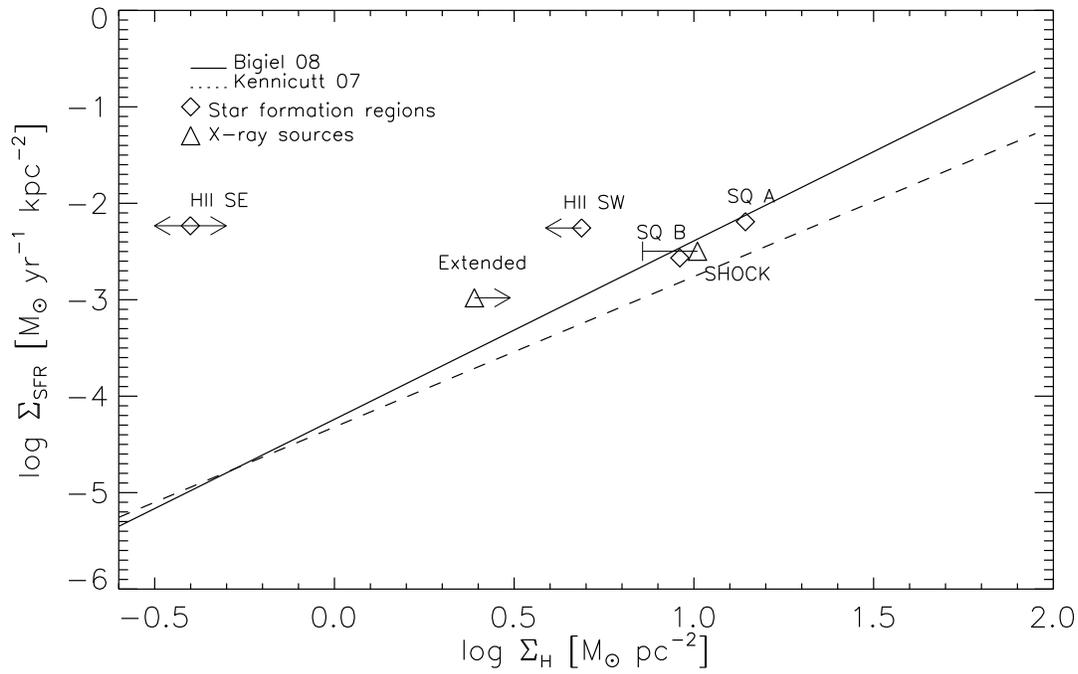}
\caption{Star formation rate per unit area $\Sigma_{\rm SFR}$ (${\rm M_\odot/yr/kpc^2}$) vs total gas mass surface density $\Sigma_{\rm H}$ (${\rm M_\odot/pc^2}$) for dust emission sources in SQ. The plotted lines are the relations found by Kennicutt et al.07 and Biegiel et al.08. The arrows pointing towards right/left for some of the plotted measurements indicate that the corresponding $\Sigma_{\rm H}$ value is a lower/upper limit (see Table 4). In case of HII SE the plotted gas surface density corresponds to the upper limit of the atomic hydrogen column density but the total gas density can also be higher since the amount of molecular gas is unknown. In the case of the shock region point, whose $\Sigma_{\rm H}$ value is the sum of both cold and hot gas components, a bar extends towards the left until the point where $\Sigma_{\rm H}$ corresponds to the cold gas mass surface density for this source.} 
\label{plot_sfr_sgas}
\end{figure}

\clearpage

\appendix
\section{Multisource fit of SQ FIR maps}
\label{ap1}

The technique we developed for the fitting of the SPITZER FIR maps is similar to the CLEAN algorithms used in radio astronomy but with some additional characteristics: 1) all the sources are fitted at the same time; 2) an intrinsic finite source width is allowed. Our basic assumption is that all the sources we model are sufficiently well described by elliptical gaussians or a combination of them. This is not necessarily true because some sources can, in principle, differ significantly from this simple shape. However this simple approach has been sufficient for a good fitting of the FIR maps. The function $F(x,y)$, describing an elliptical gaussian on a map, is defined by the following formulas: 
\begin{eqnarray}
F(x,y)=B_{\rm loc}+A\times e^{-U/2} \\
U=(x'/\sigma_x)^2+(y'/\sigma_y)^2\\
x'=(x-x_s)\cos\theta-(y-y_s)\sin\theta \\
y'=(x-x_s)\sin\theta+(y-y_s)\cos\theta 
\end{eqnarray}
The first two equations define the shape of the elliptical gaussian while the last two are the coordinate transformations between the map and the frame defined by the elliptical gaussian axis. For each source the fitting procedure has to provide 7 parameters: $A$ (the gaussian amplitude), $(x_s,y_s)$ (the position of the source center), $(\sigma_x,\sigma_y)$ (the gaussian widths in two ortogonal directions), $\Theta$ (the rotation angle of the gaussian axis from the array axis) and an offset value $B_{\rm loc}$ (the local background for each source). 

\begin{figure}
\plottwo{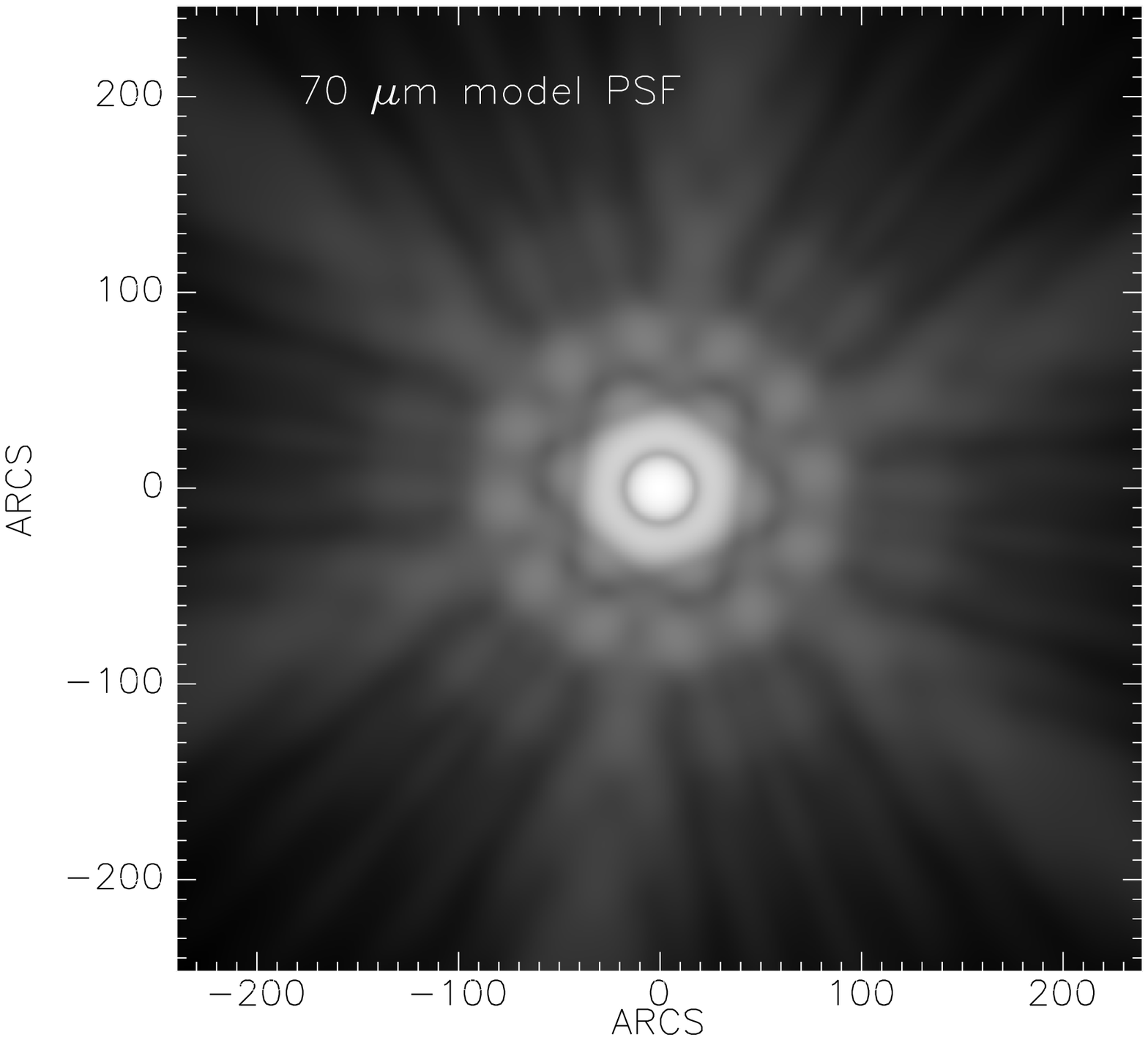}{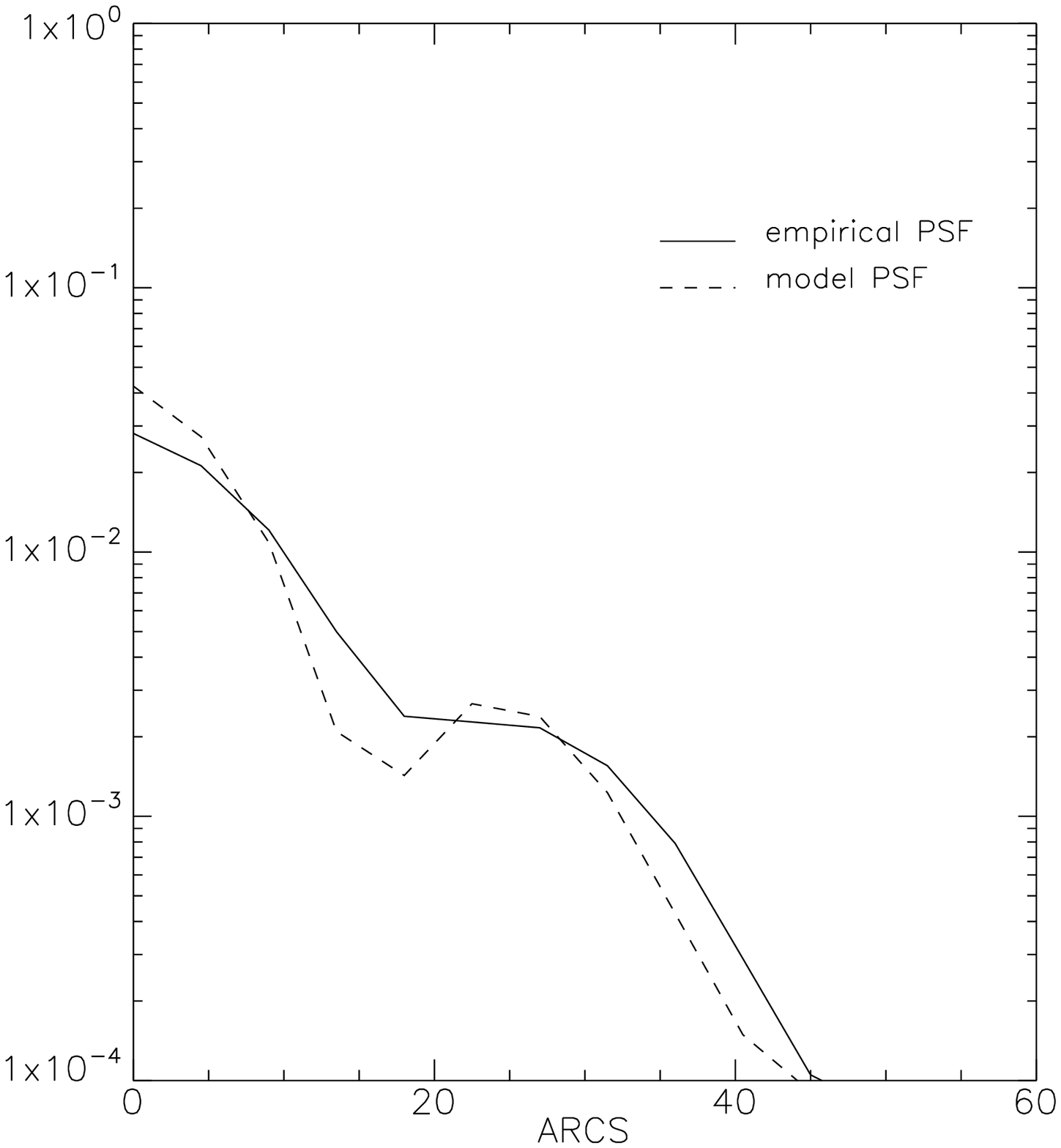}
\caption{a) Theoretical $70 {\rm \mu m}$ PSF; b) Average radial profiles empirical and theoretical  $70 {\rm \mu m}$ PSF.}
\label{psf70}
 \end{figure}

\begin{figure}
\plottwo{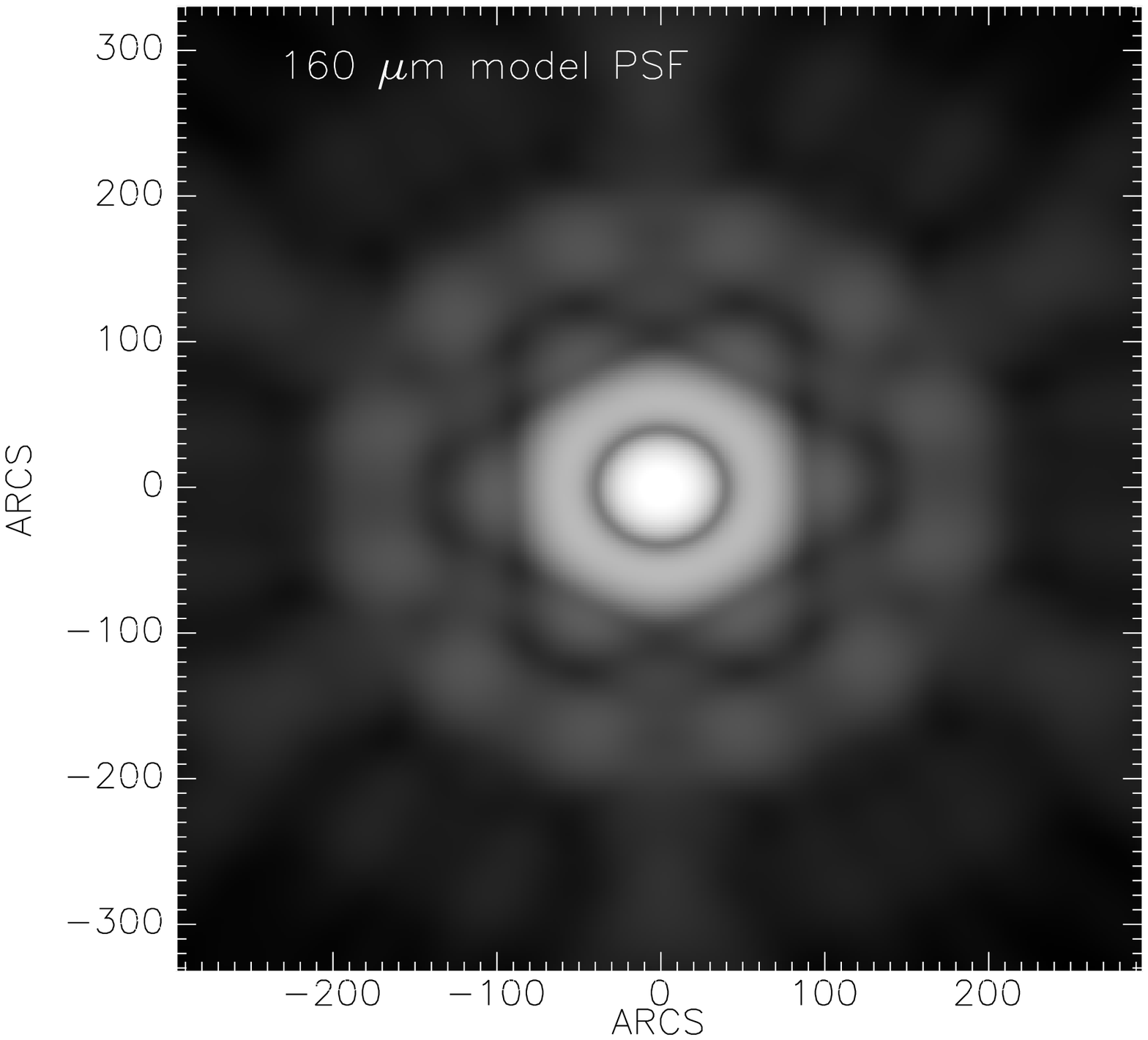}{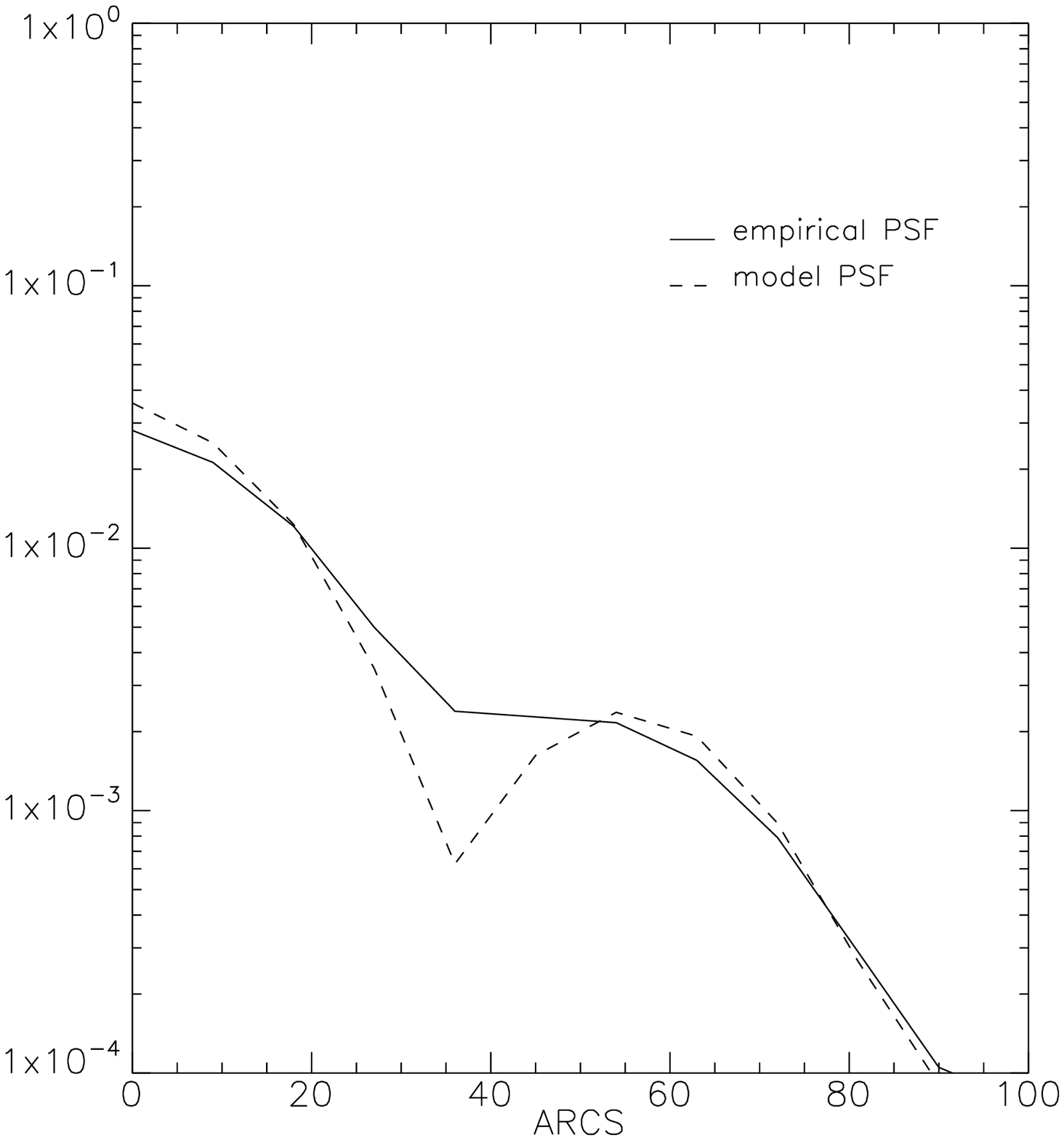}
\caption{a) Theoretical $160 {\rm \mu m}$ PSF; b) Average radial profiles empirical and theoretical  $160 {\rm \mu m}$ PSF.}
\label{psf160}
 \end{figure}

A real image is not just the sampling of the original signal, of course, but it's the sampling of the convolution of the signal with the point spread function (PSF) of the telescope optics--detector instrument. Using the program STINYTIM\footnote{by John Krist for the Spitzer Science Center: http://ssc.spitzer.caltech.edu/dataanalysistools/tools/contributed/general/stinytim/}, we obtained theoretical PSFs of the MIPS instrument for both 70 and 160 ${\rm \mu m}$ bands. Fig. \ref{psf70} and \ref{psf160} show the theoretical PSFs for the 70 ${\rm \mu m}$ and $160{\rm \mu m}$ bands and their profiles. In panel (b) of these figures, it is shown also the radial profile of an empirical PSFs (kindly provided by G. Bendo). Although there are some intrinsic differences between empirical and theoretical profiles, we decided to use theoretical PSFs because they can be sampled at any desired rate and do not contain noise that would be added to the fit (however we made some tests to determine the level of uncertainty introduced by using theoretical PSFs instead of real ones in our fitting procedure. We have found that the difference on the final results is negligible compared to other sources of error, e.g. only $\approx 5\%$ on the inferred source fluxes).

The first step of our fitting procedure is to assume a distribution of elliptical gaussian sources, with initial trial values of the parameters, in an array of the same size of the data. First from a direct inspection of the FIR maps and then after some iterations, we identified 10 main sources to be fitted using elliptical gaussians (see Fig. \ref{70mfit}): 2 for NGC 7320 (one for the disk emission plus one for a compact HII region), 1 each for SQ A and SQ B, 2 for NGC 7319 (one for the circumnuclear emission plus one for a HII region), 1 for a source located north of NGC 7319 (HII N), 2 for NGC 7318b (the 2 compact HII regions located on the southern spiral arms HII-SE and HII-SW), 1 for a source peaked in the middle of the shock region. Convolving this array with a theoretical PSF, an artificial image is obtained that can be compared with the real data. Varying the parameters of the elliptical gaussians, it is possible to improve the agreement between artificial and real images until a certain degree of accuracy is reached. To minimize $\chi ^2$ and find the best fit model parameters, we used the Levemberg--Marquardt (LM) algorithm (see \citealt{BE92}), implemented by the IDL routine MPFIT2DFUN\footnote{written by Craig Markwardt: http://cow.physics.wisc.edu/$\sim$craigm/idl/fitting.html for details on the $\chi^2$ minimization routine}. Although this algorithm is less dependent on the initial trial values of the parameters, compared with other $\chi ^2$ minimization methods, the results are inevitably dependent on these. Therefore it is a good rule to assign the initial values as close as possible to the real ones (or better to which we believe are the real ones). The criteria we used to choose the initial values for the free parameters are the following: \\
1) Amplitude: any number of the order of the peak amplitude of the source (this fitted parameter has been found almost independent from the initial value);\\
2) Position: initial positions near the peaks of the sources;\\
3) Gaussian Widths: derived approximately from the apparent width of the source;\\
4) Rotation angle: any number.\\
5) Local Background: initial value equal to zero.\\ 
As one can notice, the choice of the initial values does not require to know in advance any parameter but the approximate positions of the source centers and roughly the sizes of the gaussians. All this information can be easily derived from the data. The fitting procedure improves the quality of the results (in the sense that we obtain a smoother fit residual map and a lower reduced $\chi^2$) if we consider only the regions close to the sources for the fit. We created a mask to exclude  all the pixels that are too distant from the calculation, that is more than 2-3 times the apparent sizes, from the source centers. 

The fitting technique as described until now has been applied successfully to the $70{\rm \mu m}$ FIR map. The fitting of the $160{\rm \mu m}$ map has been performed in a slightly different way because one cannot clearly see on this map all the compact sources that are detected on the $70{\rm \mu m}$. This effect is most probably due to both lower resolution and intrinsic color differences. Since our goal was to produce consistent SEDs of the main sources in SQ, we introduced new constrains for the fit of the $160{\rm \mu m}$ map. We assumed that the centre position of each source on the $160{\rm \mu m}$ map is the same as derived from the 70 ${\rm \mu m}$ map fitting. We only allowed a common shift of all the source positions in order to correct possible small differences in the astrometry of the two maps. Then we assumed that the orientation angles and the ratio of the elliptical gaussian axial widths $(\sigma_x,\sigma_y)$ for the compact sources (SQ A, HII SE, HII SW, SQ B, HII N and the HII regions in NGC 7319 and NGC 7320) are the same as those inferred from the $70{\rm \mu m}$ map fit. Therefore we assumed that the shape of the emitting compact sources is similar at $70$ and $160{\rm \mu m}$ even if the size can be different. This is actually expected because colder emission can come from regions that are simply farther away from the central heating source in case of star formation regions. 

Figs. \ref{70mfit} and \ref{160mfit} show the results of our fitting technique for the $70{\rm \mu m}$ and $160{\rm \mu m}$ maps in four panels: a) the background subtracted Spitzer FIR map of SQ, b) the best fit image obtained with our method, c) the fit residuals and d) the "deconvolved" map (that is actually the map showing the elliptical gaussians whose PSF convolution gives the best fit map). As one can see, this method produces best fit maps with remarkable similarity to the original maps ($\chi_{70{\rm \mu m}}^2/N_{\rm free}=0.8$, $\chi_{160{\rm \mu m}}^2/N_{\rm free}=0.7$). The ``deconvolved'' map at $70{\rm \mu m}$ shows, apart from the emission of the foreground galaxy NGC 7320 and AGN galaxy NGC 7319, many discrete compact sources that are responsible for the several peaks seen on the real map. There is a rather extended source that peaks in the middle of the shock region, even though some excess of emission is still seen on the fit residual map at this position suggesting that the accuracy of the fit is not very good for this faint source. At $160{\rm \mu m}$ the emission is dominated by the AGN galaxy, the foreground galaxy and the source peaked on the shock region. It is quite interesting that the gaussian used to fit the emission at the position of the shock is very well aligned along the main axis of the shock ridge. However the east--west width of this gaussian is too large to be related only to the X-ray ridge (see Sect. \ref{shock_phot_par})). Some compact star formation regions are undetected (or only marginally detected) at $160{\rm \mu m}$, as we somehow expected because no peaks are clearly seen at their position. 

The main fitted parameters and the derived flux densities for each of the fitted sources can be found in Table \ref{FIR_sources}. The uncertainties on the integrated fluxes are derived from the covariance matrix provided by the LM $\chi^2$ minimization algorithm. For the estimate of the error on each point of the map, necessary to evaluate $\chi^2$ we considered the variance of background fluctuations on each map. On the final integrated fluxes for each source a $10\%$ error is added due to flux calibration uncertainties (MIPS data handbook). As one can notice from Table.\ref{FIR_sources}, the inferred $160{\rm \mu m}$ sizes for the AGN galaxy NGC 7319 and the compact sources HII SE, SQ B and HII N are much larger than the $70{\rm \mu m}$ size. In case of the AGN galaxy the size difference can be physically understood if one thinks that additional warm emission come from the central part of the galaxy and, therefore, the $70{\rm \mu m}$ flux is more peaked towards the center (see also discussion in Sect. \ref{agn_gal_em_disc}). The larger $160{\rm \mu m}$ size for the compact sources arises because of the confusion with fainter sources close to the main central source. For these sources we derived from the deconvolved $160{\rm \mu m}$ gaussians the amount of flux within one $70{\rm \mu m}~ FWHM$ from the peak. These reduced $160{\rm \mu m}$ fluxes are quoted within brackets in Table \ref{phot_spitzer}. 

\subsubsection{Comparison between results obtained by the fitting technique and aperture photometry}
\begin{figure}
\includegraphics[scale=0.8,angle=0]{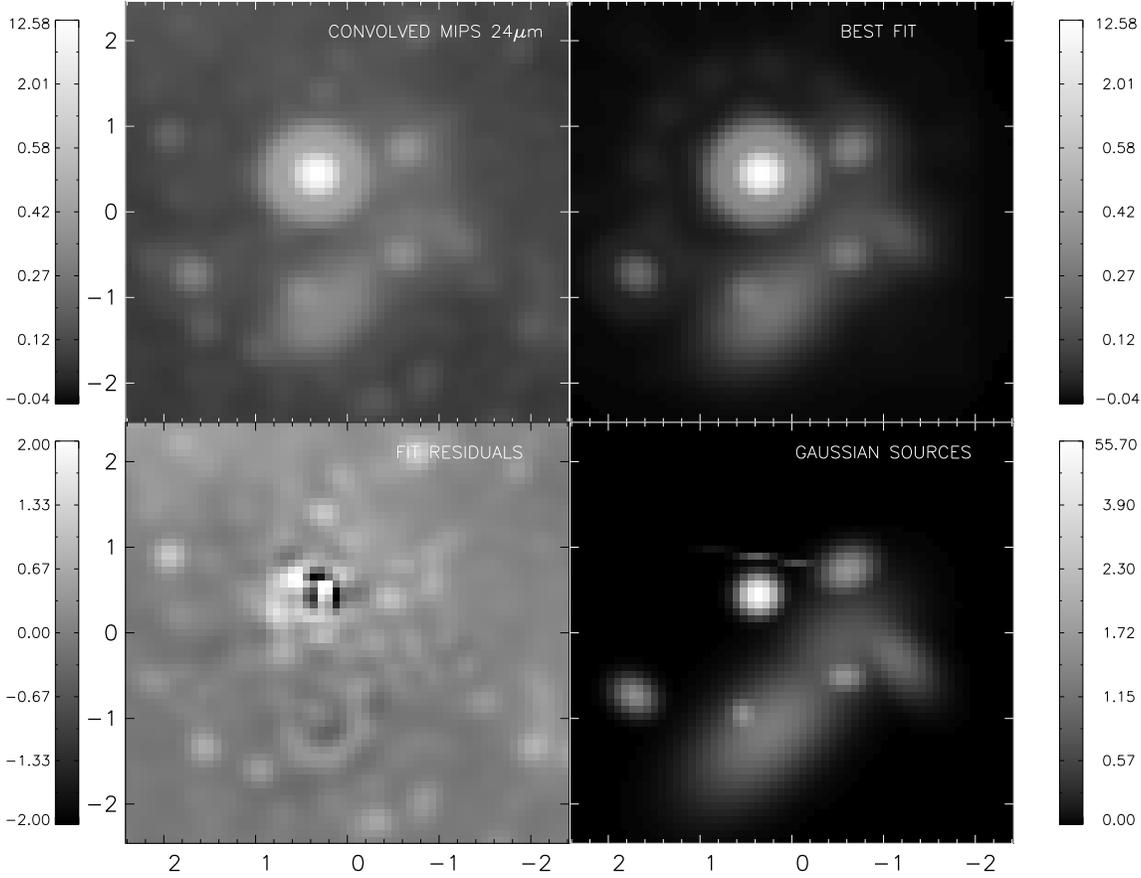}
\caption{Convolved $24{\rm \mu m}$ map fit. Top--left: convolved $24{\rm \mu m}$ map; Top--right: best fit map; Bottom--left: fit residuals; Bottom--right: deconvolved gaussian map. Units of the values aside the color bars are MJy/sr.}
\label{fit_24_res70_pix70}
\end{figure}

Since in Sect. \ref{IF_SED_fit_par} we performed source SED fitting including MIR and FIR measurements, performed respectively using aperture photometry and the map fitting technique, it is important to understand how different the fluxes obtained by the fitting procedure are compared to those we would have obtained by aperture photometry on a high resolution map. To quantify this difference, we made a test using the higher resolution $24{\rm \mu m}$ map where dust emission morphology looks remarkably similar to the $70{\rm \mu m}$ emission (at least if one considers the main sources of emission). We convolved the $24{\rm \mu m}$ MIR map to match the resolution of the $70{\rm \mu m}$ map, resampled the map to match the pixel size of the $70{\rm \mu m}$ map and then performed the gaussian fitting technique on the convolved map. The convolution has been done using the kernel function created by Carl Gordon (\citealt{Gord08}). For the fit we assumed the same distribution of gaussians as in the fit of the FIR maps, but leaving all the parameters free. The convolved map and the fit results are shown in Fig. \ref{fit_24_res70_pix70}. In Table \ref{gausfit24} and in Fig. \ref{ap_vs_fit24} we compared the results from aperture photometry (described in Sect. \ref{sfr_phot}) and gaussian fitting. For all but one source (HII SW) the fluxes derived using the two different methods are consistent within the uncertainties and also the inferred source sizes are close to those found at $70{\rm \mu m}$ (the problem with HII SW arises because of the vicinity of the nucleus of NGC 7318a that is much brighter in the MIR than in the FIR). Based on the results of this test, we are confident that the two photometric techniques give consistent results.

\section{Dust SED templates and models}
\label{sed_temp_mod}
In order to fit the observed source SEDs, we needed to model the dust emission from several environments: dust in PDR/HII regions, dust heated by diffuse radiation fields, dust in AGN tori and collisionally heated dust embedded in hot plasmas. In this section, we describe the characteristics of the SEDs we used to reproduce the emission from each of these regions.

\subsubsection{Dust in PDR/HII regions}
Dust in HII and photodissociation regions (PDR), close to young stars, is warm and emits predominantly in the MIR. To model this emission, we used an SED template  derived by the fitting of Milky Way star formation region infrared emission with the theoretical model of \cite{Groves08}. The Galactic star formation regions considered are the radio--selected sample of \cite{Conti04}. The template is the average fitted spectra of the observed emission from these regions. Further details will be given in a forthcoming paper (Popescu et al. 2010, submitted, Sect. 2.9). The template is shown in the left panel of Fig. \ref{agn_templ}. 

\subsubsection{Dust heated by diffuse radiation fields}

To fit the diffuse dust emission, we created a grid of theoretical SEDs of emission from dust heated by a uniform radiation field. The code we used to calculate dust emission, developed by Joerg Fischera (details will be given in Fischera et al. 2010, in preparation), assumes a dust composition including graphite, silicates, iron and PAH molecules. The abundances and size distributions are the same as in Table 2 of \cite{Fischera08}. Dust emission is calculated taking into account the stochastic temperature fluctuations of dust grains following the numerical method described by \cite{Guha89}, combined with the step wise analytical solution of \cite{Voit91}. We created the set of SEDs by varying the intensity and the color of the radiation field heating the dust and keeping the dust composition fixed. The standard shape of the radiation field spectra, adopted by the dust emission code, is that derived by \cite{Mathis83} to model the Galactic local ISRF. In this spectrum, the stellar contribution consists of four components: a UV emission from early type stars plus three blackbody curves at temperatures $T_2=7500~{\rm K}$, $T_3=4000~{\rm K}$ and $T_4=3000~{\rm K}$ used to reproduce, respectively, the emission from young/old disk stars and Red Giants. Each blackbody curve is multiplied by a dilution factor $W_i$ that defines its intensity. The whole spectrum is multiplied by a parameter, $\chi_{\rm isrf}$, whose value is unity when the spectrum intensity is the same as in the local radiation field. In order to consider different colors of the radiation field, we defined a second parameter, $\chi_{\rm color}$, that multiplies only the blackbody curves used to model the old disk population and the Red Giant emission (blackbody temperatures $T_3$ and $T_4$). For a range of values of $\chi_{\rm isrf}$ (0.1,0.3,0.5,1,2,4) and $\chi_{\rm color}$ (0.25,0.5,1,2,4), we calculated the corresponding dust emission and created the set of models for the SED fit. The result of increasing the $\chi_{\rm isrf}$ value is that the overall intensity of the radiation field is higher and, therefore, the equilibrium dust temperature is higher, leading to a warmer emission in the FIR (see left panel of Fig. \ref{diff_sed_int}). A value of $\chi_{\rm color}$ higher than unity increases the relative contribution of the cold stellar emission, which peaks in the optical range. These wavelengths are more efficiently absorbed by big grains than by PAH molecules (and small grains). As a consequence, by varying this parameter, one modifies the ratio between the peak intensity at the FIR, produced by big grains and the peak intensity at $8{\rm \mu m}$, due to PAH molecules which are mainly heated by UV photons (see right panel of Fig. \ref{diff_sed_int}). A similar effect would have also been produced by varying the relative abundance of PAH molecules and solid grains. 

\subsubsection{Dust emission from AGN torus}
We selected a theoretical dusty torus SED between those created by Feltre et al. (2010, in preparation), using the model presented by \cite{fritz06}. In this model, the characteristic of the torus, supposed to be homogeneous, are defined by several parameters: opening angle, external to internal radius ratio, equatorial optical depth at $9.7{\rm \mu m}$ and two parameters, $\beta$ and $\gamma$, that determine the radial and the angular density profile respectively according to the formula $\rho(r,\theta)=\alpha r^{-\beta} e^{-\gamma|\cos(\theta)|}$, where $r$ is the radial coordinate and $\theta$ is the angle between a volume element and the equatorial plane (note that $\alpha$ is determined by the torus optical depth). We selected a model with a large opening angle ($140^\circ$), with an angular gradient in the density profile ($\beta=0$, $\gamma=6$), external to internal radius ratio equal to $30$ and equatiorial optical depth $\tau(9.7{\rm \mu m})=6$. The chosen parameters are well inside the range of values found by \cite{fritz06} from the fitting of a sample of Seyfert2 galaxies. Then, we extracted the output SED in the equatorial view, according to the Seyfert 2 classification of NGC 7319, the AGN galaxy in SQ. The extracted SED is the template we used for fitting the AGN torus emission and it is shown in the right panel of Fig. \ref{agn_templ}.

\subsubsection{Dust heated by collisions in X-ray plasmas}
\label{dust_coll_code}
Finally, we needed to model the emission from collisionally heated dust. For this purpose, we used a code, created by J.Fischera, based on the works by \cite{Guha89} and \cite{Voit91} for calculating the stochastical heating of the dust grains and using the results of \cite{Draine79} and \cite{Dwek87} to determine the heating rates due to dust--plasma particle collisions. Dust composition is assumed to be a mix of graphite ($47\%$) and silicates ($53\%$). PAH molecules are not included but, since they have an extremely small destruction time scale when embedded in hot plasmas with $T>10^6~{\rm K}$, their abundance is expected to be negligible. In order to perform the calculations, one needs to specify the physical properties of the plasma, i.e. density, temperature and metallicity. The exact value of metallicity is not important because dust is predominantly heated by collisions with electrons, provided mainly by ionized hydrogen and helium. Plasma temperature and density are derived from X-ray data (see Appendix \ref{ap3}). An important phenomena that should be taken into account while preparing the SED models is dust destruction due to sputtering (\citealt{Draine79}). Because the efficiency of sputtering depends on the grain size, we expect that the size distribution of grains in hot plasmas is different from that of dust in the cold ISM, typicall assumed to be $n(a)\propto a^{-k}$ with $k=3.5$ (\citealt{Mathis77}). Therefore we created a set of SEDs varying 1) the plasma physical parameters (see Table \ref{xray_dens}) and 2) the exponent $k$ of the size distribution in the range $[1,3.5]$. A $k$ value lower than $3.5$ means that the relative number of big grains is higher compared to the standard size distribution. Since the destruction time scale due to dust sputtering is directly proportional to the grain size (\citealt{Draine79}), a higher relative abundance of big grains is indeed expected in hot plasmas. Once the heating rates due to collisions are fixed, a change in $k$ modifies the color of the emitted radiation (see Fig. \ref{coll_sed_k}). This happens because big grains are generally colder than small grains, and do not provide warm dust emission due to stochastic heating. Therefore, changing their relative abundance, one can obtain colder or warmer SEDs. 

\section{X-ray plasma temperature and density in SQ shock and halo}
\label{ap3}
As we mentioned in \ref{dust_coll_code}, the hot plasma physical parameters, needed for the calculation of the emission SED from collisionally heated dust, in both the shock and halo gas, are derived from the X-ray emission. The temperatures characteristic of the plasma have been obtained by \cite{T05} fitting XMM-NETWON X-ray data and quoted in their Table 3. They fitted the source X-ray spectra with two temperature plasma models. The fitted colder component is predominant for the shock region and has a temperature $T\approx3\times10^6~{\rm K}$ that we assumed to calculate the dust emission. For the halo region, the two plasma components have similar luminosities. In this case, we assumed the average value $T=6\times10^6~{\rm K}$ for the dust emission calculation (the fitted temperatures are $T_1=3.5\times10^6~{\rm K}$ and $T_2=8.8\times10^6~{\rm K}$). The plasma densities have been derived from the X-ray luminosity, which can be expressed as:  
\begin{equation}
L=n_t n_e \Lambda(T)V
\label{xray_lum}
\end{equation}
where $n_t$ and $n_e$ are the ion and the electron number densities respectively, $\Lambda(T)$ is the cooling function and $V$ is the volume of the X-ray emitting plasma. From a given $L$, $\Lambda(T)$ and $V$, one can calculate the gas density $n=(n_t n_e)^{0.5}$ by inverting eq.\ref{xray_lum}. Given the plasma temperatures found by \cite{T05}, we derived $\Lambda(T)$ by interpolating the CIE curve calculated by \cite{Suth93} at two different metallicities corresponding to $100\%$ and $30\%$ the solar metallicity. Then we assumed for the shock gas an emitting volume equal to the projected shock ridge area ($A=330~{\rm kpc^2}$) multiplied by a parametrized line of sight depth $l$, and for the halo we adopted an emitting sphere of radius $40~{\rm kpc}$, corresponding to $90''$ at $94~{\rm Mpc}$ distance. The luminosity values we used are those inferred by \cite{T05} ($L_{\rm X}^{\rm shock}\approx2.4\times10^{41}~{\rm erg/s}$ and $L_{\rm X}^{\rm halo}\approx2.3\times10^{41}~{\rm erg/s}$), properly scaled to the different assumed distance ($85~{\rm Mpc}$ in that paper vs $94~{\rm Mpc}$ here). The results for different choices of the emitting volumes and metallicity are shown in Table \ref{xray_dens}. For the shock region plasma we obtained densities in the range $4-18\times10^{-3}~{\rm cm^{-3}}$. As expected, the minimum density is obtained for higher metallicity (that is, higher $\Lambda$) and largest volume, while the maximum density corresponds to lower metallicity (lower $\Lambda$) and smallest volume. For the halo gas, the inferred densities are $1.1\times10^{-4}$ and $1.6\times10^{-4}~{\rm cm^{-3}}$ for  $Z=Z_\odot$ and $Z=0.3~Z_\odot$ respectively.

\section{IRAC and MIPS color corrections}
From the IRAC and MIPS data handbook, the color corrections for each Spitzer band are defined as:
\label{ap4}
\begin{eqnarray}
K=\frac{\int\left( F_\nu / F_{\nu o}\right)\left(\nu / \nu_o\right)^{-1} Rd\nu }{\int\left(\nu/\nu_o\right)^{-2}Rd\nu}~(IRAC)\\
K=\frac{\int\frac{F_\lambda}{F_{\lambda o}}R_\lambda d\lambda}{\int\left(\frac{\lambda_o}{\lambda}\right)^5 \frac{e^{\frac{hc}{\lambda_o kT_o}}-1}{e^{\frac{hc}{\lambda K T_o}}-1}\lambda R_\lambda}~(MIPS) 
\end{eqnarray}
where the subscripts $\nu_o$ and $\lambda_o$ refer to the band reference frequency or wavelength, $R$ is the instrument spectral response and $T_o=10.000~{\rm K}$ (these formulas derive from the conventions used to calculate the quoted fluxes on the Spitzer maps).

\section{Estimate of an upper limit to the dust injection rate from halo stars in SQ}
\label{ap5}
An upper limit to the dust injection rate from halo stars in SQ can be inferred from the halo R-band surface brightness, $S_{\rm R}=24.4~{\rm mag/arcs^2}$ (\citealt{Moles98}), and using theoretical predictions for the stardust injection. We assumed that all the R-band halo luminosity is produced by $3~{\rm M_\odot}$ stars.   According to table A.1 in \cite{Zhuk08}, a $3~{\rm M_\odot}$ star with metallicity $Z=0.02$ injects $M_{\rm d}=1.2\times10^{-2} ~{\rm M_\odot}$ of dust during its all life. Given the lifetime of such  star, $\tau_{\rm life}\approx 3\times 10^8~{\rm yr}$, one can derived the average dust injection rate: $\dot{M_{\rm d}}=M_{\rm d}/\tau_{\rm life}=3.5\times10^{-11}~{\rm M_\odot/yr}$. Since a star spends most of its life time in the main sequence phase, we approximate the R-band flux of a single halo star as that of a blackbody sphere with parameters characteristic of a main sequence $3~{\rm M_\odot}$ star: 
\begin{equation}
F_{\rm R}=\pi B_{\rm R}(T)\frac{R^2}{d^2}
\end{equation}
where $B_{\rm R}(T)$ is the blackbody R-band flux density at temperature $T=12000~{\rm K}$, $R=2R_\odot$ is the star radius and $d=94~{\rm Mpc}$ is the distance from SQ. The prediction for the R-band flux density produced by a single halo star and observed from the earth is $F_{\rm R}=1.35\times10^{-23}~{\rm ergs/cm^2/s/A}$. Dividing $\dot{M_{\rm d}}$ to $F_{\rm R}$, we obtain the dust injection rate per observed R-band flux density: $\dot{M_{\rm d}}/F_{\rm R}=2.6\times10^{12}~{\rm M_\odot/yr(ergs/cm^2/s/A)^{-1}}$. Combining the observed R-band surface brightness $S_{\rm R}$ and the solid angle $\Omega$ covering the halo region of SQ, one can obtain the total received flux $F_{\rm R}^{\rm obs}=2.9\times 10^{-14}~{\rm ergs/cm^2/s/A}$. Then the dust injection rate from halo stars is equal to: 
\begin{equation}
\dot{M_{\rm d}}=\left(\frac{\dot{M_{\rm d}}}{F_{\rm R}}\right)F_{\rm R}^{\rm obs}
\end{equation}
Using this simple approach we found $\dot{M_{\rm d}}=0.075 {\rm M_\odot/yr}$. This value should be considered as an upper limit since one can show, using the theoretical predictions for low--intermediate mass star dust injection rates in \cite{Zhuk08}, that $3~{\rm M_\odot}$ stars have the highest value of the dust injection rate per R-band flux $\dot{M_{\rm d}}/F_{\rm R}$.  

\begin{deluxetable}{lcccc}
\tablecolumns{5} 
\tablewidth{0pt} 
\tabletypesize{\scriptsize}
\tablecaption{Comparison of photometry results at $24{\rm \mu m}$ obtained with aperture photometry and the gaussian fitting technique}
\tablehead{\colhead{Source}&\colhead{$F_{24{\rm \mu m}}^{\rm ap}$}&\colhead{$F_{24{\rm \mu m}}^{fit}$}&\colhead{$\Delta_{70{\rm \mu m}}^{fit}$}&\colhead{$\Delta_{24{\rm \mu m}}^{\rm fit}$}\\
& \colhead{(mJy)} & \colhead{(mJy)} & \colhead{(arcs)} & \colhead{(arcs)}}
\startdata
SQ A&  $11\pm2$& $12\pm0.5$& $16$& $15$ \\
HII SE& $7.5\pm 1$& $6.7\pm0.4$& $14$& $12$ \\
HII SW\tablenotemark{a}& $2.4\pm0.35$& $6\pm0.6$& $11$& $24$ \\
SQ B&  $5.9\pm0.8$& $5.6\pm0.8$& $15$& $12$ \\
NGC 7319\tablenotemark{b}& $185\pm8$ & $195\pm8$ & $12$ & $12$ \\ 
NGC 7320\tablenotemark{b}& $38\pm2$ & $40\pm2$ & $44$ & $43$ \\
HII N & $1.1\pm 0.2$& ND\tablenotemark{c} & $22$ & ND\tablenotemark{c} \\
\enddata
\tablecomments{Col. 1: Source name; col. 2: aperture photometry fluxes; col. 3: fluxes from the gaussian fitting technique ; col. 4: average source FWHMs as derived from the $70{\rm \mu m}$ map fitting ; col. 5: average source FWHMs as derived from the $24{\rm \mu m}$ convolved map fit. }
\tablenotetext{a}{This source is the only source, detected by the fitting technique, whose inferred fluxes are considerably different. The size of the gaussian that fit the emission on the $24{\rm \mu m}$ convolved map is much larger than that found at $70{\rm \mu m}$. The reason for this is the contamination from the nearby nucleus of NGC 7318a that emits strongly in the MIR but not in the FIR. }
\tablenotetext{b}{For NGC 7319 (the AGN galaxy) and NGC 7320 (the foreground galaxy) the apertures are chosen such to cover the entire emission of the galaxies. In the fitting procedure these sources are modeled with two gaussians, one for the centrally peaked emission and one for a peripherical HII region. The sizes shown in column 4 and 5 refer to the central source that contribute most of the flux.} 
\tablenotetext{c}{ND: non detected}
\label{gausfit24}
\end{deluxetable}

\begin{deluxetable}{lccccc}
\tablecolumns{6} 
\tablewidth{0pt} 
\tabletypesize{\scriptsize}
\tablecaption{X-ray gas parameters for the shock and the halo components}
\tablehead{\colhead{Source}& \colhead{$Z$} & \colhead{$\Omega$} & \colhead{$los / R $} & \colhead{$n$} & \colhead{$M_{\rm gas}$} \\
& \colhead{($Z_\odot$)}& \colhead{$\left(10^{-23}~\frac{{\rm erg~cm}^3}{{\rm s}} \right)$} & \colhead{(kpc)} & \colhead{(${\rm cm}^{-3}$)} &\colhead{($10^9~{\rm M_\odot}$)}} 
\startdata
SHOCK & $1$ & $ 3.82$  & $5$  & $0.012 $ & $ 0.65$ \\
SHOCK & $1$ & $ 3.82$  & $10$ & $0.009$ & $ 0.92$ \\
SHOCK & $1$ & $ 3.82$  & $20$ & $0.006$ & $1.30$ \\
SHOCK & $1$ & $ 3.82$  & $40$ & $0.004$ & $1.84$ \\
SHOCK & $0.3$ & $1.92$ & $5$ &  $0.018$ & $0.92$ \\
SHOCK & $0.3$ & $1.92$ & $10$ & $0.012$ & $1.30$ \\
SHOCK & $0.3$ & $1.92$ & $20$ & $0.009$ & $1.83$ \\
SHOCK & $0.3$ & $1.92$ & $40$  &$0.006$ & $2.60$ \\
HALO  & $1$ & $2.54$ & $40$ & $0.0011$ & $10.4$ \\ 
HALO  & $0.3$ & $1.39$ & $40$ & $0.0016$ & $14.4$\\ 
\enddata
\tablecomments{Col. 1: Source; col. 2: assumed metallicity; col. 3: cooling rate; col. 4: line of sight depth (for the shock), or sphere radius (for the halo); col. 5: gas number density defined as $n=(n_{\rm t} n_{\rm e})^{1/2}$; col. 6: gas mass.}  
\label{xray_dens}
\end{deluxetable}

\end{document}